\documentclass[rd,showpacs,twocolumn]{revtex4-1}
\usepackage{amsfonts,amsmath,amssymb,amsthm}
\usepackage[bookmarks, bookmarksopen, bookmarksnumbered]{hyperref}

\usepackage{graphicx}
\usepackage{color}
\usepackage[utf8]{inputenc}
\usepackage[T1]{fontenc}
\usepackage{lmodern}
\usepackage[right]{rotating}
\usepackage{longtable}
\usepackage{array}
\usepackage{multirow}
\usepackage{paralist}

\newcommand{\codename}[1]{\texttt{#1}}
\newcommand{\beq}[1]{\begin{equation} #1 \end{equation}}

\newcommand{\GRHydro}{\codename{GRHydro}~}
\newcommand{\GRHydroNoSpace}{\codename{GRHydro}}

\graphicspath{{pdf/}{png/}}

\begin{document}
\title{Modeling Equal and Unequal Mass Binary Neutron Star Mergers Using Public Codes}
\date{\today}

\author{Roberto \surname{De Pietri}}
\affiliation{Parma University and INFN Parma, Parco Area delle Scienze 7/A, I-43124 Parma (PR), Italy}
\author{Alessandra \surname{Feo}}
\affiliation{Parma University and INFN Parma, Parco Area delle Scienze 7/A, I-43124 Parma (PR), Italy}
\author{Francesco  \surname{Maione}}
\affiliation{Parma University and INFN Parma, Parco Area delle Scienze 7/A, I-43124 Parma (PR), Italy}
\author{Frank \surname{L\"offler}}
\affiliation{Center for Computation \& Technology, Louisiana State University, Baton Rouge, LA 70803 USA}

\begin{abstract}

We present three-dimensional simulations of the dynamics of binary neutron star
(BNS) mergers from the late stage of the inspiral process up to $\sim 20$ ms
after the system has merged, either to form a hyper-massive neutron star (NS)
or a rotating black hole (BH). We investigate five equal-mass models of total
gravitational mass $2.207$, $2.373$, $2.537$, $2.697$ and $2.854\ M_\odot$,
respectively; and four unequal mass models with $M_{\mathrm{ADM}}\simeq 2.53\
M_\odot$ and $q\simeq 0.94$, $0.88$, $0.83$, and $0.77$ (where $q =
M^{(1)}/M^{(2)}$ is the mass ratio). We use a semi-realistic equation of state
(EOS), namely the seven-segment piece-wise polytropic SLyPP with a thermal
component given by $\Gamma_{th} = 1.8$. We have also compared the resulting
dynamics (for one model) using both, the BSSN-NOK and CCZ4 methods for the
evolution of the gravitational sector, and also different reconstruction
methods for the matter sector, namely PPM, WENO and MP5. Our results show
agreement and high resolution, but superiority of BSSN-NOK supplemented by WENO
reconstruction at lower resolutions.

One of the important characteristics of the present investigation is that for
the first time, it has been done using only publicly available open source
software: the Einstein Toolkit code, deployed for the dynamical evolution; and
the LORENE code, for the generation of the initial models. All of the source
code and parameters used for the simulations have been made publicly available.
This not only makes it possible to re-run and re-analyze our data, but also
enables others to directly build upon this work for future research.

\end{abstract}

\LTcapwidth=\columnwidth


\maketitle

\section{Introduction}
\label{sec:intro}

The new generation of ground-based laser interferometer gravitational wave
observatory Advanced LIGO~\cite{TheLIGOScientific:2014jea} and Advanced
Virgo~\cite{TheVirgo:2014hva} have just now opened a new window in the Universe
with the first detection~\cite{FirstDetection} of Gravitational Waves (GW)
emitted by the merger of two black holes. Among other likely sources of GW
signals in the sensitive frequency band of ($40-2000$) Hz are signals from
binary neutron star (BNS) mergers, with expected event rates reaching $\approx
0.2-200$ per year between $2016-19$~\cite{LIGOVIRGO:2013,Abadie:2010cf}.
Signals from these events are expected to contain the signature of the equation
of state (EOS) governing matter at nuclear density.

In the description of the BNS mergers are involved essentially three stages,
the inspiral, the merger and the final evolution to its final state
(post-merger stage) that would quite likely be a final black hole (BH)
surrounded by an accretion disk. The inspiral phase can be modeled with good
accuracy by post-Newtonian calculations and, in particular, using the Effective One
Body (EOB) approach~\cite{buonanno:1999effective}. They are capable of
producing accurate waveforms up to a time very close to the merger. More
recently, the EOB approach started to be used to model tidal corrections~\cite{damour:2010effective,bini:2012effective}. These analytic techniques are
useful for quickly computing waveform templates to matched filtering searches
in GW detector data analysis. The role of Numerical Relativity (NR) in this
regime is mainly to test and help improve the properties of these analytic
techniques. However, one has to keep in mind that such comparison and
improvement would need the use of NR data extrapolated at infinite resolutions
or at a resolution where discretization errors are negligible. To do this
analysis, the convergence properties of the used code during
the inspiral phase needs to be known. For the post-merger stage, NR is the only
available investigation tool to confront the experimental result that would be
obtained by a successful LIGO/VIRGO detection with the underlying physics of
neutron stars (NS). As pointed out in~\cite{Takami:2014tva,Takami:2015gxa} and
in references therein, an accurate description of the GW emission of different
model sources (models inferred by different choice of the underlying
NS physics through different choices of EOS) are useful for developing
empirical relations to be able to infer NS parameters from future GW detections
as well as to get information on the correct EOS that describe matter at these
extreme conditions.

In this work, we concentrate on the information that can be extracted from BNS
simulations using the SLy EOS~\cite{Douchin01}, where a semi-realistic
seven-segment piece-wise (isentropic)-polytropic approximant
\cite{Read:2009constraints}, namely the SLyPP EOS, is used and whose different
parameters are reported in Table~\ref{table:PPSLy}. Such an approximant covers
the whole range of density simulated, starting from the NR {\it atmosphere} of
$6.18\cdot 10^6$ $\mathrm{g}/\mathrm{cm}^3$ up to the density in the interior
of the NS supplemented by an artificial thermal component to ensure that the
hydro-dynamical evolution is consistent with the Energy-Momentum tensor
conservation. To such extent, we simulated five equal mass BNS systems and four
unequal mass systems, and for each of these systems we simulated the late-inspiral
phase (last three to five orbits, depending on the model) and post-merger
using at least three different resolutions and the state-of-the-art WENO
fifth-order reconstruction method~\cite{Harten:1987un,Shu:1999ho}. We evolved
the gravitational part of the system using the well-tested BSSN-NOK evolution
scheme~\cite{Nakamura:1987zz,Shibata:1995we,Baumgarte:1998te,Alcubierre:2000xu,Alcubierre:2002kk}.
For one BNS system, we also compared the evolution using different
reconstruction methods (PPM and MP5, in addition to WENO) and a different
spacetime evolution scheme: CCZ4~\cite{Alic:2011gg}. We checked the influence
of the EOS by simulating BNS models that have the same total baryonic mass
but different EOS, in our case two simple one-piece
(isentropic)-polytropic EOSs with different stiffness ($\Gamma=2.75$ and
$\Gamma=3.00$). We did a complete convergence analysis of the computed merger
times of the BNS as a function of the resolution in the case of WENO and PPM
reconstruction (using BSSN-NOK evolution for the gravitational sector).

The organization of the paper is as follows. In Sec.~\ref{sec:setup} we
describe the properties of the BNS systems investigated in the present work, and
we review the numerical setup used. In Sec.~\ref{sec:results} we present and
discuss our results. Conclusions that can be drawn from the present
investigation are given in Sec.~\ref{sec:conclusions} and, finally,
the computational cost of the simulations is discussed in the appendix.

Throughout this paper we use a space-like signature $-,+,+,+$, with Greek
indices running from 0 to 3, Latin indices from 1 to 3, and the standard
convention for summation over repeated indices. The computations are performed
using the standard $3+1$ split into (usually) space-like coordinates
$(x,y,z)=x^i$ and a time-like coordinate $t$. Our coordinate system
$(x^\mu)=(t,x^i)=(t,x,y,z)$ (far from the origin) has, as it can be checked,
almost isotropic coordinates and far from the origin they would have the usual
measure unit of ``time'' and ``space''. In particular, $t$ is time when
measured from an observer at infinity.

All computations have been done in normalized computational units (hereafter
denoted as CU) in which $c=G=M_\odot=1$. We report all results in cgs units
except for values of the polytropic constant $K$, whose unit of measurement
depends on the value of the dimensionless polytropic exponent $\Gamma$, so we
report $K$ in the above defined normalized unit CU). CU are also used to denote
resolutions, e.g., $dx=0.25$ CU, and there they mean the resolution on the
finest grid at initial time (which for most cases is the same for the entire
evolution). We also report masses in terms of the solar mass $M_\odot$. The
reader should note that, as is usual in most of the work on this subject, we
describe matter using the variable $\rho$ (baryon mass density), $\epsilon$
(specific internal energy) and $P$, instead of the typical notation used in
astrophysics, $\overline{\rho}$ (energy density), $\overline{n}$ (baryon number
density) and $P$. Their relation is the following: $\overline{\rho} = e = \rho
(1 + \epsilon)$ and $\overline{n}=\rho/m_B$ ($m_B$ is the baryon mass).

\section{Initial Models and Numerical Methods}
\label{sec:setup}

\begin{figure}
\begin{centering}
  \includegraphics[width=0.45\textwidth]{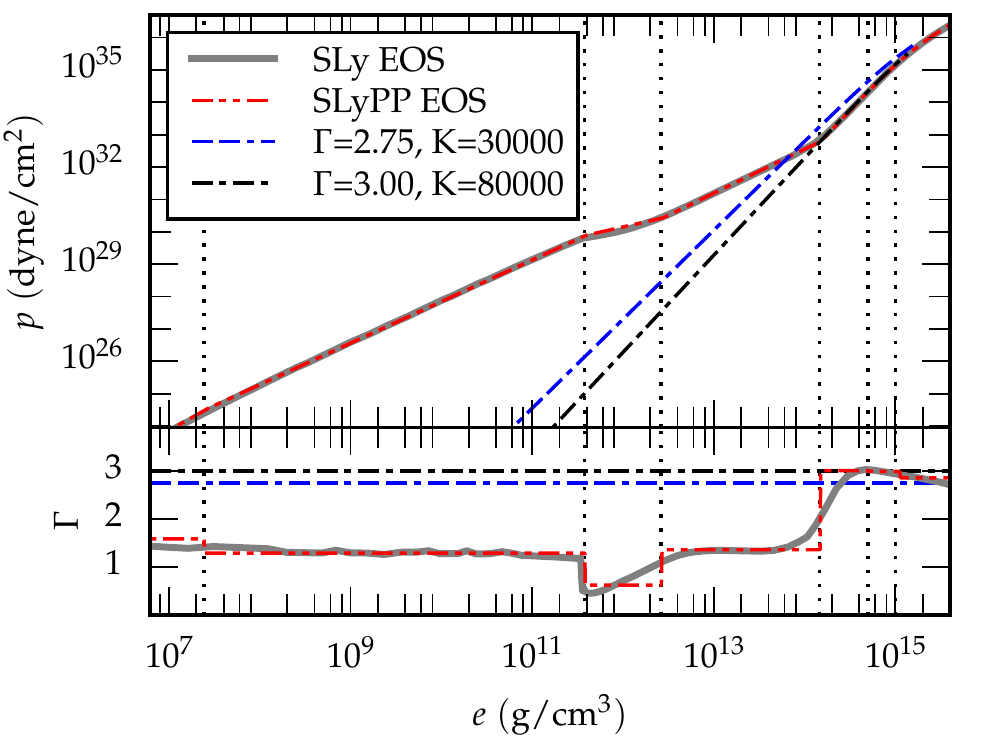}\\
\end{centering}
\vspace{-2mm}
\caption{Plot of the pressure ($P$) and of the adiabatic index
($\Gamma=d\log(P)/d\log(\rho)$) as a function of the energy density
($e=\rho(1+\epsilon)$) for the SLy EOS, its piece-wise polytropic approximation
(the one used in the present work) and two isentropic polytropic EOS $P=K
\rho^\Gamma$.}
\label{fig:EOS}
\end{figure}

\begin{table}
\begin{tabular}{lllll}
  \multirow{2}{*}{i}  & \multirow{2}{*}{$\Gamma_i$}&  $\rho_i$               & $\rho_i$   \\
     &          &   (CU)                  & ($g/cm^3$) \\
\hline
0    & 1.58425 & $1\times10^{-11}$         & $\simeq 6.18\times 10^{6}$\\
1    & 1.28733 & $3.951156\times10^{-11}$  & $\simeq 2.43\times 10^{7}$\\
2    & 0.62223 & $6.125960\times10^{-7}$   & $\simeq 3.78\times 10^{11}$\\
3    & 1.35692 & $4.254672\times10^{-6}$   & $\simeq 2.63\times 10^{12}$\\
4    & 3.005   & $2.367449\times10^{-4}$   & $\simeq 1.46\times 10^{14}$\\
5    & 2.988   & $8.114721\times10^{-4}$   & $\simeq 5.01\times 10^{14}$\\
6    & 2.851   & $1.619100\times10^{-3}$   & $\simeq 1.00\times 10^{15}$\\
\hline
\end{tabular}
\caption{Parameters used for the seven-segment piece-wise polytropic SLyPP EOS
(see \cite{Read:2009constraints}) used to represent the SLy EOS described in
\cite{Douchin01}. The thermal component is described by $\Gamma_\textrm{th} =
1.8$. The values $K_i$ are chosen using $K_0=1.685819\times10^2$ CU while the
remaining values $K_i$ (with $i>0$) are calculated to produce a continuous EOS.
The value $\rho_0$ is the setting of the atmosphere used in our simulations.
\label{table:PPSLy}}
\end{table}

In the present work we have analyzed the dynamics of the merger of two NSs from
the late stage of the inspiral process to around 20 ms after the system has
merged to form a hyper-massive NS or a rotating BH. To be properly described,
such a system needs the use of the Einstein's GR equations to
describe the metric ($g_{\mu\nu}$) of the dynamical spacetime and a proper
treatment for the description of matter (see~\cite{Loffler:2011ay} for more details
on the employed equations). For describing the matter present in
the system we use its perfect-fluid approximation, and close the system of
evolution equations using an EOS of
the form $P=P(\rho,\epsilon)$, where $P$ is the pressure, $\rho$ is the rest
mass density, and $\epsilon$ is the specific internal energy of the matter.
In the present work we used a seven-segment
isentropic polytropic approximant (what we refer to here as SLyPP) of the widely
used SLy EOS prescription~\cite{Douchin01} supplemented by a thermal component
described by $\Gamma_\textrm{th} = 1.8$ (see and Figure\ \ref{fig:EOS}).

More precisely, $P(\rho,\epsilon) = P_{\mathrm{cold}}(\rho) + P_{\mathrm{th}}(\rho,\epsilon)$, and in each density region $\rho_i \leq \rho <
\rho_{i+1}$:

\begin{eqnarray} 
 P_{\mathrm{cold}}        &=& K_i \rho^{\Gamma_i} \\
 \epsilon_{\mathrm{cold}} &=& \epsilon_i + \frac{K_i}{\Gamma_i-1}\rho^{\Gamma_i-1} \\
 P_{\mathrm{th}}          &=& \Gamma_{\mathrm{th}} \rho (\epsilon - \epsilon_{\mathrm{cold}}),
\end{eqnarray}

where $\epsilon_i$ and $K_i$ are chosen to ensure the pressure and specific
energy density continuity. Fig.~\ref{fig:EOS} shows a plot of the pressure
($P$) and of the adiabatic index ($\Gamma$) as a function of the energy density
$e$ for the SLy EOS (gray line)~\cite{Douchin01} and its piece-wise polytropic
approximation, used in the present work (dashed red line, SLyPP)
\cite{Read:2009constraints} supplemented by a thermal component
$\Gamma_\textrm{th} = 1.8$. Also shown are two often-used isentropic EOS
with $P = K \rho^{\Gamma}, \Gamma\in\{2.75, 3.00\}$. These can be seen as a
single-piece approximant of the same SLy EOS (see
\cite{DePietri:2014mea,Loffler:2014jma}), and have been used for test cases here.
The exact parameters of the
seven-segment isentropic polytropic approximant are shown in
Table~\ref{table:PPSLy}  and, as can be seen in the panels of
Fig.~\ref{fig:EOS} where the vertical dots lines shown the transition of the
approximant between the various region, they represent various physically
distinct parts of the star.

Following the discussion of~\cite{bauswein:2010testing} we have chosen to use
the arbitrary choice of $\Gamma_{\mathrm{th}}=1.8$. This choice has a few
drawbacks, the main one being that in the low-density region it should approach
$4/3$, since there the pressure is provided primarily by an ideal gas of
ultra-relativistic electrons and photons, so we overestimate the pressure
support. On the other hand (see~\cite{bauswein:2010testing}), a value of
$\Gamma_{\mathrm{th}}=2.0$ seems to be too high, while a value of
$\Gamma_{\mathrm{th}}=1.5$ might yield a too low pressure support in the core.

All equations are solved as evolution equations with respect to a coordinate
time $t$, using the common 3+1 decomposition of space-time
(see~\cite{Loffler:2011ay} and references therein for details and notation).

The initial data of our simulations is calculated using the LORENE
code~\cite{Gourgoulhon:2000nn,lorene:web} that provides the possibility to
generate arbitrary initial data for irrotational BNS.

\begin{table} 
\begin{tabular}{lccccccc}
\multirow{2}{*}{name}&$M_0^{(1)}$&$M_0^{(2)}$& $M^{(1)}$ & $M^{(2)}$ &  $\Omega$ & $\!M_\mathrm{ADM}$ & $\!J$ \\ %& $t_\mathrm{merger}^{EOB}$\\
            & $[M_\odot]$ & $[M_\odot]$ & $[M_\odot]$ & $[M_\odot]$ &  
  $\left[\frac{\mathrm{krad}}{s}\right]$ & $[M_\odot]$ & $\!\!\left[\frac{GM_\odot^2}{c}\right]$ \\ %& [ms] \\\hline
\hline
 SLy12vs12  & 1.20 & 1.20 & 1.11 & 1.11 & 1.932 & 2.207 & 5.076 \\ %% & 22.933\\
 SLy13vs13  & 1.30 & 1.30 & 1.20 & 1.20 & 1.989 & 2.373 & 5.730 \\ %% & 18.499\\
 SLy14vs14  & 1.40 & 1.40 & 1.28 & 1.28 & 2.040 & 2.536 & 6.405 \\ %% & 15.220\\
 SLy15vs15  & 1.50 & 1.50 & 1.36 & 1.36 & 2.089 & 2.697 & 7.108 \\ %% & 12.708\\
 SLy16vs16  & 1.60 & 1.60 & 1.44 & 1.44 & 2.134 & 2.854 & 7.832 \\ %% & 10.765\\
\hline
SLy135vs145 & 1.35 & 1.45 & 1.24 & 1.32 & 2.040 & 2.536 & 6.397 \\ %% & 15.243\\
SLy13vs15   & 1.30 & 1.50 & 1.20 & 1.36 & 2.040 & 2.535 & 6.376 \\ %% & 15.268\\
SLy125vs15  & 1.25 & 1.55 & 1.16 & 1.40 & 2.040 & 2.533 & 6.337 \\ %% & 15.352\\
SLy12vs16   & 1.20 & 1.60 & 1.11 & 1.44 & 2.039 & 2.531 & 6.281 \\ %% & 15.464\\
\hline
G275th14vs14& 1.40 & 1.40 & 1.29 & 1.29 & 2.053 & 2.554 & 6.513 \\ %% & 14.581 \\
G300th14vs14& 1.40 & 1.40 & 1.26 & 1.26 & 2.028 & 2.498 & 6.243 \\ %% & 15.903 \\
\end{tabular}
\caption{Properties of the initial irrotational BNS models simulated in the
present work. All these models were generated using the public LORENE code
\cite{lorene:web}. The columns show, in this order, the baryonic masses of the
two stars ($M_0^{(1)}$, $M_0^{(2)}$), their gravitational masses ($M^{(1)}$,
$M^{(2)}$) at infinite separation, the initial rotational angular velocity
($\Omega$), the total initial ADM mass ($M_{\mathrm{ADM}}$), and the angular
momentum ($J$). The notation for the model names is as in the following
example: SLy13vs15 denotes the SLyPP EOS and baryonic masses equal to $1.3
M_\odot$ and $1.5 M_\odot$, respectively, while, G275th14vs14 means polytropic
($\Gamma$-law) EOS with $\Gamma=2.75$.
\label{TAB:InitialData}}
\end{table}

The properties of the initial data we simulated are summarized in Table
\ref{TAB:InitialData} where are reported, for each model, the baryonic masses
of the two stars ($M_0^{(1)}$, $M_0^{(2)}$), their gravitational masses
($M^{(1)}$, $M^{(2)}$) at infinite separation, the initial rotational angular
velocity ($\Omega$), and the total mass ($M_\mathrm{ADM}$) and angular momentum
(J). All these models were generated to be at a relative physical distance of
$40$ km.  We generated five equal-mass models of total gravitational mass
$2.207$, $2.373$, $2.537$, $2.697$ and $2.854 M_\odot$, respectively, and four
unequal mass models with $M\simeq 2.53\ M_\odot$ and $q\simeq 0.94$, $0.88$,
$0.83$, and $0.77$ (where $q = M^{(1)}/M^{(2)}$).

One of the main characteristics of the present investigation is that it can be
reproduced using only freely available open source software. The GNU licensed
LORENE code~\cite{lorene:web} has been used for generating the initial
condition, and the Einstein Toolkit~\cite{Loffler:2011ay,EinsteinToolkit:web}
was deployed for the dynamical evolution. The Einstein Toolkit is a free,
publicly available, community-driven general relativistic (GR) code. In
particular we have chosen the eleventh release (code name ``Hilbert'',
ET\_2015\_05). Some local modification and additions were necessary, all of
which are open-source and freely available from the gravity SVN server of Parma
University, and all of which are planned to be proposed to be in the next
release of the Einstein Toolkit (see appendix).

The Einstein Toolkit is based on the \codename{Cactus} Computational
Toolkit~\cite{Cactuscode:web,Goodale:2002a,CactusUsersGuide:web}, a software
framework for high-performance computing (HPC\@).  Its main tools used in the
present study are:
\begin{itemize}
\item
     The adaptive mesh refinement (AMR) methods implemented by
     \codename{Carpet}~\cite{Schnetter:2003rb, Schnetter:2006pg,CarpetCode:web}.
\item
     Hydrodynamic evolution techniques provided by the \GRHydro
     package~\cite{Baiotti:2004wn,Hawke:2005zw,Moesta:2013dna}.
\item
     The evolution of the spacetime metric handled by the \codename{McLachlan}
     package~\cite{McLachlan:web}.  This code is auto-generated by Mathematica
     using \codename{Kranc}~\cite{Husa:2004ip,Lechner:2004cs,Kranc:web},
     implementing the Einstein equations via a $3+1-$dimensional split using
     the BSSN-NOK
     formalism~\cite{Nakamura:1987zz,Shibata:1995we,Baumgarte:1998te,
     Alcubierre:2000xu,Alcubierre:2002kk} and CCZ4 \cite{Alic:2011gg}

\end{itemize}

All the main properties used in the deployed code are described in much more
detail in~\cite{Loffler:2014jma} for the Einstein Toolkit in general,
in~\cite{Moesta:2013dna} for especially some details about the hydrodynamics,
and in~\cite{DePietri:2014mea,Loffler:2014jma} for similar usage in previous
work. In particular:

\begin{itemize}
\item 
We used a fourth-order Runge-Kutta~\cite{Runge:1895aa,Kutta:1901aa} method with
Courant factor 0.25. Kreiss-Oliger dissipation was applied to the curvature
evolution quantities in order to damp high-frequency noise.

\item 
We use fourth-order finite difference stencils for the curvature evolution,
$1+\log$~\cite{Alcubierre:2002kk} slicing, and a $\Gamma$-driver shift
condition~\cite{Alcubierre:2002kk}.  During evolution, a Sommerfeld-type
radiative boundary condition is applied to all components of the evolved
BSSN-NOK variables as described in~\cite{Alcubierre:2000xu}.

\item
The HLLE (Harten-Lax-van Leer-Einfeldt) approximate Riemann
solver~\cite{Harten:1983on,Einfeldt:1988og} is used. The main reason for this
choice is that we are evolving also the magnetic field (even if it is set to
zero).

\item
An artificial low-density atmosphere with $\rho_{\text{atm}}=10^{-11}$ CU is
used, with a threshold of $\rho_{\text{atm\_reset}}=1.01\times10^{-11}$ CU
below which regions are set to be atmosphere.  Hydrodynamical quantities are
also set to be atmosphere at the outer boundary.

\item
All evolutions presented here use a mirror symmetry across the $(x,y)$ plane,
consistent with the symmetry of the problem, which reduces the computational
cost by a factor of $2$. 

\item
We use mesh refinement, with a fixed box-in-box topology for all simulations
except otherwise noted. The boundaries of the six levels (seven for simulations
involving a final black hole) are indicated in Tab.~\ref{tab:grid}.

\item 
Different reconstruction methods have been used, and their effects on the
results are shown in Sec.~\ref{sec:results}. The reconstruction
method used for most of the simulations is the essentially non-oscillatory method WENO (5\textsuperscript{th}
order weighted-ENO)~\cite{Harten:1987un,Shu:1999ho}, while the other
reconstruction methods employed are PPM (the piece-wise parabolic
reconstruction method)~\cite{Colella:1982ee}, and MP5 (5\textsuperscript{th}
order monotonicity preserving)~\cite{suresh:97}.

\item 
In contrast to~\cite{Loffler:2011ay}
and~\cite{DePietri:2014mea,Loffler:2014jma} we reconstruct instead of the
three-velocity the product of the three-velocity and the Lorentz factor to
prevent artificial velocities exceeding the speed of light due to numerical
errors.

\end{itemize}

The computational cost of the simulation is described in the
appendix~\ref{SEC:compCOST} and shown in Table~\ref{TAB:COSTs}, and it is
almost the same in the case of using different reconstruction methods
(WENO, PPM, MP5) or in the case of the CCZ4 method for the evolution of the
gravitational sector. From the data shown in Table\ref{TAB:COSTs} one may note
that simulations with resolution $dx=0.75$, CU can be performed on a
workstation while a whole simulation of a model at resolution $dx=0.25$ CU will
need current computational resources comparable to 256 cores for 6 days.

As will be shown in the next section, the fact that our code gives a reasonable
numerical result at very coarse resolution such as $dx=0.75$, CU has been very
valuable in setting the details of the simulations and on the specific choice
of the initial model to be studied.

\begin{figure*}
\begin{centering}
  \includegraphics[width=0.32\textwidth]{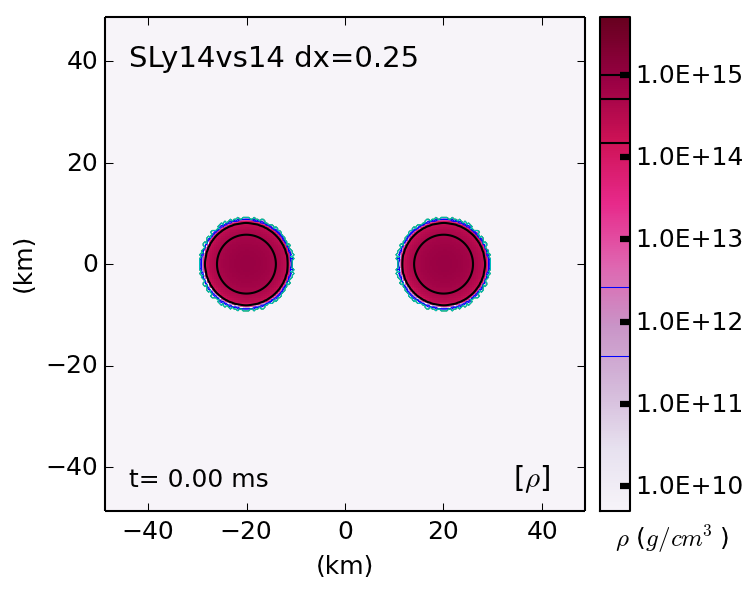}
  \includegraphics[width=0.32\textwidth]{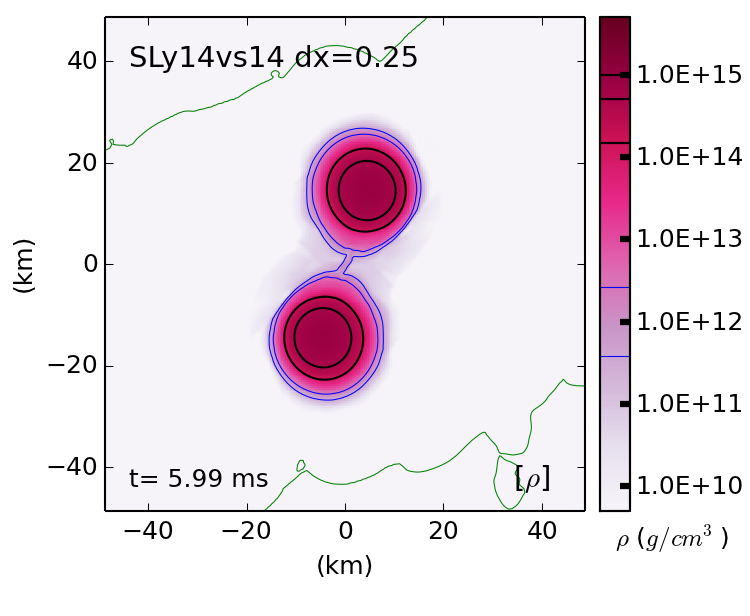}
  \includegraphics[width=0.32\textwidth]{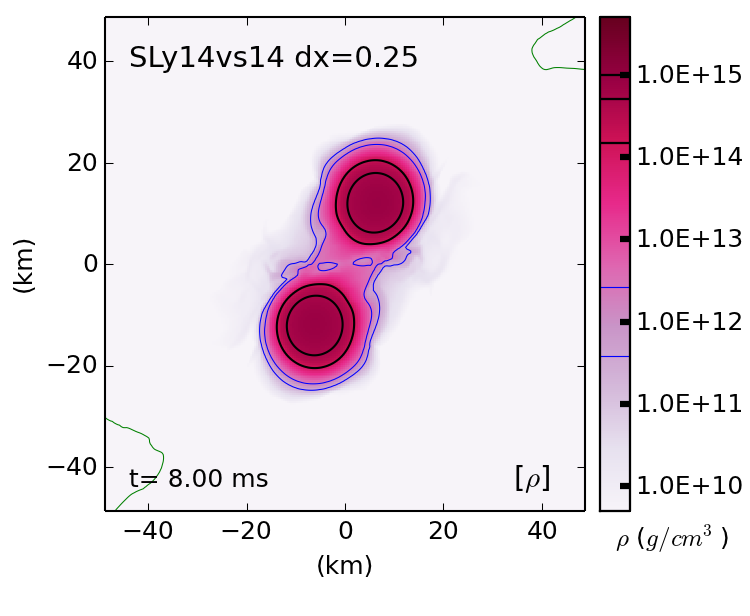}\\
  \includegraphics[width=0.32\textwidth]{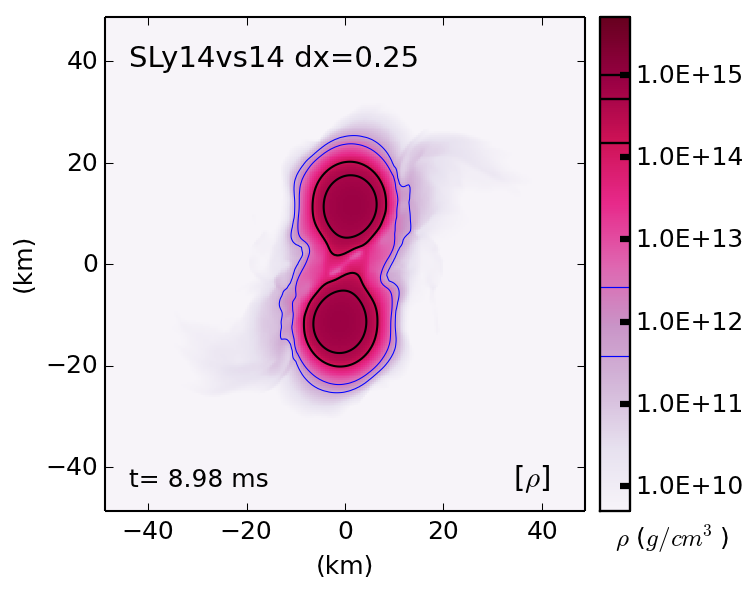}
  \includegraphics[width=0.32\textwidth]{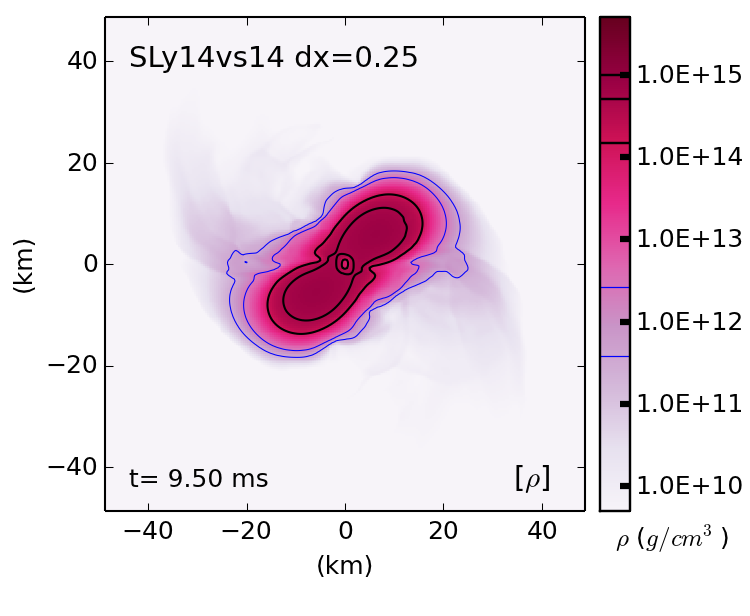}
  \includegraphics[width=0.32\textwidth]{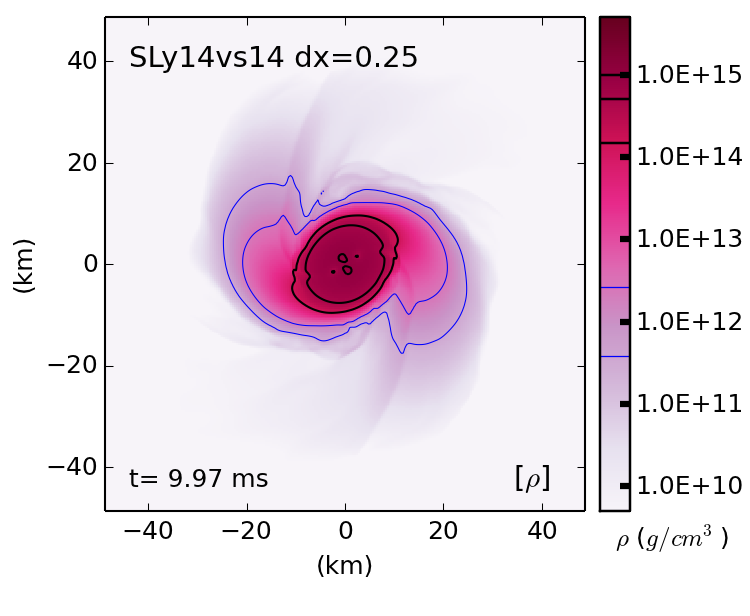}\\
  \includegraphics[width=0.32\textwidth]{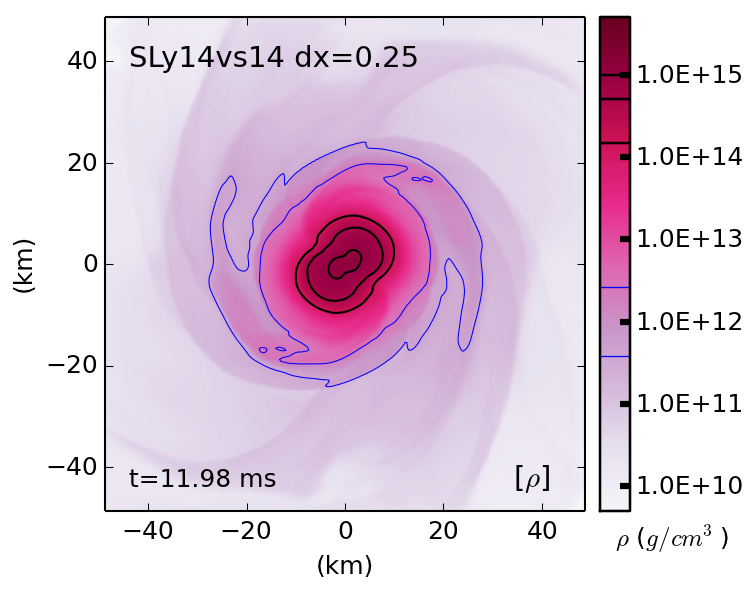}
  \includegraphics[width=0.32\textwidth]{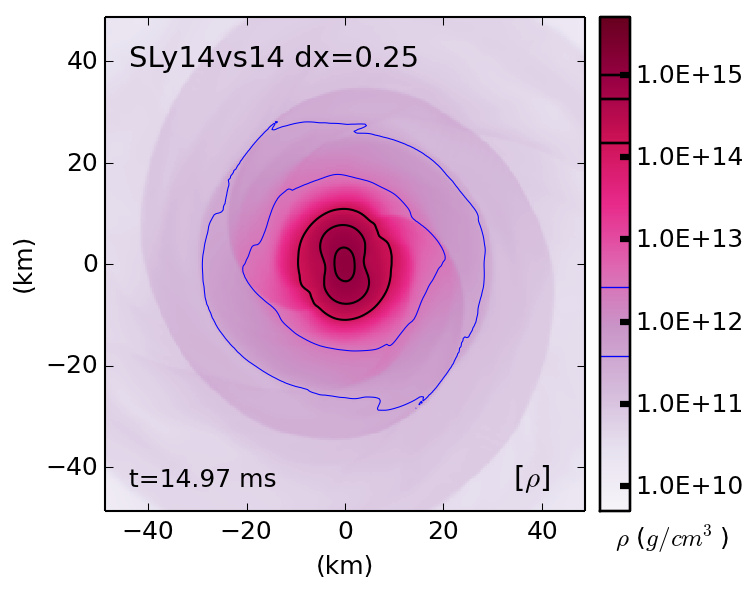}
  \includegraphics[width=0.32\textwidth]{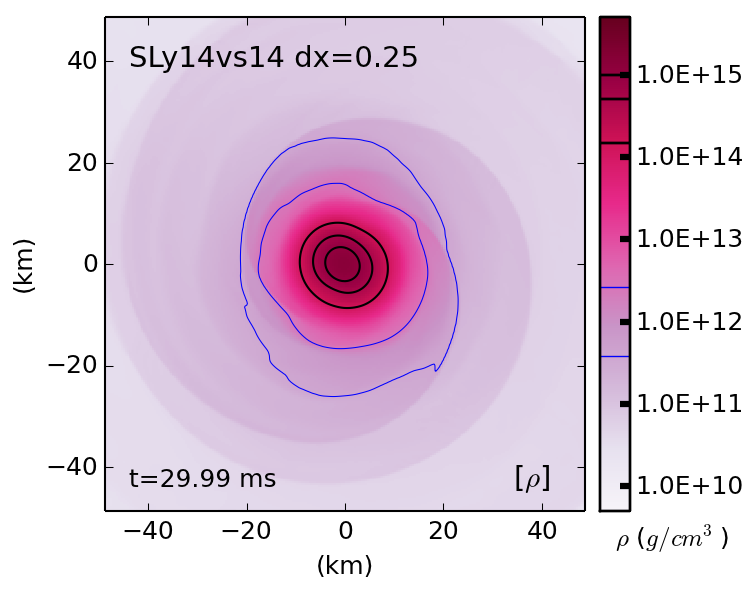}\\
\end{centering}
\vspace{-0.8mm}
\caption{
Dynamics of the evolution of model SLy14vs14. Shown using a color gradient is
the evolution of the matter density ($\rho$) at various times. The isocontour
lines denote the transition of the various pieces of the polytropic isentropic
approximant of the SLyPP EOS. The more external, blue isocontour represents the
neutron-drip ($\rho_2$) density.}
\label{fig:SNAPSHOT}
\end{figure*}

\section{Results}
\label{sec:results}

The general characteristics of the evolution of the systems is shown (for model
SLy14vs14, where 14vs14 stands for baryon masses $1.4 M_\odot$) in
Figure~\ref{fig:SNAPSHOT}, where we plotted the density in the $x-y$ plane
during various stages of the evolution. In the top panel the initial
configuration is shown, alongside a snap-shot depicting a sensible
tidal-deformation and the start of the actual merger phase, which then proceeds
on the middle panel. The after-merger relaxation phase can be seen at the
bottom panel, until the end of the simulation.

\begin{figure}
\begin{centering}
  \includegraphics[width=0.45\textwidth]{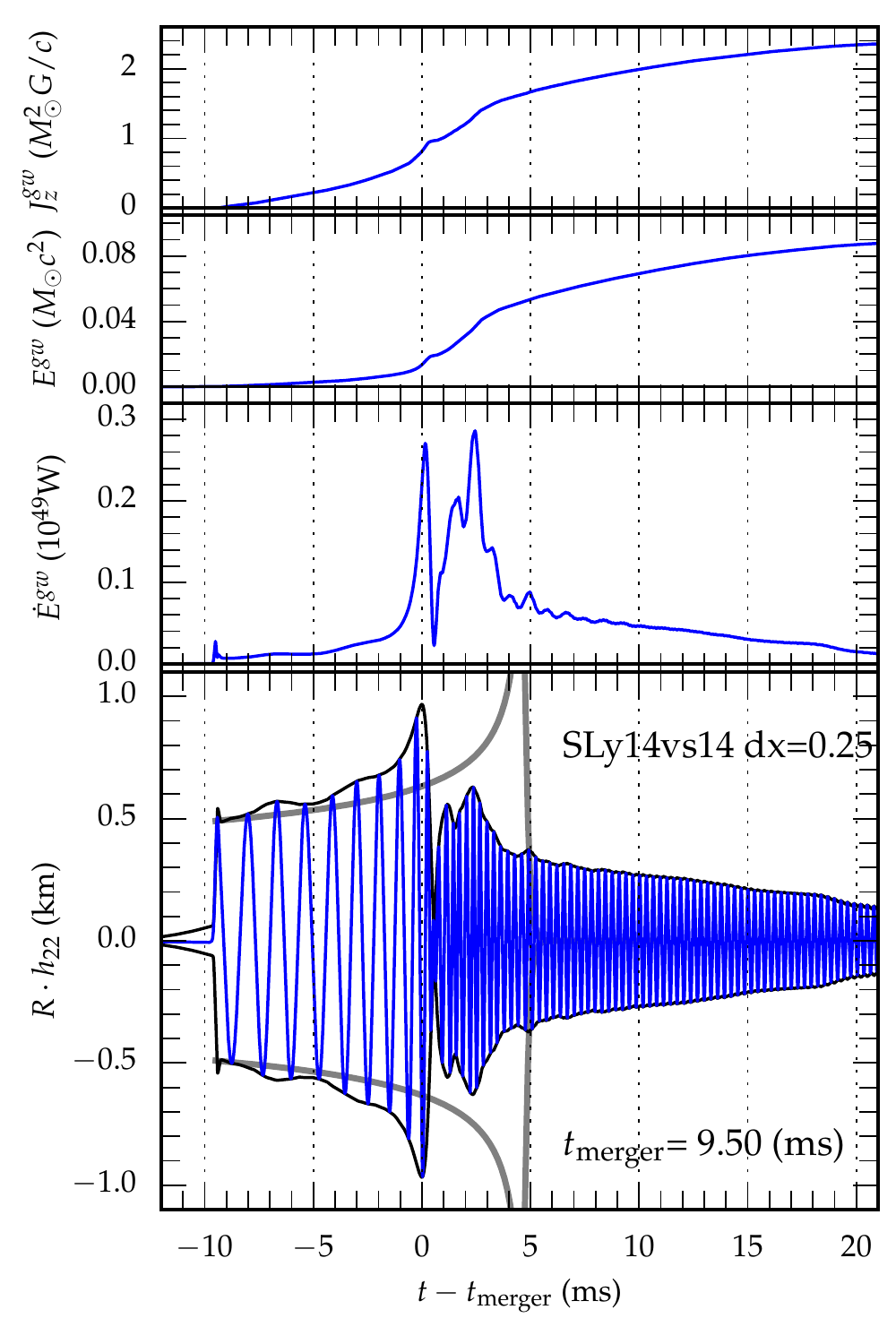}
\end{centering}
\caption{Results for the evolution for model SLy14vs14 at resolution $dx=0.25$
CU. The two top panels show the total angular momentum and total energy
momentum carried away by gravitational radiation. The third panel depicts the
total GW luminosity (energy flux). Finally, the bottom panel shows the envelop
of the GW amplitude and its real part, multiplied by the distance of the
observer to the origin. The thick gray line shows the envelop of the signal
expected from a binary black hole (BBH) system of the same masses (see
text).}\label{fig:Overview14vs14}
\end{figure}

The overall process is associated with emission of gravitational radiation and
a consequent loss of energy and angular momentum associated to such emission.
The procedure we follow to extract this information is described in detail in
the following (subsection~\ref{SEC:GW}). In the case of model SLy14vs14 we
have found that the maximum signal we obtain is in the $h_{22}$ component of
the GW signal, with the total amount of energy extracted from the GW emission
being of the order of $0.08\ M_{\odot} c^2$. The loss of angular momentum (in
the $z$ component) is of the order of $2\ M_\odot^2 G/c$ (see
Figure~\ref{fig:Overview14vs14}). Note especially that there is a clear peak on
the amplitude of the GW signal ($h_{22}$), which will later be used as an
indication of the merger time.

\subsection{Gravitational Wave Signal Extraction}
\label{SEC:GW}

From each simulation, we extracted the GW signal using both the
curvature-perturbation theory based on the Newman-Penrose scalar $\psi_4$
~\cite{Newman62} and the Regge-Wheeler~\cite{Regge:1957td} and
Zerilli~\cite{Zerilli:1970se} theory of metric-perturbations of the
Schwarzschild spacetime, in the gauge-invariant formulation by
Moncrief~\cite{Moncrief:1974am}.

The Newman-Penrose scalar $\psi_4$ is calculated by the Einstein Toolkit module
\codename{WeylScal4}, following the prescription of~\cite{Baker:2001sf}, and is
decomposed in spin-weighted spherical harmonics of spin $-2$ by the module
\codename{Multipole}:

\begin{equation}
 \psi_4(t,r,\theta,\phi)  = 
 \sum_{l=2}^{\infty} {\sum_{m=-l}^{l} {\psi_4^{lm}(t,r)\ {{}_{-2}\!}{Y}_{lm}(\theta,\phi)}}.
\end{equation}

We extracted $\psi_4$ components up to $l=5$. $\psi_4$ is linked to the GW
strain by the following relation, formally valid only at spatial infinity:
\beq{\psi_4\ =\ \ddot{h}_+\ -\ \ddot{h}_x} We therefore need to integrate each
component of $\psi_4^{lm}$ twice in time to get $h^{lm}$. We performed those
integrations numerically in the time domain using a simple trapezoidal rule.

\begin{equation}
\tilde{h}^{(0)}_{lm} (t) \ =\ \int_{0}^t{dt' \int_{0}^{t'}{dt'' \psi_4^{lm}(t'',r)}}.
\end{equation}

Please note that here we defined the $\tilde{h}^{(0)}_{lm}$ quantities in the
coordinate time of the detector and not with respect to the retarded time $t_\mathrm{ret}$
(see below for the definition of the retarded time) at the location of the
detector. Following the discussion of~\cite{Damour2008}, we 
integrated the signal twice from coordinate time $t=0$ and not from 
the zero of the retarded time ($t_\mathrm{ret}$) at the detector. 
We then perfom a linear fit of the strain $\tilde{h}^{(0)}_{lm}$ and
subtracted its result to the integrated signal to obtain
\begin{equation}
\tilde{h}_{lm}\ =\ \tilde{h}^{(0)}_{lm}\ -\ Q_1 t\ -\ Q_0,
\end{equation}
where the fit parameters have a clear physical interpretation as $Q_1 =
\dot{\tilde{h}}(t=0)$ and $Q_0 = \tilde{h}(t=0)$.

However, we found that the free choice of integration constants is not very
effective in eliminating the initial spurious radiation. The choice of starting
the integral from $t=0$ in coordinate time may be useful if complex fitting
procedures are used, or if the simulation time of the data is short; but this
is not the case for the present simulations. Starting the integral from $t=0$
also has the advantage of much smaller values of $Q_1$ and $Q_0$ to recover
with the fitting procedure, thus reducing the fit errors.

On the other hand, integrating from zero retarded time instead leads to
unphysical amplitude oscillations in the first $4 \, \mathrm{ms}$ of the
inspiral. These are not present with our final integration procedure and also
not in the signal from the Regge-Wheeler-Zerilli variables, which do not
require integrations. The resulting waveforms obtained with the procedure
described above show a left-over non-linear drift in $h(r,t)$, that was
attributed in~\cite{Reisswig:2011notes} to unresolved high-frequency noise
aliased in the low-frequency signal that does not result in a zero average in
the double integration. Instead of adopting the fixed-frequency integration
method recommended in~\cite{Reisswig:2011notes} (which we found to be too
sensible to the low-frequency cutoff choice) we performed a fit to a second
order polynomial as already done in~\cite{Baiotti:2009gravitational},
obtaining:

\begin{equation}
 \tilde{h}_{lm}\ =\ \tilde{h}^{(0)}_{lm}\ -\ Q_2 t^2\ -\ Q_1 t\ -\ Q_0.
\end{equation}

We found that this is sufficient to eliminate most of the non-physical drift in
the signal, at least for the dominant $l=2, m=\pm 2$ modes responsible for
almost all of the radiated energy and angular momentum.  For higher
multipoles, even higher-order fitting polynomials may be used to reduce the
drift, as already observed in~\cite{Berti:2007inspiral}. However, we did find
that the use of such a higher-order fitting polynomial is not sufficient to
eliminate the residual drift still present, and we find no advantage in
applying such procedure. We instead stick to use only second-order polynomials
where the additional $Q_2$ constant may be interpreted as an offset constant
value due to the fact that the $\psi_4^{lm}(t=0)$ is extracted at a finite
radius and not at or close to infinity. 

For our simulations, we also used the GW extraction from the Regge-Wheeler
(odd, $\psi^{o}$) and Zerilli (even, $\psi^{e}$) functions that are computed by
the code \codename{WaveExtract} and are related to the GW strain (in our
normalization) by:

\begin{equation}
  \hat{h}_{lm}\ =\ \frac{1}{r\sqrt{2}}\left(\psi_{lm}^e\ +\ i\psi_{lm}^o\right),
\end{equation}

where $r$ is the Schwarzschild coordinate, while our standard NR coordinates
$(R,\theta,\phi)$ are asymptotically similar to isotropic
coordinates~\cite{Nakano:2015perturbative} and we have the following expression
for

\begin{eqnarray}
  R              &=& r \left(1+\frac{M_\mathrm{ADM}}{2r} \right)^2  \quad,                              \label{EQ:isoR} \\ 
  t_\mathrm{ret} &=& R^* = R + 2 M_\mathrm{ADM} \log \left(\frac{R}{2 M_\mathrm{ADM}} -1 \right) \quad, \label{EQ:tret}
\end{eqnarray}

where $r$ is the coordinate radius, $t$ is the coordinate time and
$M_\mathrm{ADM}$ is the initial ADM mass of the system (see table
~\ref{TAB:InitialData}). We also checked that the resulting $r$ agrees with the
areal radius $R=\sqrt{A(r)/4\pi}$, where $A(r)$ is the surface of the sphere of
constant coordinate radius ($r$) that is derived using such extraction. We
found the relative difference between the measured values of $R$ and
$M_\mathrm{ADM}$ from the simulation and the values used in
Eqs.~\label{EQ:isoR,EQ:tret} at coordinate time $t=0$, to be a few parts over
$10^{-4}$, for a detector at coordinate radius $r=700$ CU.

We extracted the GW signal at 13 different coordinate radii from $r=100$ CU to
$r=700$ CU to check its convergence and whether the extraction at the outermost
radius is a sufficient approximation of the theoretically necessary extraction
at spatial infinity. In the following, we do not develop a full discussion of
the convergence properties of the extraction at different radii or the
comparison of different extraction methods; we limit our work here to the
representative model SLy14vs14. Here, the signals $R \tilde{h}_{22}(t-R*)=R
\tilde{h}_{22}(t-t_\mathrm{ret})$, computed at coordinate radii $r=700$ CU and
$r=650$ CU agree with an error of the order of 1\% during the entire
simulation. We also found that the amplitude and the phase of the signal
extracted at different radii were scaling better with $\psi_4$ than with
the Regge-Wheeler and Zerilli (RWZ) extraction. The RWZ waveforms showed also
unphysical amplitude oscillations in the final part of the signal.  For these
reasons, all the amplitudes and signals discussed in the present work will
refer to extraction based on $\psi_4$, performed at coordinate radius $r=700$
CU, and we use

\begin{equation}
  R h_{lm} (t) = R \tilde{h}_{lm}(t-t_{\mathrm{ret}})
\end{equation}

where $R$ and $t_{\mathrm{ret}}$ are the radius and retarded time at which the
GW signal is extracted, computed using Eqs.~(\ref{EQ:isoR},\ref{EQ:tret}). This
corresponds to the GW signal as measured at unit distance and as if the signal
would have been emitted at the coordinate origin.

From the GW signal we were able to compute the radiated energy, angular
momentum and linear momentum fluxes, following the procedure described
in~\cite{Brugmann:2008calibration}. We arrive at the final expression for the
complex GW signal on a spherical surface at radius $r$:

\begin{equation}
  h(t,\theta,\phi) = \sum_{l=2}^{\infty} \sum_{m=-l}^{l} h_{lm}(t) {}_{-2}\!Y_{lm}(\theta,\phi).
\end{equation}

In the results presented here, this quantity refers to the extraction radius
$r=700$ CU, and the sum over $l$ is limited to $l=2\dots5$, as that it is the
maximum $l$ for which we extract components. The above expression for the
gravitational signal allows to obtain the final equations:

\begin{eqnarray}
\frac{dE^{gw}}{dt}   &=& \frac{R^2}{16\pi}\int{d\Omega \left|\dot{h}(t,\theta,\phi)\right|^2} \label{EQ:dEdt}\\
\frac{dJ_z^{gw}}{dt} &=& \frac{R^2}{16\pi} Re \!\left[\int{\!\!d\Omega \label{EQ:dJdt}
       \left( \partial_{\phi}\; \dot{\bar{h}}(t,\theta,\phi) \right)
       h(t,\theta,\phi)}\right]
\end{eqnarray}

where  $\dot{h}(t,\theta,\phi)$ is the derivative with respect to $t$ of
$h(t,\theta,\phi)$ and $R$ is the physical radius of the extraction sphere.

The above procedure allows us to define the merger time ($t_\mathrm{merger}$)
as the time for which we have the maximum of the $22$-component of the GW
signal ($h_{22}$). Clearly, values defined in this way depend on the
resolutions used to perform the simulation and on the accuracy of the employed
simulation method. A detailed study of the convergence of the simulation
methods used is of paramount importance.

\begin{figure}
\begin{centering}
  \includegraphics[width=0.45\textwidth]{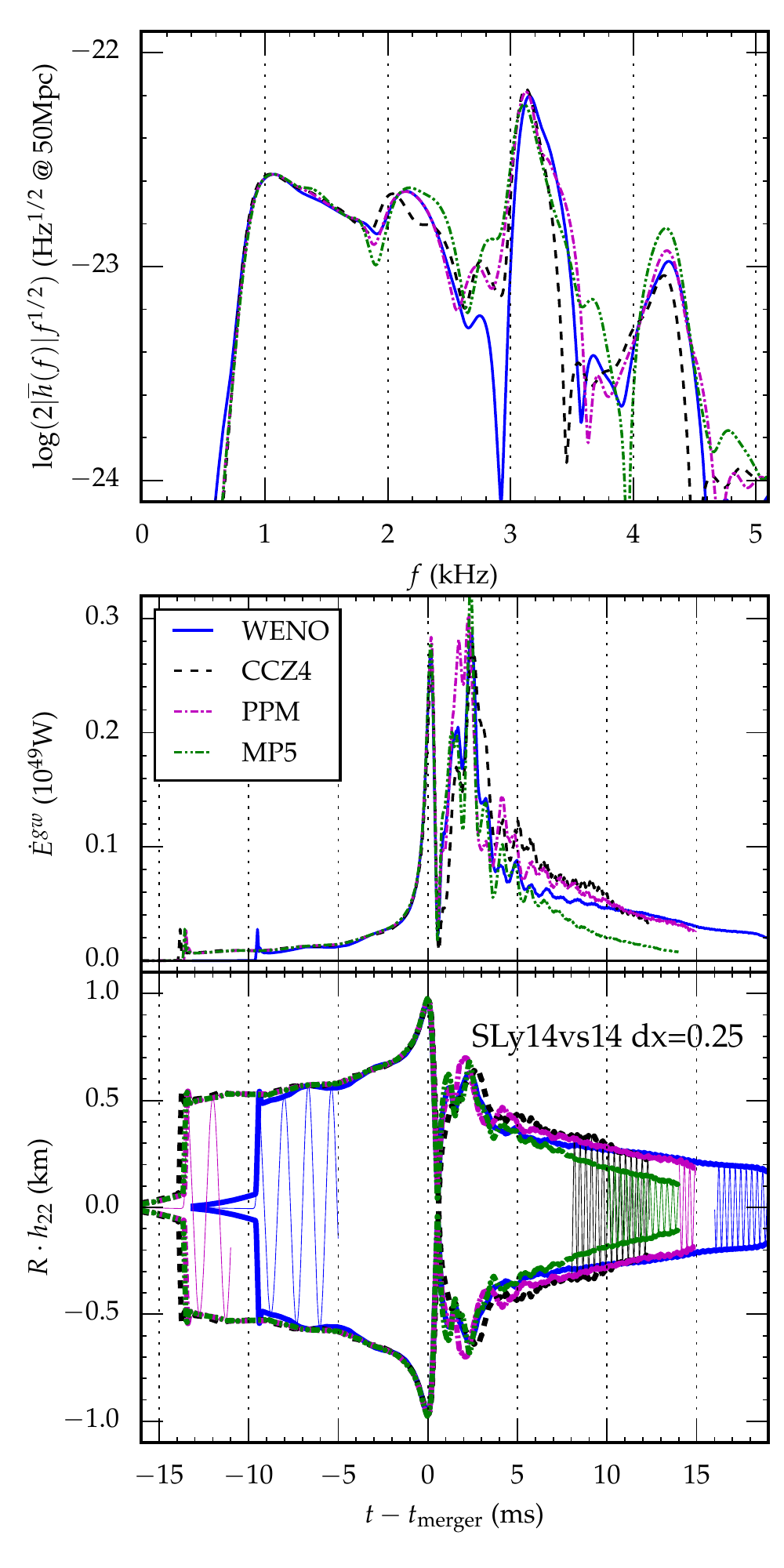}
\end{centering}
\caption{Comparison of the evolution of model SLy14vs14 at resolution $dx=0.25$
CU, using different reconstruction methods for the hydrodynamics equations and
different evolution schemes for the gravitational sector. Quantities are aligned
in time at their respective (different) merger times. 
The upper panel shows the power spectral density (PSD=$2|\overline{h}(f)|\cdot f^{1/2}$) 
of the effective GW signal in the optimally oriented case for a source at $50$ Mpc, 
where it is considered the signal from $t_{\mathrm{merger}}-9$ ms to $t_{\mathrm{merger}}+11$ ms
and a Blackman-windowing function has been applied.
The second panel shows
the total GW  luminosity (energy flux). Finally, the bottom panel shows the
envelop of the gravitational wave amplitudes and their real parts, multiplied by
the distance to the observer. Please note that the evolutions are all similar
to each other, and they differ mainly in the merger time, but not that much in
the amount of total energy carried away by GW in this stage, and also not in
the damping time of the final excitation of the merger remnant.}
\label{fig:14vs14methods} \end{figure}

\begin{figure}
\begin{center}
\hspace{-6mm}
\includegraphics[width=0.49\textwidth]{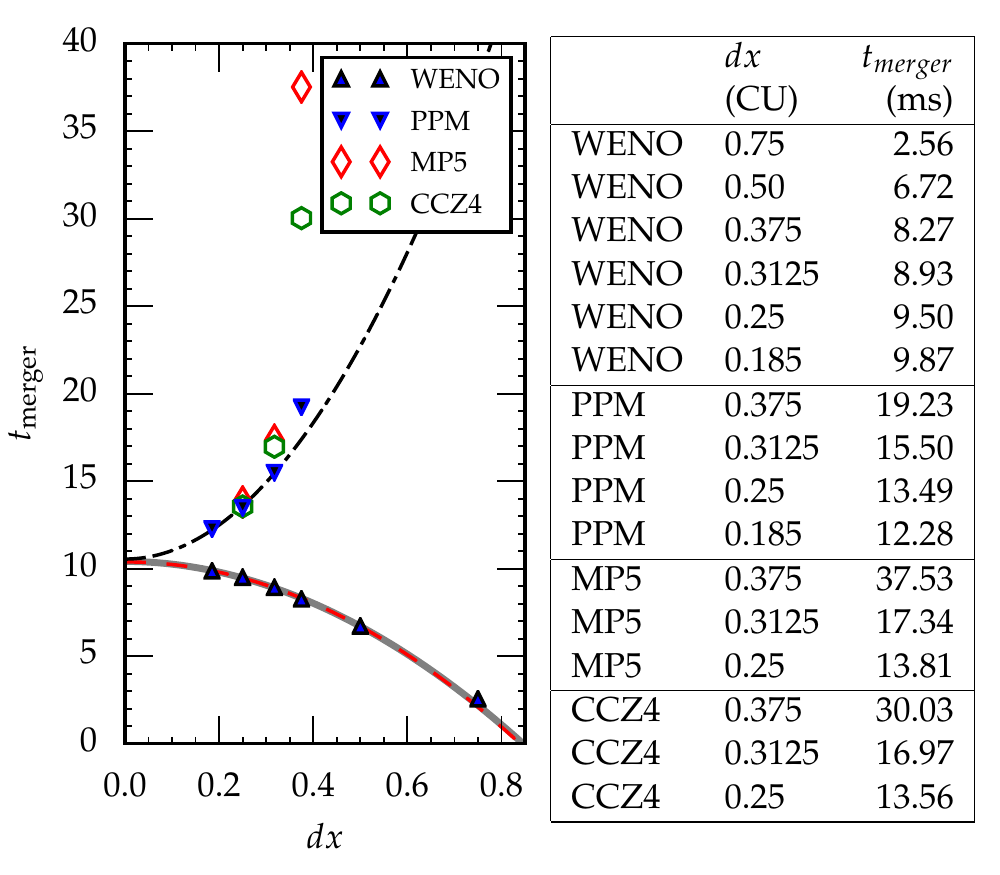}
\end{center}
\vspace{-5mm}
\caption{Fits of the merger time $t_\mathrm{merger}$ as a function of the
resolution $dx$ using the PPM (dashed-dot black line) and the WENO (dashed red
line) reconstruction methods, and the BSSN-NOK evolution method for the
gravitational variable assuming second order convergence. In the case of WENO
reconstruction we also report (thick gray line) the fit where the convergence
order $\gamma$ is computed. In this case the computed convergence order is
$\gamma=1.96\pm 0.14$. The values of the merger time extrapolated to $dx=0$ are
$10.39\pm0.03$ and $10.55\pm 0.20$ for WENO and PPM data, respectively. Please
note that the two fits corresponding to the WENO data are so close that they
are on top of each other here.
}\label{FIG:convergenceMERGER}
\end{figure}

\subsection{Accuracy in the Determination of Merger Time}
\label{SEC:mergertime}

The code we use (\GRHydroNoSpace), as discussed previously, allows the use of
different reconstruction methods for the hydrodynamics sector (among others:
WENO, PPM and MP5). We also have the potential to use an alternative scheme for
the integration of the gravitational sector, namely CCZ4.
Different choices in numerical methods may lead not only to different orders of
convergence, but also to different ranges of convergence even for methods with
the same convergence order. To investigate the effect of different numerical
methods on the errors (especially in the determination of merger time) and gain
confidence in the obtained results, we simulated the same model (SLy14vs14)
using different combinations of numerical methods, but keeping all other
parameters the same. The different combinations were:

\begin{inparaenum}[\itshape(i)]
  \item WENO reconstruction and BSSN-NOK scheme for the gravitational sector;
  \item PPM reconstruction and BSSN-NOK scheme for the gravitational sector;
  \item MP5 reconstruction and BSSN-NOK scheme for the gravitational sector;
  \item WENO reconstruction and CCZ4 scheme for the gravitational sector.
\end{inparaenum}

The first result we report here is that, as shown in
Fig.~\ref{fig:14vs14methods}, all four combinations describe the same dynamics,
for a resolution $dx=0.25$ CU. However, from a single resolution alone a
possibly different degree of accuracy cannot be deducted. Therefore, we have
examined the four different aforementioned combinations of numerical methods at
various resolutions in the range $dx=0.75,\ldots,0.185$ CU and focused the
analysis on the pre-merger stage. For each resolution, we performed a full
simulation and computed the merger time ($t_\mathrm{merger}$). Clearly all the
values we obtained are an approximation of the ``true''  merger time of the
system, but differences between numerical methods and trends for different
resolutions can be observed. For example, in the case of the standard setup
used in this work (WENO and BSSN-NOK), $t_\mathrm{merger}$ ranges from $2.95$
ms at a very coarse $dx=0.75$ CU resolution to $9.86$ ms at the highest
resolution used ($dx=0.185$ CU, about three times more expensive to perform
than the reference resolution $dx=0.25$).

We also found that simulations of model SLy14vs14 do not merge (within a
simulation time of $30$ ms), contrary to what we found using WENO and BSSN-NOK,
at all the other three combinations of evolution methods, when using resolution
coarser than or equal to $dx=0.50$ CU. A full table of the computed merger
times is shown in Figure~\ref{FIG:convergenceMERGER} where we also plot the
merger time vs.\ resolution for all four choices of numerical methods. The
figure clearly shows that the combination WENO and BSSN-NOK performs much
better at lower resolutions, at least for systems like the one we investigate
here. In practice, using our standard setup, we obtained reasonable merger
properties even at a very low resolution of $dx=0.50$ CU and, in some cases, at
least qualitative agreement using a resolution as low as $dx=0.75$ CU. This
simulation can be performed on a laptop.

The fact that a WENO-CCZ4 evolution needs a higher resolution than the
WENO-BSSN-NOK may be due to the fact that the Hamiltonian constraint (even if
damped) enters in the evolution equations of the CCZ4 system, while this is not
the case for the BSSN-NOK scheme \cite{Alic:2011gg}. It was also pointed out in
\cite{Alic:2013xsa} that CCZ4 may show instability depending on the value of
the damping parameter (we used $k_1=0.05,k_2=0$, for $k_3=1$). Our result
indicates that a deeper study of the properties of the CCZ4 formulation (and
the very similar Z4c formulation~\cite{Ruiz:2010qj}) is needed for its use at
very low resolution. At the same time, the result that the PPM scheme is less
accurate does not come as a surprise, since it is formally a third-order method
instead of a fifth-order one (like WENO). The fact that the results for the MP5
reconstruction are worse than the ones for WENO (at the resolutions used here)
needs further investigation too, and is left for future work.
\begin{figure*}
\begin{centering}
  \includegraphics[width=0.24\textwidth]{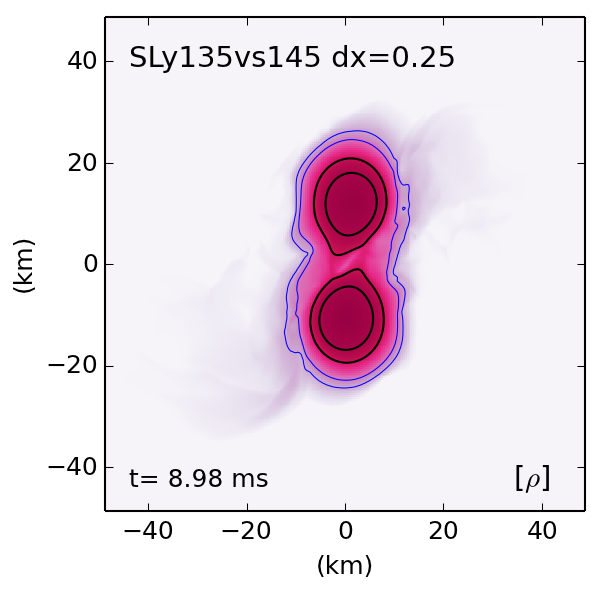}
  \includegraphics[width=0.24\textwidth]{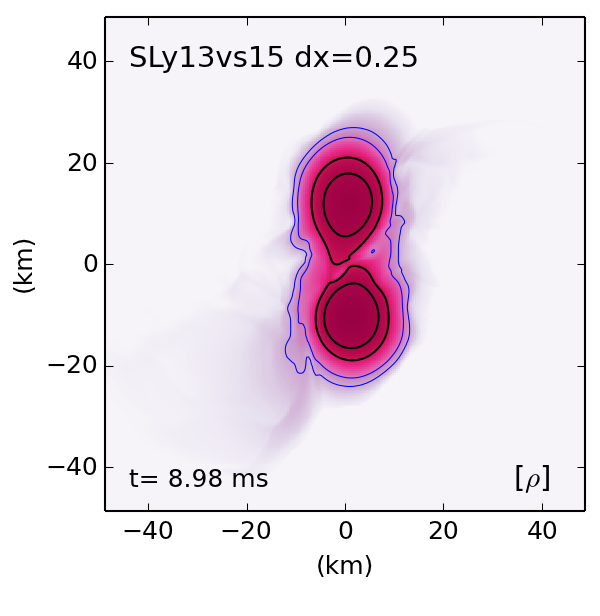}
  \includegraphics[width=0.24\textwidth]{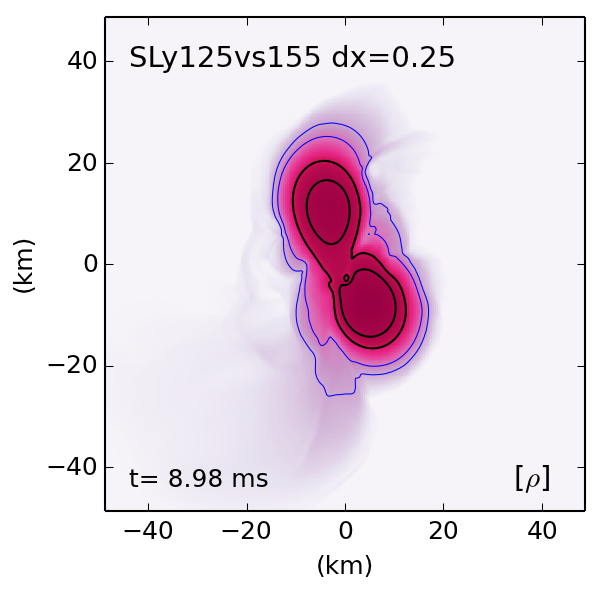}
  \includegraphics[width=0.24\textwidth]{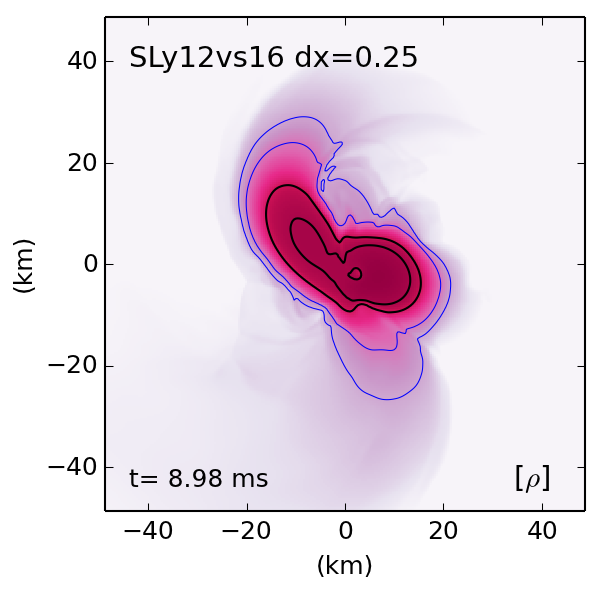}
\\
  \includegraphics[width=0.24\textwidth]{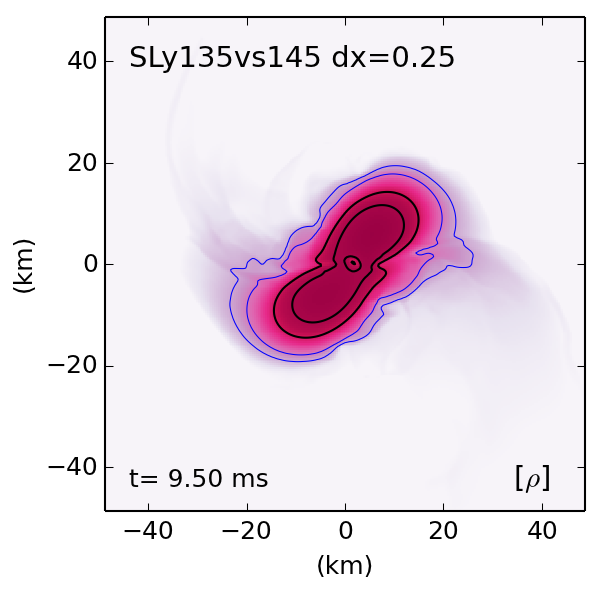}
  \includegraphics[width=0.24\textwidth]{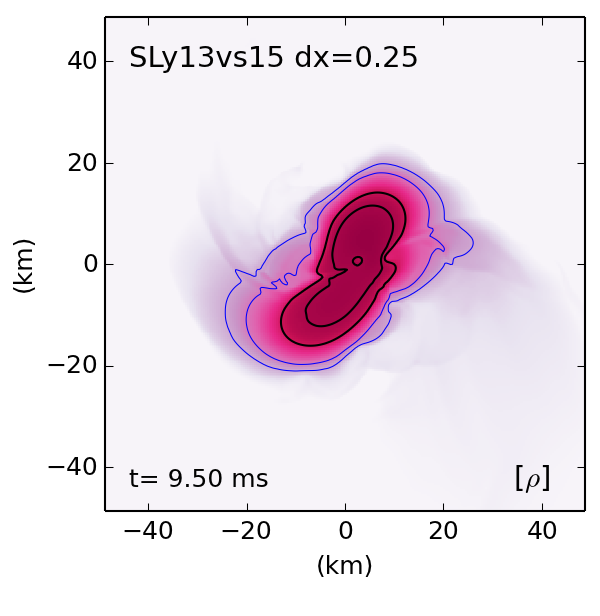}
  \includegraphics[width=0.24\textwidth]{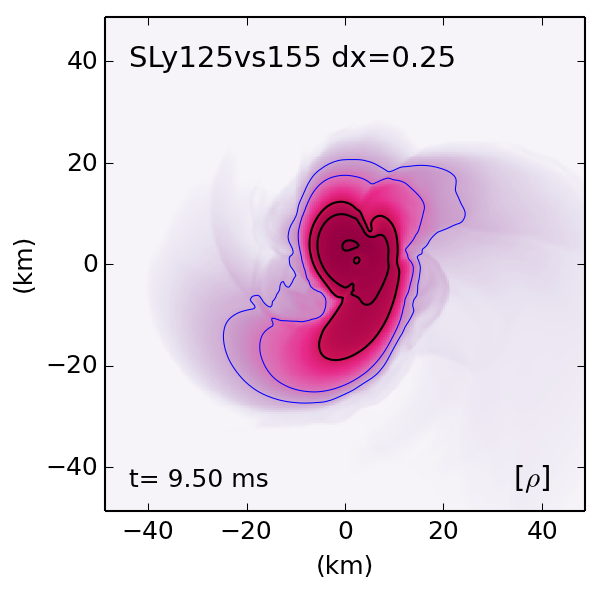}
  \includegraphics[width=0.24\textwidth]{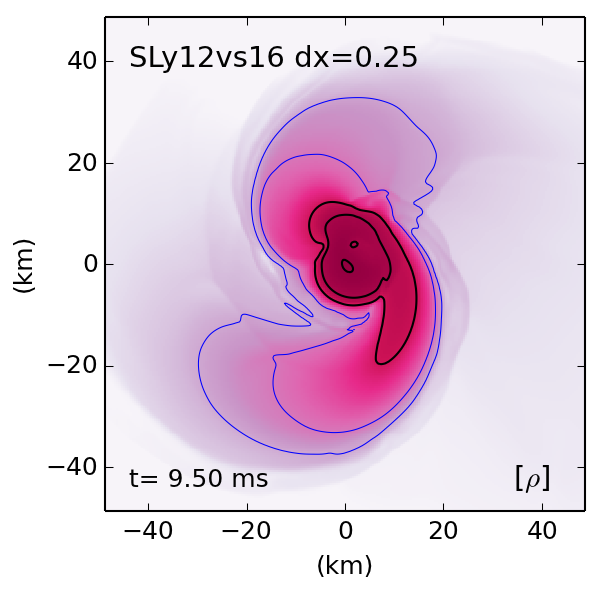}
\\
  \includegraphics[width=0.24\textwidth]{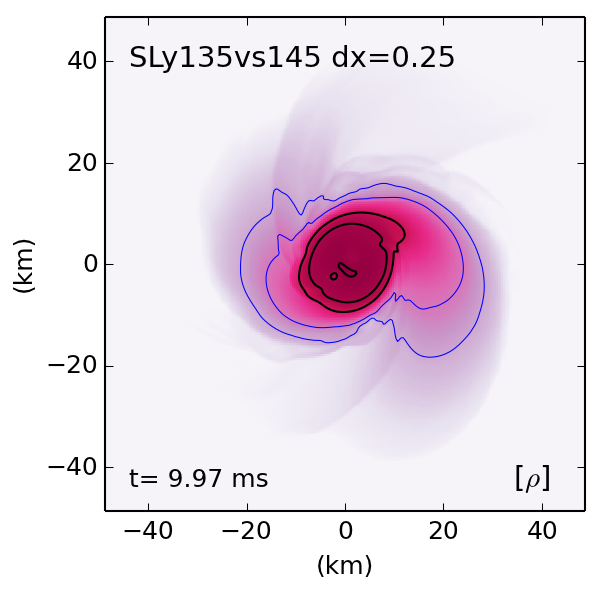}
  \includegraphics[width=0.24\textwidth]{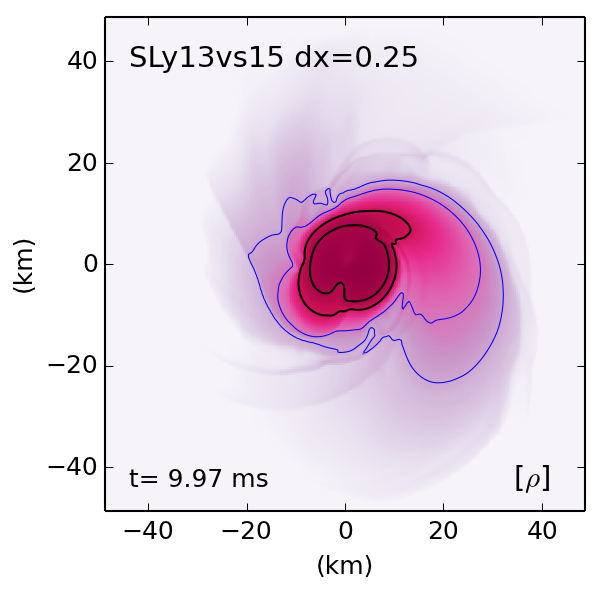}
  \includegraphics[width=0.24\textwidth]{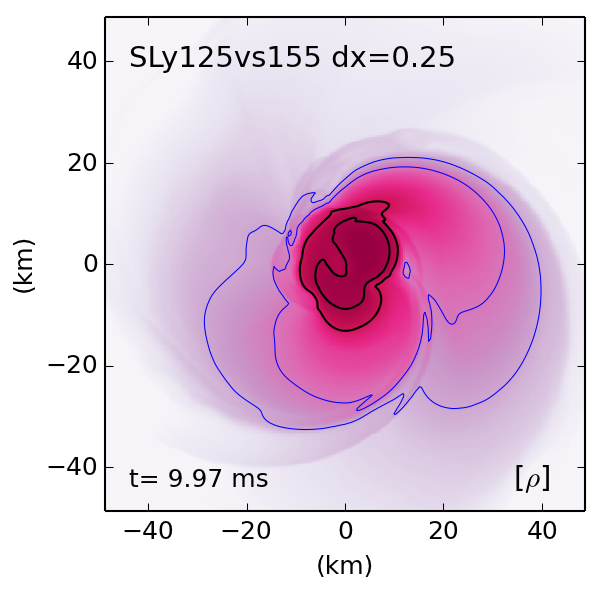}
  \includegraphics[width=0.24\textwidth]{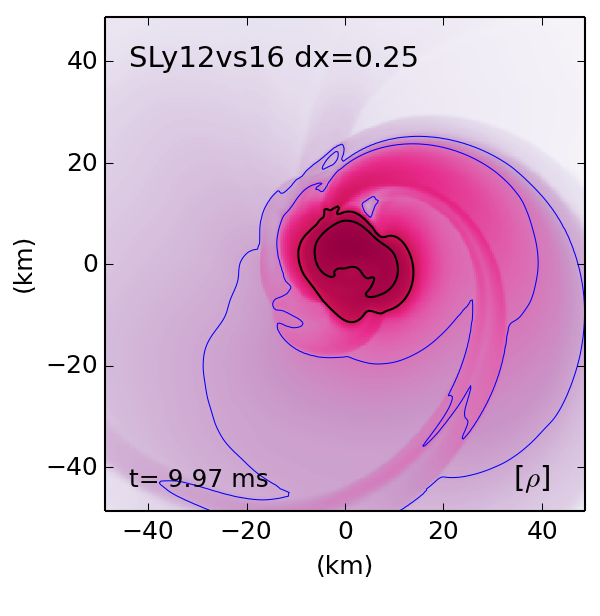}
\\
  \includegraphics[width=0.24\textwidth]{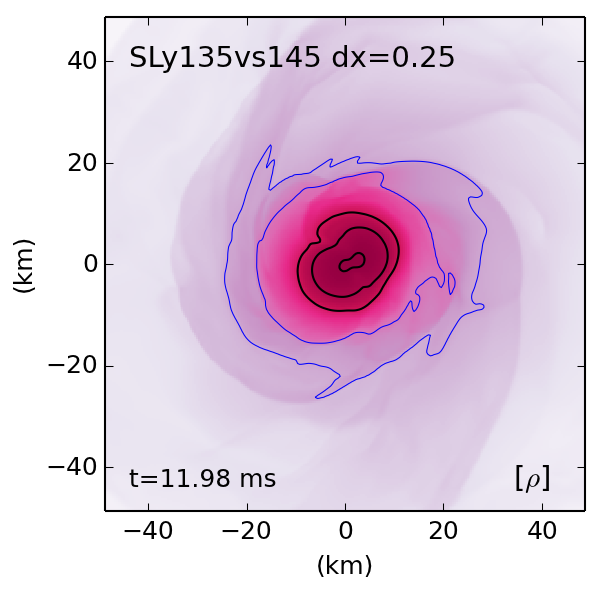}
  \includegraphics[width=0.24\textwidth]{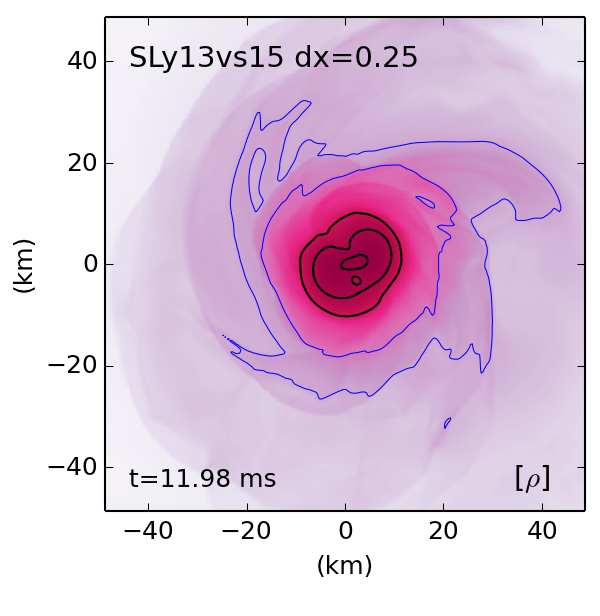}
  \includegraphics[width=0.24\textwidth]{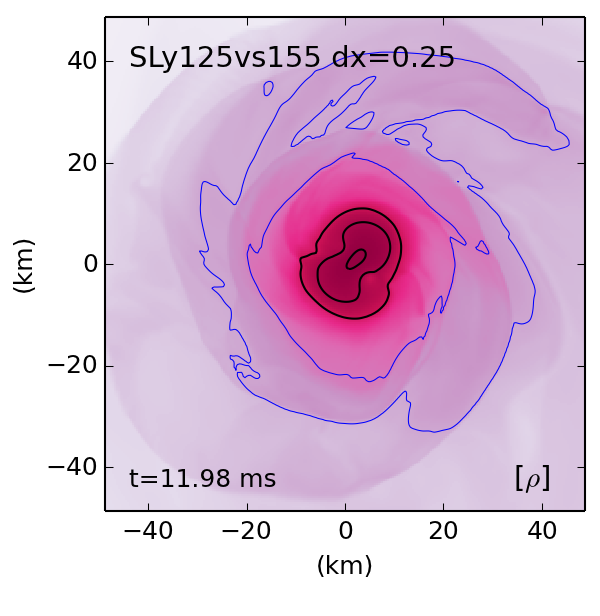}
  \includegraphics[width=0.24\textwidth]{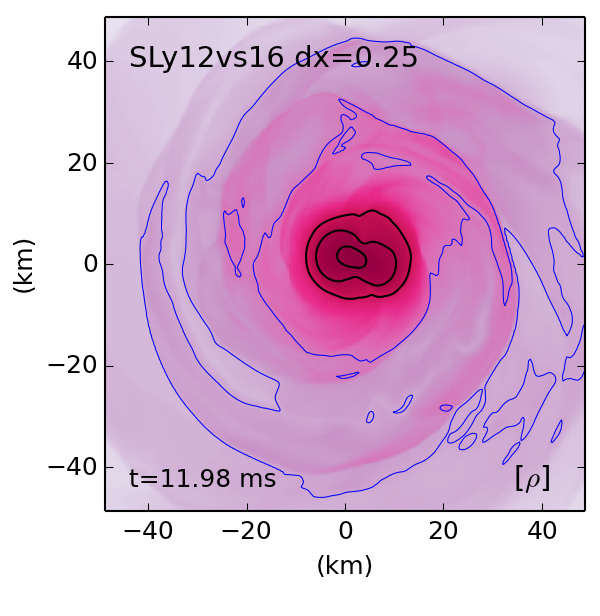}
\end{centering}
\vspace{-0.8mm}
\caption{Snapshot of the dynamics (in the merger phase) of the evolution of the 
four unequal mass models at resolution $dx=0.25$ CU. The density color codes
and the isocontours are the same of Figure~\ref{fig:SNAPSHOT}.}
\label{fig:SNAPSHOTunequalMass} 
\end{figure*}

Despite all observed differences it is important to make sure that all tested
methods lead to the same determination of the ``true'' merger time, denoted as
$t_\mathrm{merger}(dx=0)$ as it is the merger time computed using an unlimited
resolution, i.e., $dx=0$.  The results for the determination of the merger
using an extrapolation  from its values as a function of the resolution are
shown in Figure~\ref{FIG:convergenceMERGER}. We have found that all data
(except when using $dx=0.75$ CU) show a convergence order of almost exactly two
for WENO-BSSN-NOK, which seems to indicate that the main error (for the merger
phase) is dominated by a second-order method. If we fit the merger time as a
function of the used resolution ($t_\mathrm{merger}(dx)$) and assume the
following dependence: $t_\mathrm{merger}(dx)=t_\mathrm{merger}^{dx=0} + A \cdot
dx^\gamma$, we obtain $t_\mathrm{merger}^{dx=0}=(10.42 \pm 0.10)$ ms and
$\gamma=1.95\pm 0.12$, which is consistent with the assumption of a
second-order convergence for the simulation of the merger phase.
A similar analysis for WENO-BSSN-NOK results in a similar convergence rate and
merger time. However, more simulations of higher resolution would be necessary
for the other two possibilities, MP5-BSSN-NOK and WENO-CCZ4. Since we find
agreement between the extrapolated time using the combination WENO-BSSN-NOK and
PPM-BSSN-NOK already for moderate resolution, we decided not to push this
analysis forward for the other two methods due to the low impact of the result
on this work compared to the high computational cost.

We cannot currently explain with certainty why the merger-time increases with
resolution for WENO+BSSN, while this is different for other combinations of
numerical methods. A similar trend has also been reported
in~\cite{hotokezaka:2015exploring}, where an adapted version of Z4c method and
a Kurganov-Tadmor scheme~\cite{kurganov:2000new} with PPM interpolation for
matter evolution have been used. In~\cite{palenzuela:2015effects} it is
asserted that sign differences in the merger time trend for different
resolutions between different codes may be due to a different sign of the
numerical errors introduced by the various methods, leading to a different
effect (in sign) on the dependence of the merger time with respect to
resolution.

Of the numerical methods employed in the present paper, the only second-order
method is the computation of the flux-vector-splitting, which seems to dominate
(in case of WENO reconstruction) the convergence order over all other numerical
methods. Higher-order flux-vector-splitting methods exist and have been used in
simulations of BNS mergers in recent work~\cite{Radice:2015nva,Radice:2015nva}.
Our study suggests that the combined use of high-order flux-vector-splitting
with WENO reconstruction seems to be the most effective numerical method to be
used in simulating BNS mergers.

\subsection{General Results for the Properties of the Evolution.}
\label{SEC:evol}

As explained in Section~\ref{sec:setup} we evolved eleven different models of
BNS systems whose properties are summarized in Table~\ref{TAB:InitialData}. The
evolution of all models shows the same overall dynamics as that of model
SL14vs14 (see Figures~\ref{fig:SNAPSHOT} and \ref{fig:Overview14vs14}) with two
notable exceptions: the evolution of model SLy16vs16 which promptly forms a BH
after the two stars merge, and the evolution of model SLy15vs15 which shows a
delayed BH formation: $7.4$ ms after the merger, at resolution $dx=0.25$ CU. A
more detailed discussion of the BH formation properties of these models can be
found in subsection~\ref{SEC:BH}).

In all other cases, the relatively uneventful inspiral stage is followed by a
very dynamical merger phase that can last up to $8$ ms and, finally, a
relaxation stage where the remaining star shows a single bar-deformed excited
state of what will be the final configuration of the system within the simulation time. There is clear
evidence that in the merger phase the two-arms structure present in the case of
equal mass systems is transformed into a single-arm structure as the mass ratio
increases (see Figure~\ref{fig:SNAPSHOTunequalMass}).

All of the following results have been obtained using WENO reconstruction and
the BSSN-NOK scheme for the gravitational sector. We evolved all of the eleven
models using at least the three resolutions $dx=0.5$, $0.375$, and $0.25$ CU to
be sure of convergence (at least during the inspiral). The results from
different resolutions were also used to determine the merger time as an
extrapolation at $dx=0$, as detailed in section~\ref{SEC:mergertime}, using the
guess $t_\mathrm{merger}(dx) = t_\mathrm{merger}^{dx=0} + A \cdot dx^2$ and
performing a fit. The results obtained in this way are shown in
Table~\ref{TAB:mergertimes}, where only data at $dx=0.5$, $dx=0.375$ and
$dx=0.25$ CU are used for the fit. This includes the computed merger time, the
extrapolated merger time, and the statistical estimate of the error, as well as
the merger time obtained from an effective one-body (EOB) \footnote{The EOB
waveform has been obtained through the publicly available LALSUIT LIGO/Virgo
software (``git://versions.ligo.org/lalsuite.git'') using the utility:
``./lalsim-inspiral -a EOBNRv2 [options]''.} waveform, corresponding to a BBH
system of the same gravitational mass that is, at $t=0$, rotating with the
frequency of the corresponding BNS system. One can observe that the
extrapolated merger time is always consistently lower than the approximation
from an EOB for all models, with greater difference for the case of the only
simulated model using the much softer G275th14v14 EOS (isentropic EOS $P=30000
\rho^{2.75}$ EOS).

This difference could be seen as an indication for the importance of tidal
effects on the late stage of the merger, but it could also stem from possible
differences in the initial data. We did not evolve a high enough number of
orbits before the merger to draw any firm conclusion of which one is the case.
In fact, the number of orbits we used (three to four) is quite low, and any
contamination of the initial data would not have time to settle down
completely, nor would a tidal effect be stabilized. We do plan to investigate
this in future work using higher numbers of orbits.

\begin{table} 
\begin{tabular}{lccccc}  
      name & $\!\!t_\mathrm{merger}^{dx=0.50}$ 
           & $t_\mathrm{merger}^{dx=0.375}$ 
           & $t_\mathrm{merger}^{dx=0.25}$ 
           & $t_\mathrm{merger}^{dx=0\;(ext)}$ 
           & $t_\mathrm{merger}^\mathrm{EOB}$\\
           & [ms] & [ms] & [ms] & [ms] & [ms] \\
\hline
    SLy12vs12 &   9.22 & 11.76 &  13.61 &  15.07$\pm$0.03 & 21.55 \\
    SLy13vs13 &   8.21 & 10.02 &  11.25 &  12.28$\pm$0.04 & 17.25 \\
    SLy14vs14 &   6.72 &  8.27 &   9.50 &  10.39$\pm$0.08 & 14.08 \\
    SLy15vs15 &   5.93 &  6.99 &   7.71 &   8.31$\pm$0.02 & 11.64 \\
    SLy16vs16 &   5.00 &  6.13 &   6.81 &   7.44$\pm$0.08 &  9.78 \\
\hline                                                            
  SLy135vs145 &   6.66 &  8.19 &   9.45 &  10.34$\pm$0.10 & 14.09 \\
    SLy13vs15 &   6.52 &  7.91 &   9.31 &  10.14$\pm$0.25 & 14.12 \\
  SLy125vs155 &   6.19 &  7.60 &   9.09 &   9.93$\pm$0.29 & 14.21 \\
    SLy12vs16 &   5.52 &  7.26 &   8.73 &   9.75$\pm$0.13 & 14.33 \\
\hline                                                            
    G275th14vs14 &4.22 &  4.81 &   5.52 &   5.88$\pm$0.17 & 13.63 \\
    G300th14vs14 &7.63 &  9.69 &  10.55 &  11.67$\pm$0.37 & 14.78 \\
\end{tabular}
\caption{Computed merger time at different resolutions for the eleven BNS
systems studied in the present work. In the forth column is reported the
extrapolated value at infinite resolution and in the fifth column the merger for a
similar binary system with BHs instead of NSs computed using an EOB waveform.}
\label{TAB:mergertimes}
\end{table}

\begin{figure*}
\begin{centering}
  \includegraphics[width=0.32\textwidth]{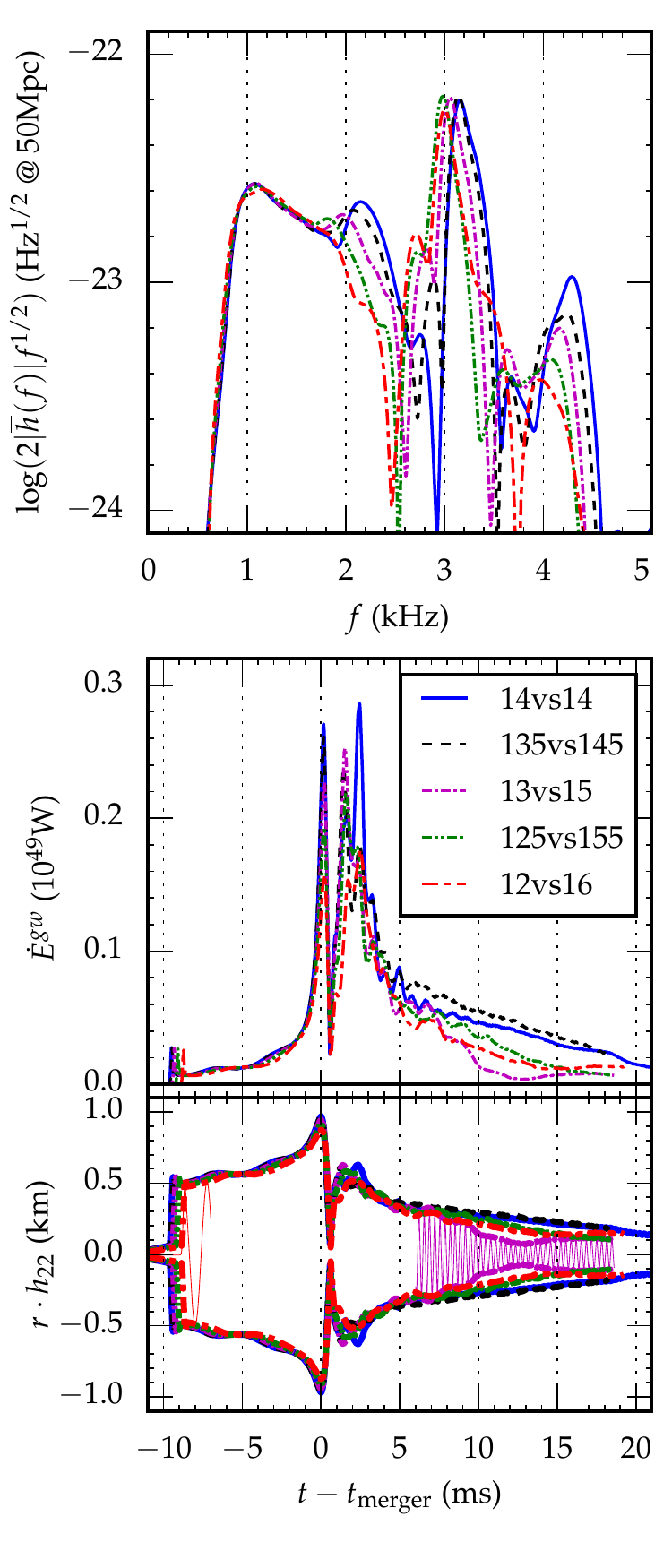}
  \includegraphics[width=0.32\textwidth]{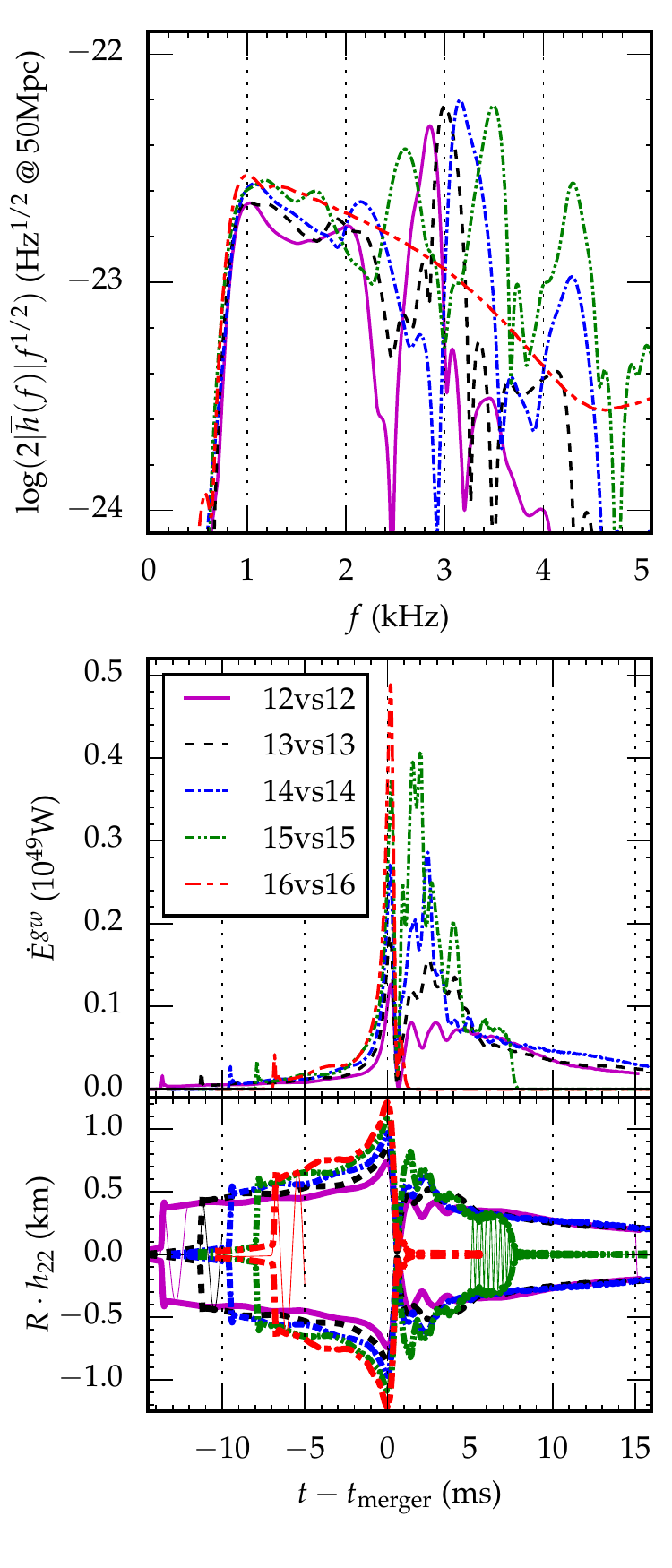}
  \includegraphics[width=0.32\textwidth]{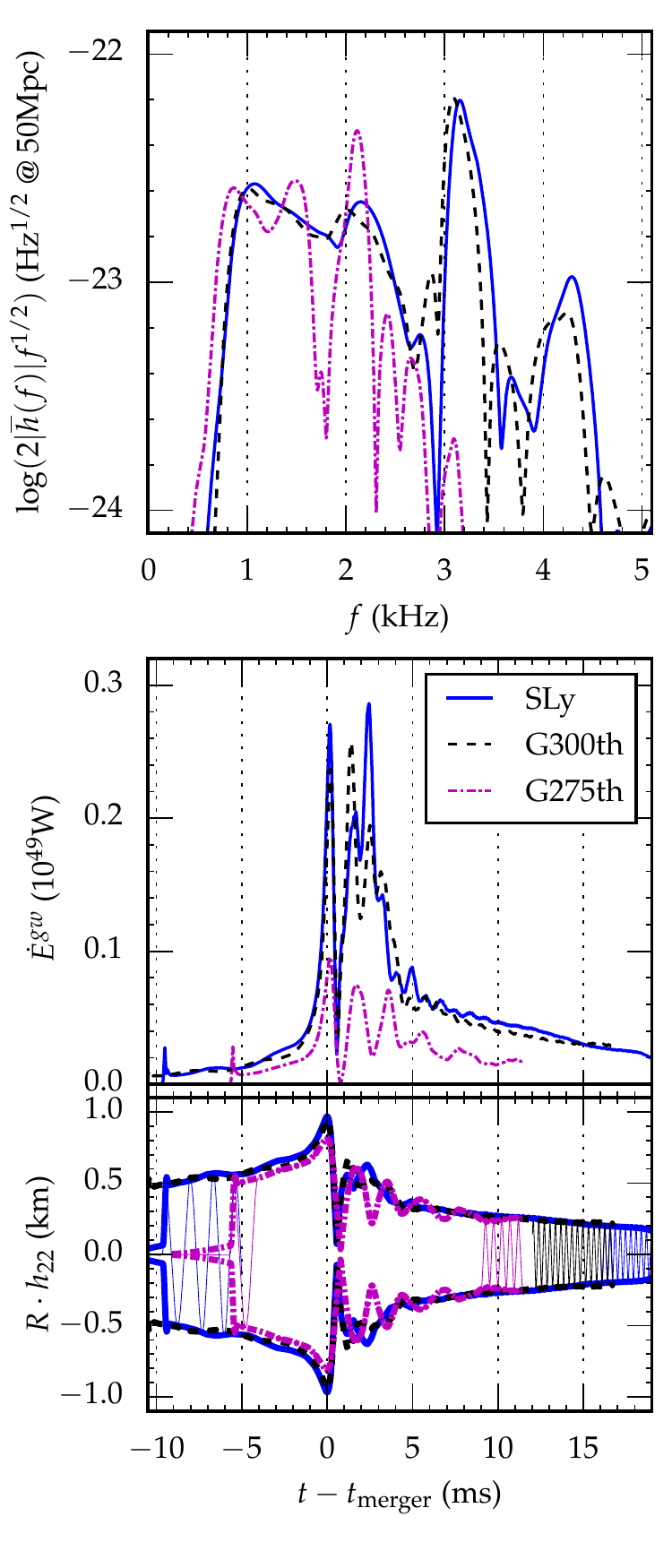}
\end{centering}
\caption{Comparison of the unequal mass models (left), the equal mass models
(center) and three models with the same baryonic mass but different EOS (on the
right). The shown quantities are the same as in Figure~\ref{fig:14vs14methods}:
As there, they are aligned in time at their respective (different) merger
times. 
The upper pannels show the power spectral density (PSD) (Fourier-transform)
of the effective GW signal  $\overline{h}(f)$
in the optimal oriented case for a source at 50 Mpc, 
where it is considered the signal from $t_{\mathrm{merger}}-9$ ms to $t_{\mathrm{merger}}+11$ ms
and a Blackman-windowing function has been applied.
The second panel shows the total GW luminosity (energy flux). Finally,
the bottom panel shows the envelop of the gravitational wave amplitudes and its
real part, multiplied by the distance to the observer.}
\label{fig:allSIMdx25}
\end{figure*}

Another interesting aspect besides the merger time is the emission of GW by the
different models. In Figure~\ref{fig:allSIMdx25} we show the overall waveform
dynamics of all models at resolution $dx=0.25$ CU. In particular, the bottom
panels show the GW signal, while the energy flux radiated in GW, calculated
according to Eq.~(\ref{EQ:dEdt}), are shown in the middle. In the top panel we
show the GW spectrum evaluated performing a Fourier transform of the data
spanning the interval between $9$ ms before and $11$ ms after the merger time,
after having multiplied the time domain signal with a Blackman window function.
Our signal sampling has a Nyquist frequency of $\simeq50.75$ kHz, and over the
specified $20$ ms interval, our spectrum has a resolution of $\simeq 45$ Hz. 

The first noticeable result is that the general picture of a non-collapsing
simulation with a $\Gamma=3$ polytropic EOS and the one using the SLyPP EOS
(with the same $\Gamma_\textrm{th} = 1.8$ thermal part) are very similar. From
the right panel can be seen that both the GW energy emission and the spectrum
nicely agree. This means that only the high-density section of the EOS is
relevant for the global behavior of the system (this is not true for models
collapsing to a BH, as explained in section~\ref{SEC:BH}).

The GW spectrum shows the presence of a dominating peak for every model except
for the promptly collapsing SLy16vs16. This peak is well-known in the
literature and corresponds to the frequency $f_p$ (also called $f_2$ or
$f_\mathrm{peak}$) of the fundamental quadrupolar $m=2$ oscillation mode of the
bar-deformed massive NS formed after the
merger~\cite{stergioulas:2011gravitational}. The frequency of this mode is
related to the compactness of the
star~\cite{hotokezaka:2013mass,bauswein:2014revealing,bauswein:2012measuring,bauswein:2015exploring,hotokezaka:2013remnant}.
It increases for the more compact systems, as shown in panel 2 for different
total mass models, and in panel 3 for different EOSs, where the $\Gamma = 2.75$
polytropic EOS is softer than the other two. It also has recently been
correlated with the star's tidal deformability~\cite{read:2013matter} and the
tidal coupling constant~\cite{bernuzzi:2015modeling}. For the unequal mass
case, we can only see a small dependence on the mass ratio, with a higher
frequency for the equal and the closest-to-equal mass models.

Secondary post-merger peaks at frequencies $f_-$ and $f_+$ (also named as $f_1$
and $f_3$ in the literature) are also present and recognizable in the spectrum,
which could be useful in extracting the NSs parameters (mass, radius) from
future GW detections~\cite{Takami:2014tva,Takami:2015gxa}. The physical
origin of these secondary frequencies is still debated. The first hypothesis
formulated was to consider them as nonlinear combinations of the $m=2$ and
$m=0$ oscillation modes of the hyper-massive NSs with the $f_p$ frequency,
since $f_p \simeq (f_-+f_+)/2$ and the difference $f_p - f_-$ is close  to the
quasi-radial oscillation frequency $f_0$~\cite{stergioulas:2011gravitational}.
More recently, in~\cite{Takami:2014tva,Takami:2015gxa}\cite{kastaun:2015properties} they were
attributed to the modulation of the main $m=2$ frequency by the nonlinear
radial oscillations of the two rotating stellar cores in the first few
milliseconds after the merger. Finally, in~\cite{bauswein:2015unified}, two
different concurrent mechanisms were proposed for the origin of the low
frequency peak: the nonlinear combination of $m=2$ and $m=0$ modes at $f_-$
(dominant in the more compact models), and the GW emission at
$f=f_\mathrm{spiral}$ from the spiral arms created by a strong deformation
during the merger. These rotate in the inertial frame slower than the central
cores, at a frequency $\frac{f_\mathrm{spiral}}{2}$, and this effect would be
dominant for less compact models. Our present data, however, are not sufficient
to clarify which hypothesis is correct.

\begin{table}
\begin{tabular}{|l||c|c|cccc|}
\hline
Model        & $f_0$ &$\tilde{f}_p$ & $f_p$ & $\hat{f}$ & $f_-$ & $f_+$ \\
             & (kHz) &(kHz) &(kHz) &   (kHz)   & (kHz)& (kHz)\\    \hline
SLy12vs12    & 1.30  & 2.80 & 2.85 &    -      &  -   & -    \\
SLy13vs13    & 1.30  & 2.91 & 3.00 &    -      & 1.88 & 4.13 \\
SLy14vs14    & 1.18  & 3.17 & 3.15 &    -      & 2.14 & 4.29 \\ \hline
G300th14v14  & 1.24  & 3.05 & 3.09 &    -      & 1.99 & 4.24 \\
G275th14v14  & 1.01  & 2.10 & 2.11 &    1.56   & -    & -    \\ \hline
SLy135vs145  & 1.17  & 3.16 & 3.14 &    2.84   & 2.05 & 4.24 \\
SLy13vs15    & 1.23  & 3.04 & 3.05 &    2.82   & 1.95 & 4.16 \\
SLy125vs155  & 1.25  & 3.01 & 2.98 &    2.72   & 1.79 & 4.07 \\
SLy12vs16    & 1.27  & 3.02 & 3.00 &    2.71   & -    & -    \\ \hline
SLy15vs15    & 0.81  & 3.50 & 3.49 &    -      & 2.59 & 4.30 \\
\hline
\end{tabular}
\caption{Main peak frequencies and damping times of the post-merger phase of
the simulated models at $dx=0.25$  $M_{\odot}$. The frequency of the $f_0$ and
$\tilde{f}_p$ peaks is determined by a single-frequency fitting procedure, while the
others are derived by an analysis of the Fourier spectrum. Both ways to obtain the
frequency have, using our current methods and resolutions, errors of about
$50-100$ Hz. The values obtained using different resolutions and 
reconstruction methods are within the error on the estimation 
of frequency differences. Indeed the data do not explicitly show any 
dependency of the peak frequency on resolution and methods.}
\label{TAB:peakFrequencies}
\end{table}

In Table~\ref{TAB:peakFrequencies} we report the frequencies of all clearly
recognizable spectral peaks for models shown in
Figure~\ref{fig:allSIMdx25}. We also report the frequency $f_0$ of the
quasi-radial mode, calculated using a fit starting at the time of the minimum
value of the lapse $\alpha$, and using a quadratic trial function with a
superimposed damped oscillation: 

  \begin{equation}
    \alpha_\mathrm{min} = a_2 t^2 + a_1 t + a_0 + A_0 
     e^{-\frac{t}{\tau_0}}\sin\left(2\pi f_0 t + \phi_0\right).
  \end{equation}
The frequency of the dominant mode $\tilde{f}_p$ of the star after the merger
is obtained by fitting the GW signal $h_+$ with a single oscillation mode 
with exponentially decaying amplitude
  \begin{equation}
    h_+ = A_{p} e^{-\frac{t}{\tau_{p}}} \sin\left(2\pi \tilde{f}_p t + \phi\right),
  \end{equation}
starting $5$ ms after the merger, and waiting for the damping of sub-dominant
modes, which happens on a timescale $\tau_0 \simeq 2-3$ ms (except for model
SLy15vs15, for which we started the fit at $2$ ms after the merger to have a
large enough time window before the collapse). 

The reported frequencies for the dominant and sub-dominant modes are obtained by 
interpolating the Fourier transform of the signal with a cubic spline, in order 
to obtain a frequency resolution of less than $50$ Hz (i.e., of the resolution
of the Fourier transform on a $20$ ms interval), and then taking its maximum in an interval 
close to each visible peak.

Most models show the sub-dominant peaks at frequencies $f_-$ and $f_+$, close
to the central $f_p$, with a maximum difference of the order of $\simeq 1$ kHz. At
the same time, those frequency differences are of the same order of $f_0$, with
an error $\left|\left|f_p - f_{+/-}\right| - f_0\right|$ around $100-200$ Hz.

We can see that in the unequal mass models the secondary peaks $f_-, f_+$ have
a lower frequency and contain less power with increasing mass difference, as
was previously already reported with only one model
in~\cite{dietrich:2015numerical}. We do not find a symmetric double-core
structure with spiral arms right after the merger, but instead a tidal tail,
generated from the capture of the less massive star outside and the heavier
star's core rotating around the center of symmetry, as can be seen in
Figure~\ref{fig:SNAPSHOTunequalMass}. Additionally, in those unequal mass
models another low frequency peak (indicated with $\hat{f}$ in
Table~\ref{TAB:peakFrequencies}) is present, closer to the dominant frequency
$f_p$. This could be related to the $f_\mathrm{spiral}$ peak
of~\cite{bauswein:2015unified}, but further analysis is necessary to clarify
its physical origin. For the equal mass case, increasing power in the secondary
frequencies can be seen for the more massive systems (in accordance
with~\cite{Takami:2014tva}), while for the less compact models (SLy12vs12) the
post-merger GW  emission is characterized only by the single $f_p$ frequency.
Finally, the softer EOS model G275th14vs14 shows an important secondary peak at
$\hat{f} = 1.56$ kHz, while no corresponding peak at $f_+ \simeq 2f_p -
\hat{f}$ is present, and the difference $f_p - \hat{f}$ is not close to $f_0$.
This is again consistent with the $f_\mathrm{spiral}$ picture predicted
in~\cite{bauswein:2015unified} for less compact stars.

\subsection{Models Resulting in Collapse to a Black Hole}
\label{SEC:BH}

\begin{figure}
  \includegraphics[width=0.45\textwidth]{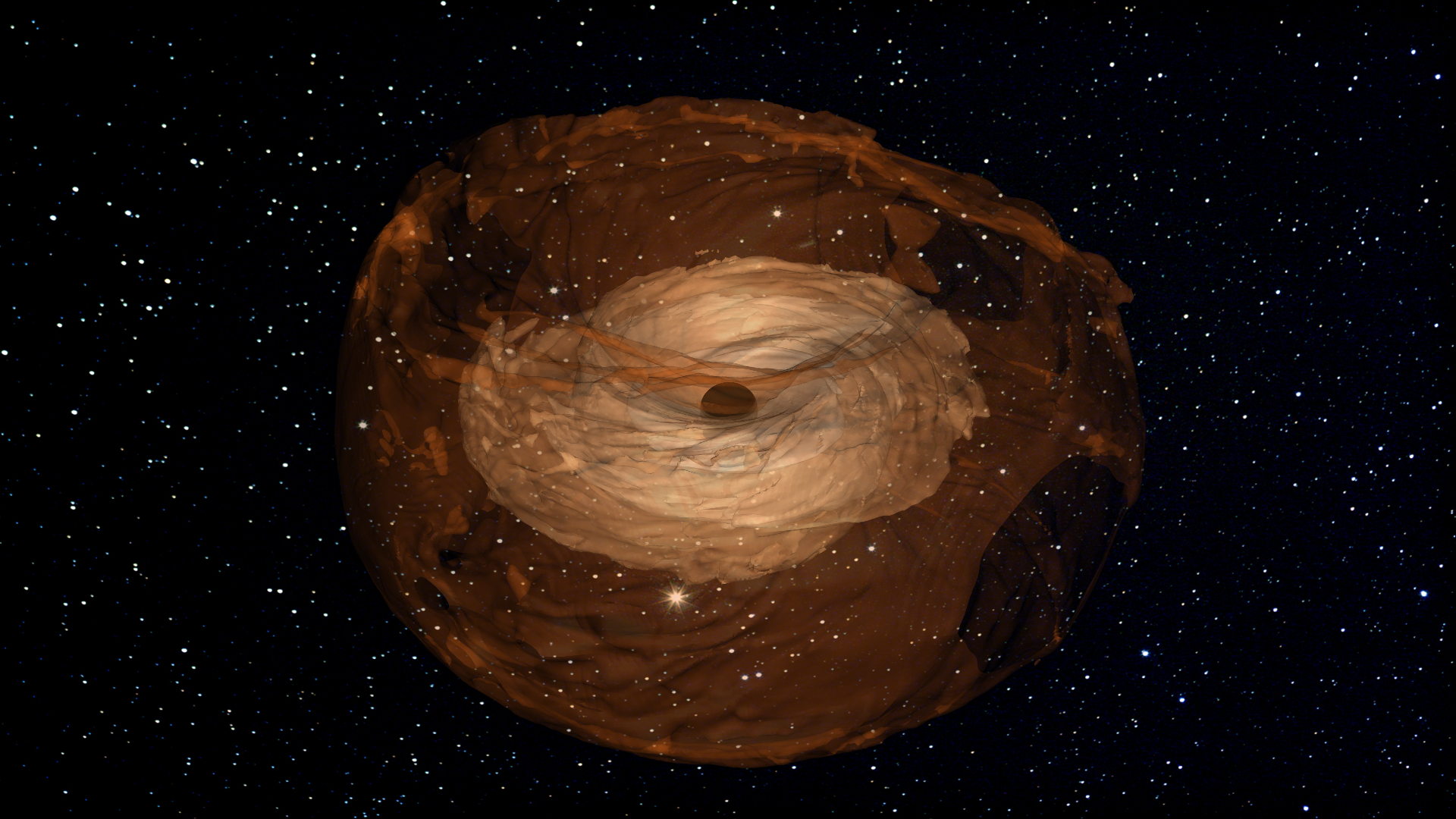}
\caption{Three-dimensional visualization of the disk-like remnant just after black hole
formation ($t=15.84$ ms). At this stage, the matter shown here is still above 
the neutron-drip threshold, but it will later drop to lower densities
(See Fig.~\ref{fig:SNAPSHOT_DISKs}-left).}
\label{fig:SNAPSHOT_DISKs3d}
\end{figure}

\begin{figure*}
\begin{centering}
  \includegraphics[width=0.32\textwidth]{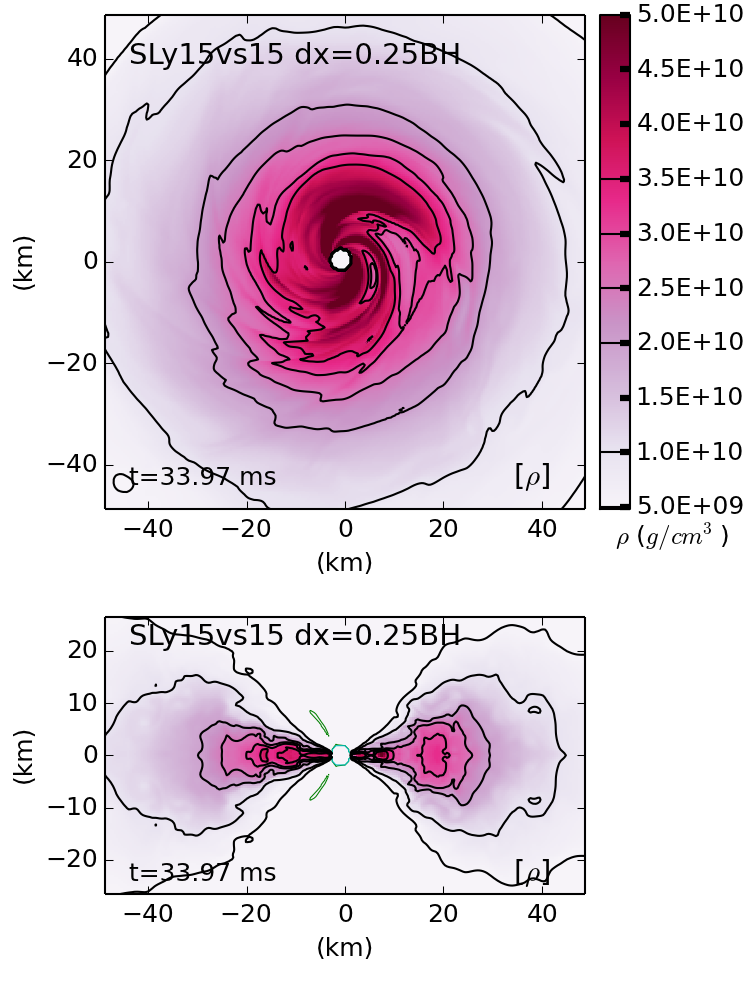}
  \includegraphics[width=0.32\textwidth]{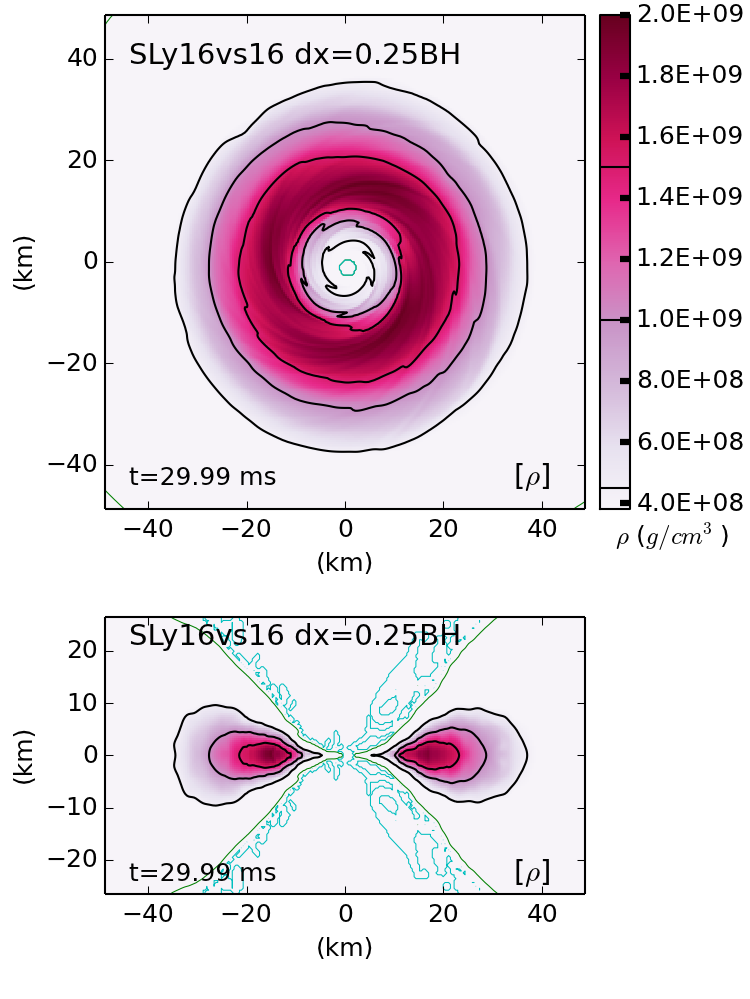}
  \includegraphics[width=0.32\textwidth]{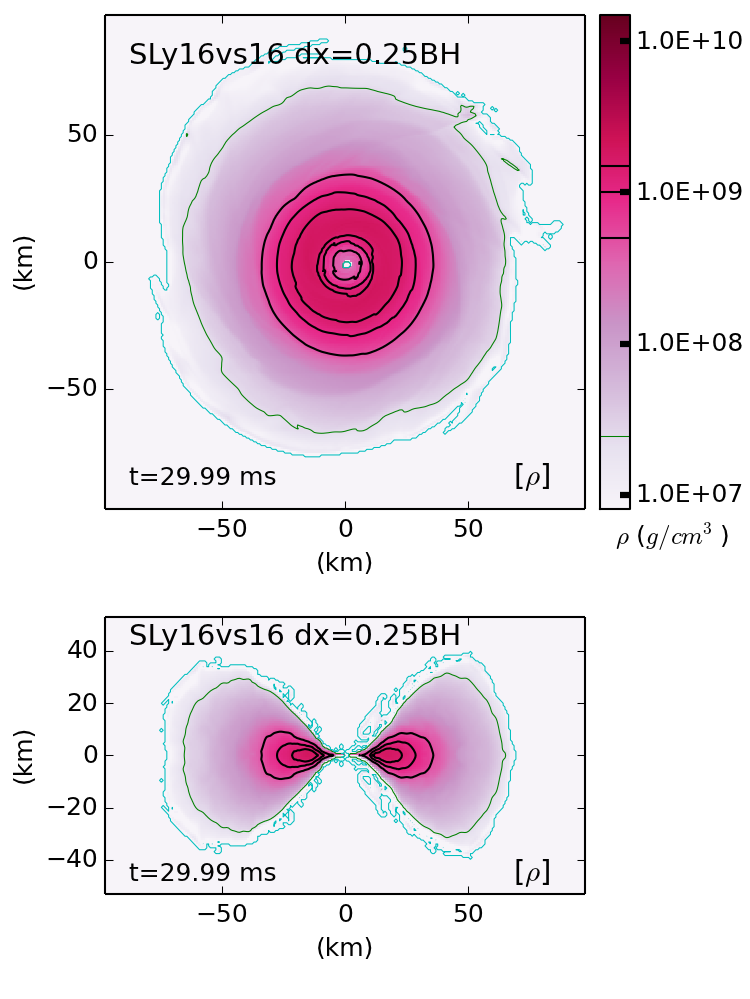}
\end{centering}
\vspace{-0.8mm}
\caption{Snapshot of the disk-like matter distribution in the $x-y$ plane (top)
and $x-z$ plane (bottom) around the back hole at the final stage of the
simulation. Model SLy15vs15 is shown on the left, while for model SLy16vs16 two
different zoom levels are chosen in the center and on the right. The cyan
outline in these two plots denotes the boundary of the atmosphere for this
model: all matter from the initial binary system is still inside this region.
The same cannot be said for model SLy15vs15, where a very low-density matter
flow reaches the outer boundary and leaves the computational domain.}
\label{fig:SNAPSHOT_DISKs}
\end{figure*}

\begin{figure}
  \includegraphics[width=0.45\textwidth]{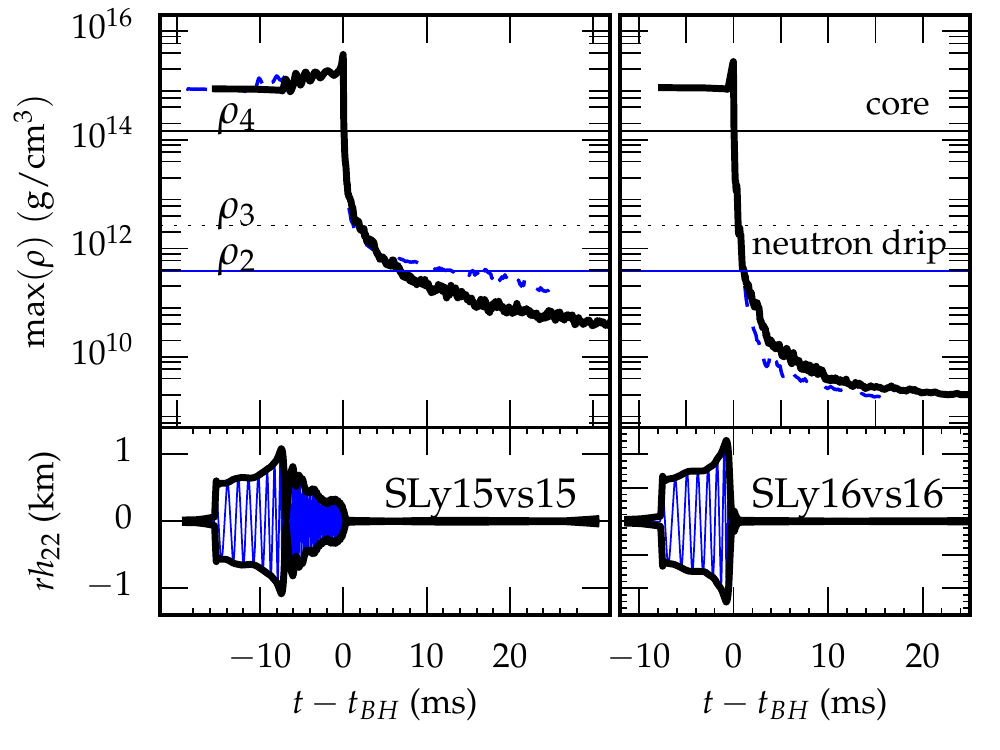}
\caption{Top panels: maximum density $\max(\rho)$ of the matter remaining on
the numerical grid outside of the horizon as a function of the time, for both
models that show BH formation. As can be seen, after the formation of the BH,
densities drop quickly below neutron drip, which an EOS has to describe
appropriately to obtain reasonable results. Bottom panel: GW envelope for waves
emitted by both models. While an extended emission from the hyper-massive
neutron star is visible for model SLy15vs15, the prompt collapse in model
SLy16vs16 cuts off emission shortly after merger besides a comparatively short
and weak BH ring-down.}
\label{fig:maxrohBH}
\end{figure}

As previously noted, the evolution of model SLy16vs16 results in a direct
collapse and the formation of a BH just after the merger. Model SLy15vs15,
instead, is characterized by a delayed collapse to a BH a few milliseconds
after merger, as shown in Figures~\ref{fig:SNAPSHOT_DISKs3d}
and~\ref{fig:SNAPSHOT_DISKs}. One of the main characteristics of our results
for these two models is that after the merger and subsequent collapse to a BH,
the remaining matter density drops quickly below neutron drip. Our simulations
do not include a proper neutron-proton treatment and not even Neutrino
emission. Nevertheless, the fact that the remnant matter is below the
neutron drip threshold is an indication that a noticeable emission 
of neutrinos should occur within the $10$ ms that follows the merger~(see
Figure~\ref{fig:maxrohBH}) and of the importance of the study of the  
r-process that may occour in neutron star mergers and on the necessity 
to use real Nuclear Physics EOS in the study of the post-merger phase.

In more detail, we observe prompt BH formation for model SLy16vs16 with the
appearance of an apparent horizon at $t=6.94$ ms and $t=7.60$ ms for
resolutions $dx=0.375$ and $dx=0.25$ CU, respectively, less than $1$ ms after
merger. For model SLy15vs15, the BH formation is delayed, at simulation times
of $t=18.81$ ms and $t=15.35$ ms for resolutions $dx=0.375$ and $dx=0.25$ CU,
respectively, resulting in a delay of $11.8$ ms and $7.4$ ms after the merger.
This model (with a total baryonic mass of $3.0 M_{\odot}$ and an initial ADM
mass of $2.697 M_{\odot}$, as reported in Table~\ref{TAB:InitialData}) is close
to one often analyzed in the published literature (with total baryonic mass
$2.989 M_{\odot}$ and initial ADM mass $2.675
M_{\odot}$)~\cite{dietrich:2015numerical,Takami:2014tva,hotokezaka:2013remnant},
so a direct comparison of the collapse times is possible.
In~\cite{Takami:2014tva,Takami:2015gxa}, also using a BSSN-NOK spacetime evolution scheme, but
employing PPM reconstruction, using only the four highest density polytropic
pieces of the SLyPP EOS (see Table~\ref{table:PPSLy}) used in the present work,
and setting $\Gamma_{th} = 2.0$, no collapse to BH was found in the first $25$
ms after the merger. Using the same setup, with the only difference of
$\Gamma_{th}=1.8$, in~\cite{hotokezaka:2013remnant} a collapse to a BH was
found about $10$ ms after the merger. In the same paper, the authors also note
that for a different model an increase of $\Gamma_{th}$ significantly delays
the collapse. In~\cite{dietrich:2015numerical}, using the Z4c formulation of
Einstein's equations and WENO reconstruction, with the same EOS
of~\cite{hotokezaka:2013remnant}, a quite long NS lifetime of $13$ ms after the
merger is reported. These differences indicate the need for a careful study of
the influence of the numerical evolution schemes, but more so the dependency
of the collapse on the EOS details, especially the adiabatic index at low
densities and the choice of $\Gamma_{th}$. Even small differences in these
choices could potentially lead to different predictions for the estimated total
energy emission in gravitational waves, and with that, their detectability.

Unlike for the inspiral phase (see Section~\ref{SEC:mergertime}), we are not
able to do a complete study of the convergence properties during the
post-merger and collapse phase, e.g. for the neutron star life time, without
using a higher resolution (incurring a higher computational cost).  Model
SLy16vs16 is promptly collapsing after the merger at all resolutions, even as
low as $dx=0.75$ CU. Model SLy15vs15, instead, shows no collapse for $dx=0.75$
CU even $49$ ms after the merger, while the neutron star collapses at $6.11$
ms, $11.81$ ms, and $7.36$ ms after merger at resolutions $dx=0.50$,
$dx=0.375$, and $dx=0.25$ CU, respectively. Further analysis would be necessary
to fully establish robust convergence for the collapse time in similar models.
We will leave a detailed study of this point to future work.

\begin{figure}
  \includegraphics[width=0.45\textwidth]{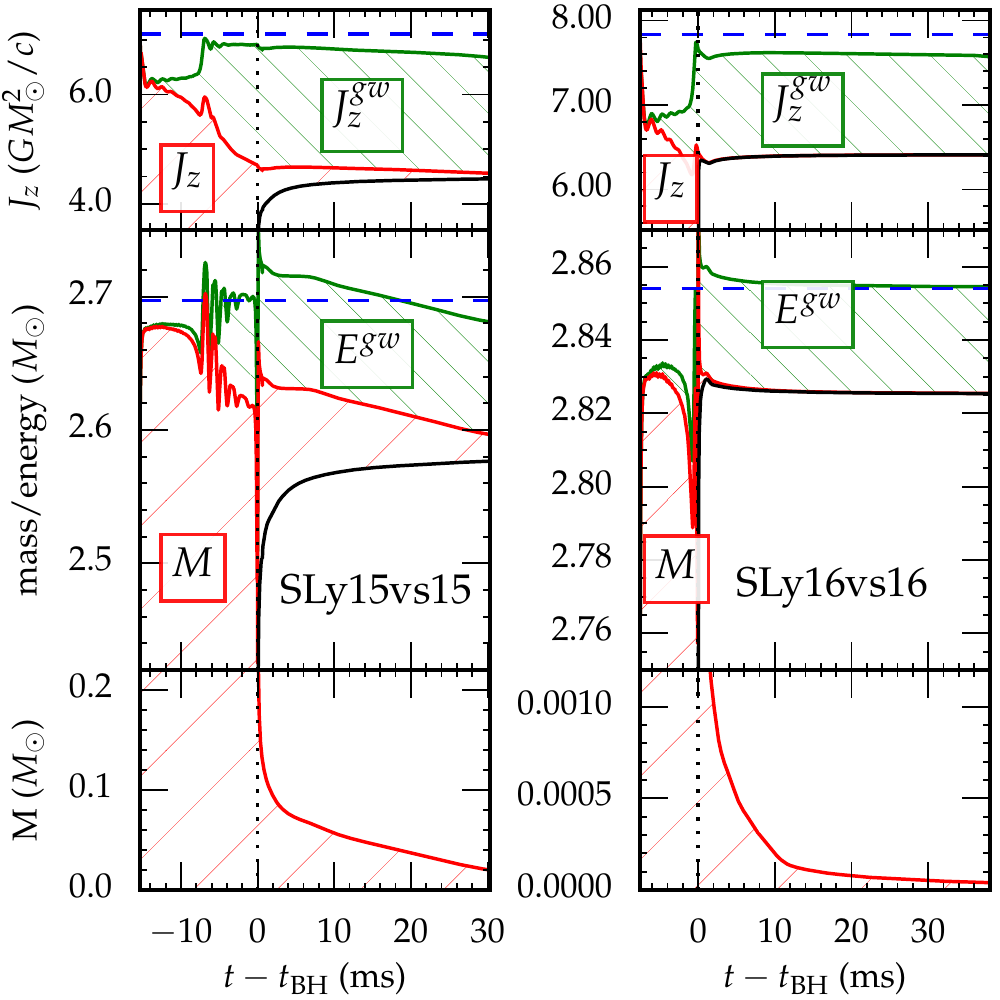}
\caption{Angular momentum and mass budget for the evolution of models SLy15vs15
(left) and SLy16vs16 (right). The blue, dashed, horizontal line indicates the
ADM mass of the initial data, as calculated by Lorene. The red area (hatched
from the bottom left to the top right) shows the contribution from matter left
on the grid, while the green area, hatched from the top left to the bottom
right, shows contributions from emitted GW, plotted in retarded time with
regard to the position of the detector, assuming emission from the origin. The
bottom-most, black, solid line indicates the angular momentum and mass of the
BH. For both models, the sum of these three contributions remains close during
the entire evolution.}
\label{fig:MJbudget}
\end{figure}

In order to characterize the dynamics of these configurations we found it
useful to analyze two quantities that can be easily extracted from simulation
data: the mass and/or energy, and the angular momentum along the $z$-axis,
assuming an axial and temporal symmetry. The latter condition is indeed
approximately reached at the end of the presented simulations. These quantities
are defined in terms of the two vectors $\xi_{(t)}^\mu=(1,0,0,0)$ and
$\xi_{(\phi)}^\mu=(0,\cos(\phi),\sin(\phi),0)$
(see~\ref{fig:SNAPSHOT_DISKs3d}), and are the same that were also used
in~\cite{DePietri:2014mea,Loffler:2014jma} to characterize the properties of
isolated rotating stars. Namely:

  \begin{eqnarray}
    M   & = & - \int (2 T^\mu_\nu -\delta^\mu_\nu T^\alpha_\alpha ) \xi^\nu_{(t)} d^3\Sigma_\mu\\
    J_z & = & \int T^\mu_\nu \xi^\nu_{(\phi)} d^3\Sigma_\mu,
  \end{eqnarray}

where $\Sigma_\mu$ are the 3-dimensional slices of fixed coordinate time $t$
($d^3\Sigma_\mu= \sqrt{\gamma} d^3\!x$). To obtain the energy and angular
momentum budgets of the collapsed models we used the energy ($E^{gw}$) and
angular momentum ($J_z^{gw}$) carried away by gravitational radiation (see
Eqs.~(\ref{EQ:dEdt}) and (\ref{EQ:dJdt})), and the mass $M_{bh}$ and angular
momentum of the BH $J_{bh}$, computed using the isolated horizon
formalism~\cite{Ashtekar:2000sz,Ashtekar:2001jb,Ashtekar:2004cn} (provided by
the module~\codename{QuasiLocalMeasures}~\cite{Dreyer:2002mx}) on the the
apparent horizon (located by the module
\codename{AHFinderDirect}~\cite{Thornburg:2003sf}). The results are shown in
Fig.~\ref{fig:MJbudget}.

For the quickly collapsing model SLy16vs16 we found, $25$ ms after the
collapse, a mass of $\approx2.83M_{\odot}$ and an angular momentum of
$\approx6.41$ $G M_{\odot}^2/c$, while the total (gravitational) mass still
present in the disk is only around $6.19\cdot 10^{-5} M_{\odot}$. In contrast,
we found for model SLy15vs15, also $25$ ms after the collapse, a BH mass of
$2.57 M_{\odot}$ and an angular momentum of $4.45$ $G M_{\odot}^2/c$, while the
remaining matter is either escaping the grid (very low density), or is still
forming and driving an accretion disk. The total remaining mass on the grid
amounts to about $0.028 M_{\odot}$, and its density is below the neutron drip
threshold.

\section{Conclusions}
\label{sec:conclusions}

We have presented a study of BNS mergers for equal and unequal mass systems
with the semi-realistic seven-segment piece-wise polytropic SLyPP EOS with a
thermal component of $\Gamma_{th} = 1.8$. All our results have been produced
using open source and freely available software, the Einstein Toolkit for the
dynamical evolution and the LORENE library for generating the initial models.
Namely, we studied five equal-mass and four unequal-mass models with total ADM
mass in the range between $2.207$ $M_\odot$ and $2.854$ $M_\odot$, and a mass ratio $q =
M^{(1)}/M^{(2)}$ as low as  $0.77$. The dynamics of all simulated models
started with the stars at a separation of  $40$ km. We followed this
late-inspiral phase and the after-merger phase for more than $20$ ms after
merger.
 
We analyzed the merger and post-merger phase of BNS evolution in detail,
studying their gravitational wave emission, the radiated energy and angular
momentum and the gravitational waves spectrum. We investigated the differences
between equal-mass models and unequal mass ones. In the first case a
bar-deformed hyper-massive NS is produced after the merger (for all models
except SLy16vs16 which promptly collapses to a BH) with symmetrical spiral arms
and a rotating double-core structure. After a relaxation period of some
milliseconds the star reaches its final state where only the fundamental $m=2$
mode is excited. Unequal mass models, instead, show after the merger a single
arm structure created by the tidal deformation of the less massive star. This
qualitative difference leads also to differences in the GW spectrum, with the
suppression of modes linked to the quasi-radial oscillations and the presence
of a new sub-dominant mode, closer in frequency to the dominant $m=2$
excitation.

Of all simulated models two collapse to a BH, one right after the merger
(SLy16vs16), and one after a short hyper-massive NS phase (SLy15vs15). For both
models a leftover accretion disk is present, with matter density under the
neutron drip limit. This means that a correct form of the low density-section
of the EOS should be important for the description of the disk and matter
surrounding the remnant BH.

We studied the convergence properties of the numerical evolution in the
inspiral phase, focusing on the value of the merger time computed for
simulations with different resolutions and different numerical methods. We
tried three different methods for the reconstruction of hydrodynamical
variables (WENO, PPM and MP5) and two different evolution schemes for the
gravitational sector (BSSN-NOK and CCZ4). The merger time showed second order
convergence, and our results for the different methods, for resolution
$dx=0.25$ CU and better, do agree. We have explicitly shown that we achieve
second-order convergence in the inspiral phase when PPM or WENO reconstruction
is combined with the BSSN-NOK methods. In the other two cases we find that a
much higher resolution is needed to explicitly show that a similar order-two
convergence is present.  Our result seems to indicate that to achieve reliable
results at lower resolutions the combined use of BSSN-NOK and WENO seems the
best setup, out of the ones we tested. This analysis allowed us to extrapolate
the merger time for all simulated models for infinite resolution $dx=0$.

In addition, we compared to two BNS models of the same total baryonic mass
($M_0=2.8$ $M_\odot$) as model SLy14vs14 but with only one piece-wise
component, and the same thermal part but different stiffness. The first model,
G300th14vs14, has the same average stiffness ($\Gamma=3.00$) as the SLyPP EOS,
while the other model, G275th14vs14, shows a softer average stiffness
($\Gamma=2.75$), which is similar to Shen EOS
(see~\cite{Loffler:2014jma}). Using these, we find that the merger time
increased only by a fraction (from $10.4$ ms to $11.7$ ms) in the case of
G300th14vs14, while it is significantly decreased (from $10.4$ ms to $5.9$ ms)
in the case of the softer G275th14vs14 model, which we attribute to the
expectation that tidal effects are greater when the deformability of the star
is increased (like in the case of softer stiffness). In the after-merger stage
we observed only a small variation in the position of the peak for the
gravitational wave emission in the first case, while the later model shows
substantial differences to the SLy14vs14 comparison model.

\acknowledgments

This project greatly benefited from the availability of public software that
enabled us to conduct all simulations, namely ``LORENE'' and the ``Einstein
Toolkit''. We express our gratitude the many people that contributed to
their realization.

This work would have not been possible without the support of the SUMA INFN
project that provided the financial support of the work of AF and the computer
resources of the CINECA ``GALILEO'' HPC Machine, where most of the simulations
were performed. Other computational resources were provided by the Louisiana
Optical Network Initiative (QB2, allocations loni\_hyrel, loni\_numrel and
loni\_cactus), and by the LSU HPC facilities (SuperMike II, allocations
hpc\_hyrel and hpc\_numrel). FL is directly supported by, and this project
heavily used infrastructure developed using support from the National Science
Foundation in the USA (Grants No. 1212401, No. 1212426, No. 1212433, No.
1212460). Partial support from INFN ``Iniziativa Specifica TEONGRAV'' and by
the ``NewCompStar'', COST Action MP1304, are kindly acknowledged.

We also graciously thank Dennis Castleberry for proofreading the manuscript.

\appendix
\section{Computational Cost}
\label{SEC:compCOST}

\begin{table}
 \centering
 \begin{tabular}{rcccccc}
  Level & min($x$/$y$)& max($x$/$y$)& min($z$)& max($z$)& $(N_x,N_y,N_z)$ \\
        &  (CU)       &     (CU)    &   (CU)  &  (CU)   & ${dx=0.25}$\\
\hline
  $1$ & $-720$ & $720$ & $0$ & $720$  & (185,185,96)\\
  $2$ & $-360$ & $360$ & $0$ & $360$  & (205,205,106)\\
  $3$ & $-180$ & $180$ & $0$ & $180$  & (205,205,106)\\
  $4$ & $ -90$ & $ 90$ & $0$ & $ 90$  & (205,205,106)\\
  $5$ & $ -60$ & $ 60$ & $0$ & $ 30$  & (265,265,76)\\
  $6$ & $ -30$ & $ 30$ & $0$ & $ 15$  & (265,265,76)\\
 ($7$ & $ -15$ & $ 15$ & $0$ & $7.5$) & (265,265,76)\\
 \end{tabular}
 \caption{Simulation grid boundaries of refinement levels. Level 7 is
          only used for simulations forming a BH, once the minimum
          of the lapse $\alpha < 0.5$. Resolutions as reported in this paper
          always refer to grid 6.}
 \label{tab:grid}
\end{table}
\begin{table} 
\begin{tabular}{lrrrrrr}  
$\bigtriangleup x$ (CU) & $0.75$ & $ 0.50$ & $0.375$ & $0.25$ & $0.185$& $0.125$  \\     
\# threads               &   16 &   64 &  128 &  256 &  512 & 2048 \\
\# MPI                   &    2 &    8 &   16 &   32 &   64 &  256 \\
\hline
Memory (GBytes)          &  3.8 &   19 &   40 &  108 & 237  &   768 \\
speed (CU/h)             &  252 &  160 &  124 &   53 &   36 &    16 \\ 
speed (ms/h)             & 1.24 & 0.78 & 0.61 & 0.26 & 0.18 & 0.08  \\ 
cost (SU/ms)             &   13 &   81 &  209 &  974 & 2915 & 26053 \\\hline
total cost (kSU, $50$ ms) & 0.65 &    4 & 10.5 &   49 &  146 & 1300  \\
\end{tabular}
\caption{Computational cost of the simulations, for the example of using
BSSN-NOK, with WENO reconstruction for the hydrodynamics. SU stands for
service unit: one hour on one CPU core.
The reported
values refers to the ``GALILEO'' PRACE-Tier1 machine locate at CINECA 
(Bologna, Italy) equipped with 521 nodes, two-8 cores Haswell 2.40 GHz,
with 128 GBytes/node memory and 4xQDR Infiniband interconnect.
Also, these are only correct for evolutions that 
do not end with the formation of a BH, as an additional refinement
level was used to resolve the BH surroundings, and more analysis
quantities had to be computed (e.g., the apparent horizon had to be found).
In addition, the simulations resulting in a BH were performed on facilities
at Louisiana State University: SuperMike II (LSU HPC) and QB2 (Loni).}
\label{TAB:COSTs}
\end{table}

The implementation of any computational intensive research program like the one
developed in the present work needs a careful analysis of the computational
cost of the simulations (the number of CPU core hours needed to perform each
simulation) and a careful management of resources. In that, speed, and with
that total runtime, however, are not the only variables to consider. Required
memory puts a lower bound on the size of the employed resources, while an upper
bound is present at the breakdown of strong scaling.

To quantify these needs, the resolution and the size of the computational grid
are most important. Table~\ref{tab:grid} shows the characteristics of the grid
we used for the present work. In particular we use a fixed structure of
mesh-refined, centered grids, with the exception of an additional refinement
level for simulations resulting in an apparent horizon, and then only starting
shortly before the
merger (when the minimum of the lapse $\alpha$ on the grid dropped below $0.5$).
In the last column of  Table~\ref{tab:grid} we show the actual grid-size in
computation-points of each level, for resolution $dx=0.25$ CU.  Clearly the
actual grid size (including ghost-zones and buffer-zones) changes varying with
resolution, and is not explicitly shown here for that reason.

With the computational domain completely specified, the next step of an
analysis of the computational cost is to asses the cost for a full simulation
of a particular model at the desired resolution. Table~\ref{TAB:COSTs} shows
the actual simulation cost as function of resolution, for a particular
High-Performance-Computer (HPC) system used in the present research program,
namely the ``GALILEO'' system installed at the Italian CINECA supercomputer
center. As it was
discussed in the conclusion, our result show that the combined use of BSSN-NOK
and WENO allows the possibility to find qualitatively accurate results in
agreement with high-resolutions simulations. This is a very desirable feature
since it allows researchers to quickly scan numerous different models in order
to select the most interesting for further study using higher resolution.

All of our results have been produced using open source and freely available
software, the Einstein Toolkit for the dynamical evolution and the LORENE
library for generating the initial models. That means that the whole set of our
result can be reproduced and re-analyzed by re-running the simulation from a
common code-base. Some modifications of the above mentioned software were
necessary, but these changes are also open source, and are available 
for download from the University of Parma WEB web server
of the gravitational group \footnote{ 
See the University of Parma Gravity group WEB page: 
\href{http://www.fis.unipr.it/gravity/Research/BNS2015.html}{www.fis.unipr.it/gravity/Research/BNS2015.html}.
}.
We kindly ask to cite this work if you find any of the material there useful
for your own research.


\begin{thebibliography}{81}%
\makeatletter
\providecommand \@ifxundefined [1]{%
 \@ifx{#1\undefined}
}%
\providecommand \@ifnum [1]{%
 \ifnum #1\expandafter \@firstoftwo
 \else \expandafter \@secondoftwo
 \fi
}%
\providecommand \@ifx [1]{%
 \ifx #1\expandafter \@firstoftwo
 \else \expandafter \@secondoftwo
 \fi
}%
\providecommand \natexlab [1]{#1}%
\providecommand \enquote  [1]{``#1''}%
\providecommand \bibnamefont  [1]{#1}%
\providecommand \bibfnamefont [1]{#1}%
\providecommand \citenamefont [1]{#1}%
\providecommand \href@noop [0]{\@secondoftwo}%
\providecommand \href [0]{\begingroup \@sanitize@url \@href}%
\providecommand \@href[1]{\@@startlink{#1}\@@href}%
\providecommand \@@href[1]{\endgroup#1\@@endlink}%
\providecommand \@sanitize@url [0]{\catcode `\\12\catcode `\$12\catcode
  `\&12\catcode `\#12\catcode `\^12\catcode `\_12\catcode `\%12\relax}%
\providecommand \@@startlink[1]{}%
\providecommand \@@endlink[0]{}%
\providecommand \url  [0]{\begingroup\@sanitize@url \@url }%
\providecommand \@url [1]{\endgroup\@href {#1}{\urlprefix }}%
\providecommand \urlprefix  [0]{URL }%
\providecommand \Eprint [0]{\href }%
\providecommand \doibase [0]{http://dx.doi.org/}%
\providecommand \selectlanguage [0]{\@gobble}%
\providecommand \bibinfo  [0]{\@secondoftwo}%
\providecommand \bibfield  [0]{\@secondoftwo}%
\providecommand \translation [1]{[#1]}%
\providecommand \BibitemOpen [0]{}%
\providecommand \bibitemStop [0]{}%
\providecommand \bibitemNoStop [0]{.\EOS\space}%
\providecommand \EOS [0]{\spacefactor3000\relax}%
\providecommand \BibitemShut  [1]{\csname bibitem#1\endcsname}%
\let\auto@bib@innerbib\@empty
\bibitem [{\citenamefont {Aasi}\ \emph {et~al.}(2015)\citenamefont {Aasi} \emph
  {et~al.}}]{TheLIGOScientific:2014jea}%
  \BibitemOpen
  \bibfield  {author} {\bibinfo {author} {\bibfnamefont {J.}~\bibnamefont
  {Aasi}} \emph {et~al.} (\bibinfo {collaboration} {LIGO Scientific}),\ }\href
  {\doibase 10.1088/0264-9381/32/7/074001} {\bibfield  {journal} {\bibinfo
  {journal} {Class. Quant. Grav.}\ }\textbf {\bibinfo {volume} {32}},\ \bibinfo
  {pages} {074001} (\bibinfo {year} {2015})},\ \Eprint
  {http://arxiv.org/abs/1411.4547} {arXiv:1411.4547 [gr-qc]} \BibitemShut
  {NoStop}%
\bibitem [{\citenamefont {Acernese}\ \emph {et~al.}(2015)\citenamefont
  {Acernese} \emph {et~al.}}]{TheVirgo:2014hva}%
  \BibitemOpen
  \bibfield  {author} {\bibinfo {author} {\bibfnamefont {F.}~\bibnamefont
  {Acernese}} \emph {et~al.} (\bibinfo {collaboration} {VIRGO}),\ }\href
  {\doibase 10.1088/0264-9381/32/2/024001} {\bibfield  {journal} {\bibinfo
  {journal} {Class. Quant. Grav.}\ }\textbf {\bibinfo {volume} {32}},\ \bibinfo
  {pages} {024001} (\bibinfo {year} {2015})},\ \Eprint
  {http://arxiv.org/abs/1408.3978} {arXiv:1408.3978 [gr-qc]} \BibitemShut
  {NoStop}%
\bibitem [{\citenamefont {Abbott}\ \emph {et~al.}(2016)\citenamefont {Abbott}
  \emph {et~al.}}]{FirstDetection}%
  \BibitemOpen
  \bibfield  {author} {\bibinfo {author} {\bibfnamefont {B.~P.}\ \bibnamefont
  {Abbott}} \emph {et~al.} (\bibinfo {collaboration} {Virgo, LIGO
  Scientific}),\ }\href {\doibase 10.1103/PhysRevLett.116.061102} {\bibfield
  {journal} {\bibinfo  {journal} {Phys. Rev. Lett.}\ }\textbf {\bibinfo
  {volume} {116}},\ \bibinfo {pages} {061102} (\bibinfo {year} {2016})},\
  \Eprint {http://arxiv.org/abs/1602.03837} {arXiv:1602.03837 [gr-qc]}
  \BibitemShut {NoStop}%
\bibitem [{\citenamefont {{LIGO Scientific Collaboration}}\ \emph
  {et~al.}(2013)\citenamefont {{LIGO Scientific Collaboration}}, \citenamefont
  {{Virgo Collaboration}}, \citenamefont {{Aasi}}, \citenamefont {{Abadie}},
  \citenamefont {{Abbott}}, \citenamefont {{Abbott}}, \citenamefont {{Abbott}},
  \citenamefont {{Abernathy}}, \citenamefont {{Accadia}}, \citenamefont
  {{Acernese}},\ and\ \citenamefont {et~al.}}]{LIGOVIRGO:2013}%
  \BibitemOpen
  \bibfield  {author} {\bibinfo {author} {\bibnamefont {{LIGO Scientific
  Collaboration}}}, \bibinfo {author} {\bibnamefont {{Virgo Collaboration}}},
  \bibinfo {author} {\bibfnamefont {J.}~\bibnamefont {{Aasi}}}, \bibinfo
  {author} {\bibfnamefont {J.}~\bibnamefont {{Abadie}}}, \bibinfo {author}
  {\bibfnamefont {B.~P.}\ \bibnamefont {{Abbott}}}, \bibinfo {author}
  {\bibfnamefont {R.}~\bibnamefont {{Abbott}}}, \bibinfo {author}
  {\bibfnamefont {T.~D.}\ \bibnamefont {{Abbott}}}, \bibinfo {author}
  {\bibfnamefont {M.}~\bibnamefont {{Abernathy}}}, \bibinfo {author}
  {\bibfnamefont {T.}~\bibnamefont {{Accadia}}}, \bibinfo {author}
  {\bibfnamefont {F.}~\bibnamefont {{Acernese}}}, \ and\ \bibinfo {author}
  {\bibnamefont {et~al.}},\ }\href@noop {} {\bibfield  {journal} {\bibinfo
  {journal} {ArXiv e-prints}\ } (\bibinfo {year} {2013})},\ \Eprint
  {http://arxiv.org/abs/1304.0670} {arXiv:1304.0670 [gr-qc]} \BibitemShut
  {NoStop}%
\bibitem [{\citenamefont {Abadie}\ \emph {et~al.}(2010)\citenamefont {Abadie}
  \emph {et~al.}}]{Abadie:2010cf}%
  \BibitemOpen
  \bibfield  {author} {\bibinfo {author} {\bibfnamefont {J.}~\bibnamefont
  {Abadie}} \emph {et~al.} (\bibinfo {collaboration} {VIRGO, LIGO
  Scientific}),\ }\href {\doibase 10.1088/0264-9381/27/17/173001} {\bibfield
  {journal} {\bibinfo  {journal} {Class. Quant. Grav.}\ }\textbf {\bibinfo
  {volume} {27}},\ \bibinfo {pages} {173001} (\bibinfo {year} {2010})},\
  \Eprint {http://arxiv.org/abs/1003.2480} {arXiv:1003.2480 [astro-ph.HE]}
  \BibitemShut {NoStop}%
\bibitem [{\citenamefont {Buonanno}\ and\ \citenamefont
  {Damour}(1999)}]{buonanno:1999effective}%
  \BibitemOpen
  \bibfield  {author} {\bibinfo {author} {\bibfnamefont {A.}~\bibnamefont
  {Buonanno}}\ and\ \bibinfo {author} {\bibfnamefont {T.}~\bibnamefont
  {Damour}},\ }\href {\doibase 10.1103/PhysRevD.59.084006} {\bibfield
  {journal} {\bibinfo  {journal} {Phys. Rev.}\ }\textbf {\bibinfo {volume}
  {D59}},\ \bibinfo {pages} {084006} (\bibinfo {year} {1999})},\ \Eprint
  {http://arxiv.org/abs/gr-qc/9811091} {arXiv:gr-qc/9811091 [gr-qc]}
  \BibitemShut {NoStop}%
\bibitem [{\citenamefont {Damour}\ and\ \citenamefont
  {Nagar}(2010)}]{damour:2010effective}%
  \BibitemOpen
  \bibfield  {author} {\bibinfo {author} {\bibfnamefont {T.}~\bibnamefont
  {Damour}}\ and\ \bibinfo {author} {\bibfnamefont {A.}~\bibnamefont {Nagar}},\
  }\href@noop {} {\bibfield  {journal} {\bibinfo  {journal} {Physical Review
  D}\ }\textbf {\bibinfo {volume} {81}},\ \bibinfo {pages} {084016} (\bibinfo
  {year} {2010})}\BibitemShut {NoStop}%
\bibitem [{\citenamefont {Bini}\ \emph {et~al.}(2012)\citenamefont {Bini},
  \citenamefont {Damour},\ and\ \citenamefont {Faye}}]{bini:2012effective}%
  \BibitemOpen
  \bibfield  {author} {\bibinfo {author} {\bibfnamefont {D.}~\bibnamefont
  {Bini}}, \bibinfo {author} {\bibfnamefont {T.}~\bibnamefont {Damour}}, \ and\
  \bibinfo {author} {\bibfnamefont {G.}~\bibnamefont {Faye}},\ }\href@noop {}
  {\bibfield  {journal} {\bibinfo  {journal} {Physical Review D}\ }\textbf
  {\bibinfo {volume} {85}},\ \bibinfo {pages} {124034} (\bibinfo {year}
  {2012})}\BibitemShut {NoStop}%
\bibitem [{\citenamefont {Takami}\ \emph
  {et~al.}(2015{\natexlab{a}})\citenamefont {Takami}, \citenamefont
  {Rezzolla},\ and\ \citenamefont {Baiotti}}]{Takami:2014tva}%
  \BibitemOpen
  \bibfield  {author} {\bibinfo {author} {\bibfnamefont {K.}~\bibnamefont
  {Takami}}, \bibinfo {author} {\bibfnamefont {L.}~\bibnamefont {Rezzolla}}, \
  and\ \bibinfo {author} {\bibfnamefont {L.}~\bibnamefont {Baiotti}},\ }\href
  {\doibase 10.1103/PhysRevD.91.064001} {\bibfield  {journal} {\bibinfo
  {journal} {Phys. Rev.}\ }\textbf {\bibinfo {volume} {D91}},\ \bibinfo {pages}
  {064001} (\bibinfo {year} {2015}{\natexlab{a}})},\ \Eprint
  {http://arxiv.org/abs/1412.3240} {arXiv:1412.3240 [gr-qc]} \BibitemShut
  {NoStop}%
\bibitem [{\citenamefont {Takami}\ \emph
  {et~al.}(2015{\natexlab{b}})\citenamefont {Takami}, \citenamefont
  {Rezzolla},\ and\ \citenamefont {Baiotti}}]{Takami:2015gxa}%
  \BibitemOpen
  \bibfield  {author} {\bibinfo {author} {\bibfnamefont {K.}~\bibnamefont
  {Takami}}, \bibinfo {author} {\bibfnamefont {L.}~\bibnamefont {Rezzolla}}, \
  and\ \bibinfo {author} {\bibfnamefont {L.}~\bibnamefont {Baiotti}},\
  }\bibfield  {booktitle} {\emph {\bibinfo {booktitle} {{Proceedings, Spanish
  Relativity Meeting: Almost 100 years after Einstein Revolution (ERE
  2014)}}},\ }\href {\doibase 10.1088/1742-6596/600/1/012056} {\bibfield
  {journal} {\bibinfo  {journal} {J. Phys. Conf. Ser.}\ }\textbf {\bibinfo
  {volume} {600}},\ \bibinfo {pages} {012056} (\bibinfo {year}
  {2015}{\natexlab{b}})}\BibitemShut {NoStop}%
\bibitem [{\citenamefont {{Douchin}}\ and\ \citenamefont
  {{Haensel}}(2001)}]{Douchin01}%
  \BibitemOpen
  \bibfield  {author} {\bibinfo {author} {\bibfnamefont {F.}~\bibnamefont
  {{Douchin}}}\ and\ \bibinfo {author} {\bibfnamefont {P.}~\bibnamefont
  {{Haensel}}},\ }\href {\doibase 10.1051/0004-6361:20011402} {\bibfield
  {journal} {\bibinfo  {journal} {Astron. Astrophys.}\ }\textbf {\bibinfo
  {volume} {380}},\ \bibinfo {pages} {151} (\bibinfo {year} {2001})},\ \Eprint
  {http://arxiv.org/abs/arXiv:astro-ph/0111092} {arXiv:astro-ph/0111092}
  \BibitemShut {NoStop}%
\bibitem [{\citenamefont {Read}\ \emph {et~al.}(2009)\citenamefont {Read},
  \citenamefont {Lackey}, \citenamefont {Owen},\ and\ \citenamefont
  {Friedman}}]{Read:2009constraints}%
  \BibitemOpen
  \bibfield  {author} {\bibinfo {author} {\bibfnamefont {J.~S.}\ \bibnamefont
  {Read}}, \bibinfo {author} {\bibfnamefont {B.~D.}\ \bibnamefont {Lackey}},
  \bibinfo {author} {\bibfnamefont {B.~J.}\ \bibnamefont {Owen}}, \ and\
  \bibinfo {author} {\bibfnamefont {J.~L.}\ \bibnamefont {Friedman}},\
  }\href@noop {} {\bibfield  {journal} {\bibinfo  {journal} {Physical Review
  D}\ }\textbf {\bibinfo {volume} {79}},\ \bibinfo {pages} {124032} (\bibinfo
  {year} {2009})}\BibitemShut {NoStop}%
\bibitem [{\citenamefont {{Harten}}\ \emph {et~al.}(1987)\citenamefont
  {{Harten}}, \citenamefont {{Engquist}}, \citenamefont {{Osher}},\ and\
  \citenamefont {{Chakravarthy}}}]{Harten:1987un}%
  \BibitemOpen
  \bibfield  {author} {\bibinfo {author} {\bibfnamefont {A.}~\bibnamefont
  {{Harten}}}, \bibinfo {author} {\bibfnamefont {B.}~\bibnamefont
  {{Engquist}}}, \bibinfo {author} {\bibfnamefont {S.}~\bibnamefont {{Osher}}},
  \ and\ \bibinfo {author} {\bibfnamefont {S.~R.}\ \bibnamefont
  {{Chakravarthy}}},\ }\href {\doibase 10.1016/0021-9991(87)90031-3} {\bibfield
   {journal} {\bibinfo  {journal} {J. Comp. Phys.}\ }\textbf {\bibinfo {volume}
  {71}},\ \bibinfo {pages} {231} (\bibinfo {year} {1987})}\BibitemShut
  {NoStop}%
\bibitem [{\citenamefont {Shu}(1999)}]{Shu:1999ho}%
  \BibitemOpen
  \bibfield  {author} {\bibinfo {author} {\bibfnamefont {C.~W.}\ \bibnamefont
  {Shu}},\ }in\ \href@noop {} {\emph {\bibinfo {booktitle} {High order methods
  for computational physics}}},\ \bibinfo {editor} {edited by\ \bibinfo
  {editor} {\bibfnamefont {T.~J.}\ \bibnamefont {Barth}}\ and\ \bibinfo
  {editor} {\bibfnamefont {H.~A.}\ \bibnamefont {Deconinck}}}\ (\bibinfo
  {publisher} {Springer},\ \bibinfo {address} {New York},\ \bibinfo {year}
  {1999})\ pp.\ \bibinfo {pages} {439--582}\BibitemShut {NoStop}%
\bibitem [{\citenamefont {Nakamura}\ \emph {et~al.}(1987)\citenamefont
  {Nakamura}, \citenamefont {Oohara},\ and\ \citenamefont
  {Kojima}}]{Nakamura:1987zz}%
  \BibitemOpen
  \bibfield  {author} {\bibinfo {author} {\bibfnamefont {T.}~\bibnamefont
  {Nakamura}}, \bibinfo {author} {\bibfnamefont {K.}~\bibnamefont {Oohara}}, \
  and\ \bibinfo {author} {\bibfnamefont {Y.}~\bibnamefont {Kojima}},\
  }\href@noop {} {\bibfield  {journal} {\bibinfo  {journal} {Prog. Theor. Phys.
  Suppl.}\ }\textbf {\bibinfo {volume} {90}},\ \bibinfo {pages} {1} (\bibinfo
  {year} {1987})}\BibitemShut {NoStop}%
\bibitem [{\citenamefont {Shibata}\ and\ \citenamefont
  {Nakamura}(1995)}]{Shibata:1995we}%
  \BibitemOpen
  \bibfield  {author} {\bibinfo {author} {\bibfnamefont {M.}~\bibnamefont
  {Shibata}}\ and\ \bibinfo {author} {\bibfnamefont {T.}~\bibnamefont
  {Nakamura}},\ }\href {\doibase 10.1103/PhysRevD.52.5428} {\bibfield
  {journal} {\bibinfo  {journal} {Phys. Rev. D}\ }\textbf {\bibinfo {volume}
  {52}},\ \bibinfo {pages} {5428} (\bibinfo {year} {1995})}\BibitemShut
  {NoStop}%
\bibitem [{\citenamefont {Baumgarte}\ and\ \citenamefont
  {Shapiro}(1999)}]{Baumgarte:1998te}%
  \BibitemOpen
  \bibfield  {author} {\bibinfo {author} {\bibfnamefont {T.~W.}\ \bibnamefont
  {Baumgarte}}\ and\ \bibinfo {author} {\bibfnamefont {S.~L.}\ \bibnamefont
  {Shapiro}},\ }\href {\doibase 10.1103/PhysRevD.59.024007} {\bibfield
  {journal} {\bibinfo  {journal} {Phys. Rev. D}\ }\textbf {\bibinfo {volume}
  {59}},\ \bibinfo {pages} {024007} (\bibinfo {year} {1999})},\ \Eprint
  {http://arxiv.org/abs/arXiv:gr-qc/9810065} {arXiv:gr-qc/9810065} \BibitemShut
  {NoStop}%
\bibitem [{\citenamefont {Alcubierre}\ \emph {et~al.}(2000)\citenamefont
  {Alcubierre}, \citenamefont {Br{\"u}gmann}, \citenamefont {Dramlitsch},
  \citenamefont {Font}, \citenamefont {Papadopoulos}, \citenamefont {Seidel},
  \citenamefont {Stergioulas},\ and\ \citenamefont
  {Takahashi}}]{Alcubierre:2000xu}%
  \BibitemOpen
  \bibfield  {author} {\bibinfo {author} {\bibfnamefont {M.}~\bibnamefont
  {Alcubierre}}, \bibinfo {author} {\bibfnamefont {B.}~\bibnamefont
  {Br{\"u}gmann}}, \bibinfo {author} {\bibfnamefont {T.}~\bibnamefont
  {Dramlitsch}}, \bibinfo {author} {\bibfnamefont {J.~A.}\ \bibnamefont
  {Font}}, \bibinfo {author} {\bibfnamefont {P.}~\bibnamefont {Papadopoulos}},
  \bibinfo {author} {\bibfnamefont {E.}~\bibnamefont {Seidel}}, \bibinfo
  {author} {\bibfnamefont {N.}~\bibnamefont {Stergioulas}}, \ and\ \bibinfo
  {author} {\bibfnamefont {R.}~\bibnamefont {Takahashi}},\ }\href {\doibase
  10.1103/PhysRevD.62.044034} {\bibfield  {journal} {\bibinfo  {journal} {Phys.
  Rev. D}\ }\textbf {\bibinfo {volume} {62}},\ \bibinfo {pages} {044034}
  (\bibinfo {year} {2000})},\ \Eprint
  {http://arxiv.org/abs/arXiv:gr-qc/0003071} {arXiv:gr-qc/0003071} \BibitemShut
  {NoStop}%
\bibitem [{\citenamefont {Alcubierre}\ \emph {et~al.}(2003)\citenamefont
  {Alcubierre}, \citenamefont {Br{\"u}gmann}, \citenamefont {Diener},
  \citenamefont {Koppitz}, \citenamefont {Pollney}, \citenamefont {Seidel},\
  and\ \citenamefont {Takahashi}}]{Alcubierre:2002kk}%
  \BibitemOpen
  \bibfield  {author} {\bibinfo {author} {\bibfnamefont {M.}~\bibnamefont
  {Alcubierre}}, \bibinfo {author} {\bibfnamefont {B.}~\bibnamefont
  {Br{\"u}gmann}}, \bibinfo {author} {\bibfnamefont {P.}~\bibnamefont
  {Diener}}, \bibinfo {author} {\bibfnamefont {M.}~\bibnamefont {Koppitz}},
  \bibinfo {author} {\bibfnamefont {D.}~\bibnamefont {Pollney}}, \bibinfo
  {author} {\bibfnamefont {E.}~\bibnamefont {Seidel}}, \ and\ \bibinfo {author}
  {\bibfnamefont {R.}~\bibnamefont {Takahashi}},\ }\href {\doibase
  10.1103/PhysRevD.67.084023} {\bibfield  {journal} {\bibinfo  {journal} {Phys.
  Rev. D}\ }\textbf {\bibinfo {volume} {67}},\ \bibinfo {pages} {084023}
  (\bibinfo {year} {2003})},\ \Eprint
  {http://arxiv.org/abs/arXiv:gr-qc/0206072} {arXiv:gr-qc/0206072} \BibitemShut
  {NoStop}%
\bibitem [{\citenamefont {Alic}\ \emph {et~al.}(2012)\citenamefont {Alic},
  \citenamefont {Bona-Casas}, \citenamefont {Bona}, \citenamefont {Rezzolla},\
  and\ \citenamefont {Palenzuela}}]{Alic:2011gg}%
  \BibitemOpen
  \bibfield  {author} {\bibinfo {author} {\bibfnamefont {D.}~\bibnamefont
  {Alic}}, \bibinfo {author} {\bibfnamefont {C.}~\bibnamefont {Bona-Casas}},
  \bibinfo {author} {\bibfnamefont {C.}~\bibnamefont {Bona}}, \bibinfo {author}
  {\bibfnamefont {L.}~\bibnamefont {Rezzolla}}, \ and\ \bibinfo {author}
  {\bibfnamefont {C.}~\bibnamefont {Palenzuela}},\ }\href {\doibase
  10.1103/PhysRevD.85.064040} {\bibfield  {journal} {\bibinfo  {journal} {Phys.
  Rev.}\ }\textbf {\bibinfo {volume} {D85}},\ \bibinfo {pages} {064040}
  (\bibinfo {year} {2012})},\ \Eprint {http://arxiv.org/abs/1106.2254}
  {arXiv:1106.2254 [gr-qc]} \BibitemShut {NoStop}%
\bibitem [{\citenamefont {L{\"{o}}ffler}\ \emph {et~al.}(2012)\citenamefont
  {L{\"{o}}ffler}, \citenamefont {Faber}, \citenamefont {Bentivegna},
  \citenamefont {Bode}, \citenamefont {Diener}, \citenamefont {Haas},
  \citenamefont {Hinder}, \citenamefont {Mundim}, \citenamefont {Ott},
  \citenamefont {Schnetter}, \citenamefont {Allen}, \citenamefont
  {Campanelli},\ and\ \citenamefont {Laguna}}]{Loffler:2011ay}%
  \BibitemOpen
  \bibfield  {author} {\bibinfo {author} {\bibfnamefont {F.}~\bibnamefont
  {L{\"{o}}ffler}}, \bibinfo {author} {\bibfnamefont {J.}~\bibnamefont
  {Faber}}, \bibinfo {author} {\bibfnamefont {E.}~\bibnamefont {Bentivegna}},
  \bibinfo {author} {\bibfnamefont {T.}~\bibnamefont {Bode}}, \bibinfo {author}
  {\bibfnamefont {P.}~\bibnamefont {Diener}}, \bibinfo {author} {\bibfnamefont
  {R.}~\bibnamefont {Haas}}, \bibinfo {author} {\bibfnamefont {I.}~\bibnamefont
  {Hinder}}, \bibinfo {author} {\bibfnamefont {B.~C.}\ \bibnamefont {Mundim}},
  \bibinfo {author} {\bibfnamefont {C.~D.}\ \bibnamefont {Ott}}, \bibinfo
  {author} {\bibfnamefont {E.}~\bibnamefont {Schnetter}}, \bibinfo {author}
  {\bibfnamefont {G.}~\bibnamefont {Allen}}, \bibinfo {author} {\bibfnamefont
  {M.}~\bibnamefont {Campanelli}}, \ and\ \bibinfo {author} {\bibfnamefont
  {P.}~\bibnamefont {Laguna}},\ }\href {\doibase
  doi:10.1088/0264-9381/29/11/115001} {\bibfield  {journal} {\bibinfo
  {journal} {Class. Quantum Grav.}\ }\textbf {\bibinfo {volume} {29}},\
  \bibinfo {pages} {115001} (\bibinfo {year} {2012})},\ \Eprint
  {http://arxiv.org/abs/arXiv:1111.3344 [gr-qc]} {arXiv:1111.3344 [gr-qc]}
  \BibitemShut {NoStop}%
\bibitem [{\citenamefont {De~Pietri}\ \emph {et~al.}(2014)\citenamefont
  {De~Pietri}, \citenamefont {Feo}, \citenamefont {Franci},\ and\ \citenamefont
  {Löffler}}]{DePietri:2014mea}%
  \BibitemOpen
  \bibfield  {author} {\bibinfo {author} {\bibfnamefont {R.}~\bibnamefont
  {De~Pietri}}, \bibinfo {author} {\bibfnamefont {A.}~\bibnamefont {Feo}},
  \bibinfo {author} {\bibfnamefont {L.}~\bibnamefont {Franci}}, \ and\ \bibinfo
  {author} {\bibfnamefont {F.}~\bibnamefont {Löffler}},\ }\href {\doibase
  10.1103/PhysRevD.90.024034} {\bibfield  {journal} {\bibinfo  {journal} {Phys.
  Rev.}\ }\textbf {\bibinfo {volume} {D90}},\ \bibinfo {pages} {024034}
  (\bibinfo {year} {2014})},\ \Eprint {http://arxiv.org/abs/1403.8066}
  {arXiv:1403.8066 [gr-qc]} \BibitemShut {NoStop}%
\bibitem [{\citenamefont {Löffler}\ \emph {et~al.}(2015)\citenamefont
  {Löffler}, \citenamefont {De~Pietri}, \citenamefont {Feo}, \citenamefont
  {Maione},\ and\ \citenamefont {Franci}}]{Loffler:2014jma}%
  \BibitemOpen
  \bibfield  {author} {\bibinfo {author} {\bibfnamefont {F.}~\bibnamefont
  {Löffler}}, \bibinfo {author} {\bibfnamefont {R.}~\bibnamefont {De~Pietri}},
  \bibinfo {author} {\bibfnamefont {A.}~\bibnamefont {Feo}}, \bibinfo {author}
  {\bibfnamefont {F.}~\bibnamefont {Maione}}, \ and\ \bibinfo {author}
  {\bibfnamefont {L.}~\bibnamefont {Franci}},\ }\href {\doibase
  10.1103/PhysRevD.91.064057} {\bibfield  {journal} {\bibinfo  {journal} {Phys.
  Rev.}\ }\textbf {\bibinfo {volume} {D91}},\ \bibinfo {pages} {064057}
  (\bibinfo {year} {2015})},\ \Eprint {http://arxiv.org/abs/1411.1963}
  {arXiv:1411.1963 [gr-qc]} \BibitemShut {NoStop}%
\bibitem [{\citenamefont {Bauswein}\ \emph {et~al.}(2010)\citenamefont
  {Bauswein}, \citenamefont {Janka},\ and\ \citenamefont
  {Oechslin}}]{bauswein:2010testing}%
  \BibitemOpen
  \bibfield  {author} {\bibinfo {author} {\bibfnamefont {A.}~\bibnamefont
  {Bauswein}}, \bibinfo {author} {\bibfnamefont {H.-T.}\ \bibnamefont {Janka}},
  \ and\ \bibinfo {author} {\bibfnamefont {R.}~\bibnamefont {Oechslin}},\
  }\href@noop {} {\bibfield  {journal} {\bibinfo  {journal} {Physical Review
  D}\ }\textbf {\bibinfo {volume} {82}},\ \bibinfo {pages} {084043} (\bibinfo
  {year} {2010})}\BibitemShut {NoStop}%
\bibitem [{\citenamefont {Gourgoulhon}\ \emph {et~al.}(2001)\citenamefont
  {Gourgoulhon}, \citenamefont {Grandclement}, \citenamefont {Taniguchi},
  \citenamefont {Marck},\ and\ \citenamefont {Bonazzola}}]{Gourgoulhon:2000nn}%
  \BibitemOpen
  \bibfield  {author} {\bibinfo {author} {\bibfnamefont {E.}~\bibnamefont
  {Gourgoulhon}}, \bibinfo {author} {\bibfnamefont {P.}~\bibnamefont
  {Grandclement}}, \bibinfo {author} {\bibfnamefont {K.}~\bibnamefont
  {Taniguchi}}, \bibinfo {author} {\bibfnamefont {J.-A.}\ \bibnamefont
  {Marck}}, \ and\ \bibinfo {author} {\bibfnamefont {S.}~\bibnamefont
  {Bonazzola}},\ }\href {\doibase 10.1103/PhysRevD.63.064029} {\bibfield
  {journal} {\bibinfo  {journal} {Phys. Rev. D}\ }\textbf {\bibinfo {volume}
  {63}},\ \bibinfo {pages} {064029} (\bibinfo {year} {2001})},\ \Eprint
  {http://arxiv.org/abs/arXiv:gr-qc/0007028} {arXiv:gr-qc/0007028} \BibitemShut
  {NoStop}%
\bibitem [{LORENE()}]{lorene:web}%
  \BibitemOpen
  LORENE,\ \href {http://www.lorene.obspm.fr/} {\enquote {\bibinfo {title}
  {{LORENE}: {L}angage {O}bjet pour la {RE}lativit\'e {N}um\'eriqu{E}},}\ }\url
  {http://www.lorene.obspm.fr/}\BibitemShut {NoStop}%
\bibitem [{EinsteinToolkit()}]{EinsteinToolkit:web}%
  \BibitemOpen
  EinsteinToolkit,\ \href {http://einsteintoolkit.org/} {\enquote {\bibinfo
  {title} {{Einstein Toolkit}: Open software for relativistic astrophysics},}\
  }\url {http://einsteintoolkit.org/}\BibitemShut {NoStop}%
\bibitem [{Cactus developers()}]{Cactuscode:web}%
  \BibitemOpen
  Cactus developers,\ \href {http://www.cactuscode.org/} {\enquote {\bibinfo
  {title} {{Cactus Computational Toolkit}},}\ }\url
  {http://www.cactuscode.org/}\BibitemShut {NoStop}%
\bibitem [{\citenamefont {Goodale}\ \emph {et~al.}(2003)\citenamefont
  {Goodale}, \citenamefont {Allen}, \citenamefont {Lanfermann}, \citenamefont
  {Mass{\'o}}, \citenamefont {Radke}, \citenamefont {Seidel},\ and\
  \citenamefont {Shalf}}]{Goodale:2002a}%
  \BibitemOpen
  \bibfield  {author} {\bibinfo {author} {\bibfnamefont {T.}~\bibnamefont
  {Goodale}}, \bibinfo {author} {\bibfnamefont {G.}~\bibnamefont {Allen}},
  \bibinfo {author} {\bibfnamefont {G.}~\bibnamefont {Lanfermann}}, \bibinfo
  {author} {\bibfnamefont {J.}~\bibnamefont {Mass{\'o}}}, \bibinfo {author}
  {\bibfnamefont {T.}~\bibnamefont {Radke}}, \bibinfo {author} {\bibfnamefont
  {E.}~\bibnamefont {Seidel}}, \ and\ \bibinfo {author} {\bibfnamefont
  {J.}~\bibnamefont {Shalf}},\ }in\ \href {http://edoc.mpg.de/3341} {\emph
  {\bibinfo {booktitle} {Vector and Parallel Processing -- VECPAR'2002, 5th
  International Conference, Lecture Notes in Computer Science}}}\ (\bibinfo
  {publisher} {Springer},\ \bibinfo {address} {Berlin},\ \bibinfo {year}
  {2003})\BibitemShut {NoStop}%
\bibitem [{\citenamefont {Allen}\ \emph {et~al.}(2011)\citenamefont {Allen},
  \citenamefont {Goodale}, \citenamefont {Lanfermann}, \citenamefont {Radke},
  \citenamefont {Rideout},\ and\ \citenamefont
  {Thornburg}}]{CactusUsersGuide:web}%
  \BibitemOpen
  \bibfield  {author} {\bibinfo {author} {\bibfnamefont {G.}~\bibnamefont
  {Allen}}, \bibinfo {author} {\bibfnamefont {T.}~\bibnamefont {Goodale}},
  \bibinfo {author} {\bibfnamefont {G.}~\bibnamefont {Lanfermann}}, \bibinfo
  {author} {\bibfnamefont {T.}~\bibnamefont {Radke}}, \bibinfo {author}
  {\bibfnamefont {D.}~\bibnamefont {Rideout}}, \ and\ \bibinfo {author}
  {\bibfnamefont {J.}~\bibnamefont {Thornburg}},\ }\href
  {http://www.cactuscode.org/Guides/Stable/UsersGuide/UsersGuideStable.pdf}
  {\emph {\bibinfo {title} {{C}actus Users' Guide}}} (\bibinfo {year}
  {2011})\BibitemShut {NoStop}%
\bibitem [{\citenamefont {Schnetter}\ \emph {et~al.}(2004)\citenamefont
  {Schnetter}, \citenamefont {Hawley},\ and\ \citenamefont
  {Hawke}}]{Schnetter:2003rb}%
  \BibitemOpen
  \bibfield  {author} {\bibinfo {author} {\bibfnamefont {E.}~\bibnamefont
  {Schnetter}}, \bibinfo {author} {\bibfnamefont {S.~H.}\ \bibnamefont
  {Hawley}}, \ and\ \bibinfo {author} {\bibfnamefont {I.}~\bibnamefont
  {Hawke}},\ }\href {\doibase 10.1088/0264-9381/21/6/014} {\bibfield  {journal}
  {\bibinfo  {journal} {Class. Quantum Grav.}\ }\textbf {\bibinfo {volume}
  {21}},\ \bibinfo {pages} {1465} (\bibinfo {year} {2004})},\ \Eprint
  {http://arxiv.org/abs/arXiv:gr-qc/0310042} {arXiv:gr-qc/0310042} \BibitemShut
  {NoStop}%
\bibitem [{\citenamefont {Schnetter}\ \emph {et~al.}(2006)\citenamefont
  {Schnetter}, \citenamefont {Diener}, \citenamefont {Dorband},\ and\
  \citenamefont {Tiglio}}]{Schnetter:2006pg}%
  \BibitemOpen
  \bibfield  {author} {\bibinfo {author} {\bibfnamefont {E.}~\bibnamefont
  {Schnetter}}, \bibinfo {author} {\bibfnamefont {P.}~\bibnamefont {Diener}},
  \bibinfo {author} {\bibfnamefont {E.~N.}\ \bibnamefont {Dorband}}, \ and\
  \bibinfo {author} {\bibfnamefont {M.}~\bibnamefont {Tiglio}},\ }\href
  {\doibase 10.1088/0264-9381/23/16/S14} {\bibfield  {journal} {\bibinfo
  {journal} {Class. Quantum Grav.}\ }\textbf {\bibinfo {volume} {23}},\
  \bibinfo {pages} {S553} (\bibinfo {year} {2006})},\ \Eprint
  {http://arxiv.org/abs/arXiv:gr-qc/0602104} {arXiv:gr-qc/0602104} \BibitemShut
  {NoStop}%
\bibitem [{Carpet()}]{CarpetCode:web}%
  \BibitemOpen
  Carpet,\ \href {http://www.carpetcode.org/} {}\bibinfo {note} {{Carpet}:
  Adaptive Mesh Refinement for the {Cactus} Framework},\ \url
  {http://www.carpetcode.org/}\BibitemShut {NoStop}%
\bibitem [{\citenamefont {Baiotti}\ \emph {et~al.}(2005)\citenamefont
  {Baiotti}, \citenamefont {Hawke}, \citenamefont {Montero}, \citenamefont
  {L{\"o}ffler}, \citenamefont {Rezzolla}, \citenamefont {Stergioulas},
  \citenamefont {Font},\ and\ \citenamefont {Seidel}}]{Baiotti:2004wn}%
  \BibitemOpen
  \bibfield  {author} {\bibinfo {author} {\bibfnamefont {L.}~\bibnamefont
  {Baiotti}}, \bibinfo {author} {\bibfnamefont {I.}~\bibnamefont {Hawke}},
  \bibinfo {author} {\bibfnamefont {P.~J.}\ \bibnamefont {Montero}}, \bibinfo
  {author} {\bibfnamefont {F.}~\bibnamefont {L{\"o}ffler}}, \bibinfo {author}
  {\bibfnamefont {L.}~\bibnamefont {Rezzolla}}, \bibinfo {author}
  {\bibfnamefont {N.}~\bibnamefont {Stergioulas}}, \bibinfo {author}
  {\bibfnamefont {J.~A.}\ \bibnamefont {Font}}, \ and\ \bibinfo {author}
  {\bibfnamefont {E.}~\bibnamefont {Seidel}},\ }\href {\doibase
  10.1103/PhysRevD.71.024035} {\bibfield  {journal} {\bibinfo  {journal} {Phys.
  Rev. D}\ }\textbf {\bibinfo {volume} {71}},\ \bibinfo {pages} {024035}
  (\bibinfo {year} {2005})},\ \Eprint
  {http://arxiv.org/abs/arXiv:gr-qc/0403029} {arXiv:gr-qc/0403029} \BibitemShut
  {NoStop}%
\bibitem [{\citenamefont {Hawke}\ \emph {et~al.}(2005)\citenamefont {Hawke},
  \citenamefont {L{\"o}ffler},\ and\ \citenamefont {Nerozzi}}]{Hawke:2005zw}%
  \BibitemOpen
  \bibfield  {author} {\bibinfo {author} {\bibfnamefont {I.}~\bibnamefont
  {Hawke}}, \bibinfo {author} {\bibfnamefont {F.}~\bibnamefont {L{\"o}ffler}},
  \ and\ \bibinfo {author} {\bibfnamefont {A.}~\bibnamefont {Nerozzi}},\ }\href
  {\doibase 10.1103/PhysRevD.71.104006} {\bibfield  {journal} {\bibinfo
  {journal} {Phys. Rev. D}\ }\textbf {\bibinfo {volume} {71}},\ \bibinfo
  {pages} {104006} (\bibinfo {year} {2005})},\ \Eprint
  {http://arxiv.org/abs/arXiv:gr-qc/0501054} {arXiv:gr-qc/0501054} \BibitemShut
  {NoStop}%
\bibitem [{\citenamefont {M{\"o}sta}\ \emph {et~al.}(2014)\citenamefont
  {M{\"o}sta}, \citenamefont {Mundim}, \citenamefont {Faber}, \citenamefont
  {Haas}, \citenamefont {Noble}, \citenamefont {Bode}, \citenamefont
  {L{\"o}ffler}, \citenamefont {Ott}, \citenamefont {Reisswig},\ and\
  \citenamefont {Schnetter}}]{Moesta:2013dna}%
  \BibitemOpen
  \bibfield  {author} {\bibinfo {author} {\bibfnamefont {P.}~\bibnamefont
  {M{\"o}sta}}, \bibinfo {author} {\bibfnamefont {B.~C.}\ \bibnamefont
  {Mundim}}, \bibinfo {author} {\bibfnamefont {J.~A.}\ \bibnamefont {Faber}},
  \bibinfo {author} {\bibfnamefont {R.}~\bibnamefont {Haas}}, \bibinfo {author}
  {\bibfnamefont {S.~C.}\ \bibnamefont {Noble}}, \bibinfo {author}
  {\bibfnamefont {T.}~\bibnamefont {Bode}}, \bibinfo {author} {\bibfnamefont
  {F.}~\bibnamefont {L{\"o}ffler}}, \bibinfo {author} {\bibfnamefont {C.~D.}\
  \bibnamefont {Ott}}, \bibinfo {author} {\bibfnamefont {C.}~\bibnamefont
  {Reisswig}}, \ and\ \bibinfo {author} {\bibfnamefont {E.}~\bibnamefont
  {Schnetter}},\ }\href {\doibase 10.1088/0264-9381/31/1/015005} {\bibfield
  {journal} {\bibinfo  {journal} {Classical and Quantum Gravity}\ }\textbf
  {\bibinfo {volume} {31}},\ \bibinfo {pages} {015005} (\bibinfo {year}
  {2014})},\ \Eprint {http://arxiv.org/abs/arXiv:1304.5544 [gr-qc]}
  {arXiv:1304.5544 [gr-qc]} \BibitemShut {NoStop}%
\bibitem [{McLachlan()}]{McLachlan:web}%
  \BibitemOpen
  McLachlan,\ \href {http://www.cct.lsu.edu/~eschnett/McLachlan/} {\enquote
  {\bibinfo {title} {{McLachlan}, a public {BSSN} code},}\ }\url
  {http://www.cct.lsu.edu/~eschnett/McLachlan/}\BibitemShut {NoStop}%
\bibitem [{\citenamefont {Husa}\ \emph {et~al.}(2006)\citenamefont {Husa},
  \citenamefont {Hinder},\ and\ \citenamefont {Lechner}}]{Husa:2004ip}%
  \BibitemOpen
  \bibfield  {author} {\bibinfo {author} {\bibfnamefont {S.}~\bibnamefont
  {Husa}}, \bibinfo {author} {\bibfnamefont {I.}~\bibnamefont {Hinder}}, \ and\
  \bibinfo {author} {\bibfnamefont {C.}~\bibnamefont {Lechner}},\ }\href@noop
  {} {\bibfield  {journal} {\bibinfo  {journal} {Comput. Phys. Commun.}\
  }\textbf {\bibinfo {volume} {174}},\ \bibinfo {pages} {983} (\bibinfo {year}
  {2006})},\ \Eprint {http://arxiv.org/abs/arXiv:gr-qc/0404023}
  {arXiv:gr-qc/0404023} \BibitemShut {NoStop}%
\bibitem [{\citenamefont {{Lechner}}\ \emph {et~al.}(2004)\citenamefont
  {{Lechner}}, \citenamefont {{Alic}},\ and\ \citenamefont
  {{Husa}}}]{Lechner:2004cs}%
  \BibitemOpen
  \bibfield  {author} {\bibinfo {author} {\bibfnamefont {C.}~\bibnamefont
  {{Lechner}}}, \bibinfo {author} {\bibfnamefont {D.}~\bibnamefont {{Alic}}}, \
  and\ \bibinfo {author} {\bibfnamefont {S.}~\bibnamefont {{Husa}}},\
  }\href@noop {} {\bibfield  {journal} {\bibinfo  {journal} {Analele
  Universitatii de Vest din Timisoara, Seria Matematica-Informatica}\ }\textbf
  {\bibinfo {volume} {42}} (\bibinfo {year} {2004})},\ \Eprint
  {http://arxiv.org/abs/arXiv:cs/0411063} {arXiv:cs/0411063} \BibitemShut
  {NoStop}%
\bibitem [{Kranc()}]{Kranc:web}%
  \BibitemOpen
  Kranc,\ \href {http://kranccode.org/} {\enquote {\bibinfo {title} {{Kranc}:
  {Kranc} assembles numerical code},}\ }\url
  {http://kranccode.org/}\BibitemShut {NoStop}%
\bibitem [{\citenamefont {Runge}(1895)}]{Runge:1895aa}%
  \BibitemOpen
  \bibfield  {author} {\bibinfo {author} {\bibfnamefont {C.}~\bibnamefont
  {Runge}},\ }\href {\doibase 10.1007/BF01446807} {\bibfield  {journal}
  {\bibinfo  {journal} {Mathematische Annalen}\ }\textbf {\bibinfo {volume}
  {46}},\ \bibinfo {pages} {167} (\bibinfo {year} {1895})}\BibitemShut
  {NoStop}%
\bibitem [{\citenamefont {Kutta}(1901)}]{Kutta:1901aa}%
  \BibitemOpen
  \bibfield  {author} {\bibinfo {author} {\bibfnamefont {W.}~\bibnamefont
  {Kutta}},\ }\href@noop {} {\bibfield  {journal} {\bibinfo  {journal} {Z.
  Math. Phys.}\ }\textbf {\bibinfo {volume} {46}},\ \bibinfo {pages} {435}
  (\bibinfo {year} {1901})}\BibitemShut {NoStop}%
\bibitem [{\citenamefont {Harten}\ \emph {et~al.}(1983)\citenamefont {Harten},
  \citenamefont {Lax},\ and\ \citenamefont {van Leer}}]{Harten:1983on}%
  \BibitemOpen
  \bibfield  {author} {\bibinfo {author} {\bibfnamefont {A.}~\bibnamefont
  {Harten}}, \bibinfo {author} {\bibfnamefont {P.~D.}\ \bibnamefont {Lax}}, \
  and\ \bibinfo {author} {\bibfnamefont {B.}~\bibnamefont {van Leer}},\
  }\href@noop {} {\bibfield  {journal} {\bibinfo  {journal} {SIAM review}\
  }\textbf {\bibinfo {volume} {25}},\ \bibinfo {pages} {35} (\bibinfo {year}
  {1983})}\BibitemShut {NoStop}%
\bibitem [{\citenamefont {Einfeldt}(1988)}]{Einfeldt:1988og}%
  \BibitemOpen
  \bibfield  {author} {\bibinfo {author} {\bibfnamefont {B.}~\bibnamefont
  {Einfeldt}},\ }\href {\doibase 10.1137/0725021} {\bibfield  {journal}
  {\bibinfo  {journal} {SIAM J. Numer. Anal.}\ }\textbf {\bibinfo {volume}
  {25}},\ \bibinfo {pages} {294} (\bibinfo {year} {1988})}\BibitemShut
  {NoStop}%
\bibitem [{\citenamefont {Colella}\ and\ \citenamefont
  {Woodward}(1984)}]{Colella:1982ee}%
  \BibitemOpen
  \bibfield  {author} {\bibinfo {author} {\bibfnamefont {P.}~\bibnamefont
  {Colella}}\ and\ \bibinfo {author} {\bibfnamefont {P.~R.}\ \bibnamefont
  {Woodward}},\ }\href {\doibase 10.1016/0021-9991(84)90143-8} {\bibfield
  {journal} {\bibinfo  {journal} {J. Comp. Phys.}\ }\textbf {\bibinfo {volume}
  {54}},\ \bibinfo {pages} {174} (\bibinfo {year} {1984})}\BibitemShut
  {NoStop}%
\bibitem [{\citenamefont {Radice}\ and\ \citenamefont
  {Rezzolla}(1997)}]{suresh:97}%
  \BibitemOpen
  \bibfield  {author} {\bibinfo {author} {\bibfnamefont {D.}~\bibnamefont
  {Radice}}\ and\ \bibinfo {author} {\bibfnamefont {L.}~\bibnamefont
  {Rezzolla}},\ }\href@noop {} {\bibfield  {journal} {\bibinfo  {journal}
  {Journal of Computational Physics}\ }\textbf {\bibinfo {volume} {136}},\
  \bibinfo {pages} {83–99} (\bibinfo {year} {1997})}\BibitemShut {NoStop}%
\bibitem [{\citenamefont {Newman}\ and\ \citenamefont
  {Penrose}(1962)}]{Newman62}%
  \BibitemOpen
  \bibfield  {author} {\bibinfo {author} {\bibfnamefont {E.}~\bibnamefont
  {Newman}}\ and\ \bibinfo {author} {\bibfnamefont {R.}~\bibnamefont
  {Penrose}},\ }\href@noop {} {\bibfield  {journal} {\bibinfo  {journal}
  {Journal of Mathematical Physics}\ }\textbf {\bibinfo {volume} {3}},\
  \bibinfo {pages} {566} (\bibinfo {year} {1962})}\BibitemShut {NoStop}%
\bibitem [{\citenamefont {Regge}\ and\ \citenamefont
  {Wheeler}(1957)}]{Regge:1957td}%
  \BibitemOpen
  \bibfield  {author} {\bibinfo {author} {\bibfnamefont {T.}~\bibnamefont
  {Regge}}\ and\ \bibinfo {author} {\bibfnamefont {J.~A.}\ \bibnamefont
  {Wheeler}},\ }\href {\doibase 10.1103/PhysRev.108.1063} {\bibfield  {journal}
  {\bibinfo  {journal} {Phys. Rev.}\ }\textbf {\bibinfo {volume} {108}},\
  \bibinfo {pages} {1063} (\bibinfo {year} {1957})}\BibitemShut {NoStop}%
\bibitem [{\citenamefont {Zerilli}(1970)}]{Zerilli:1970se}%
  \BibitemOpen
  \bibfield  {author} {\bibinfo {author} {\bibfnamefont {F.~J.}\ \bibnamefont
  {Zerilli}},\ }\href {\doibase 10.1103/PhysRevLett.24.737} {\bibfield
  {journal} {\bibinfo  {journal} {Phys. Rev. Lett.}\ }\textbf {\bibinfo
  {volume} {24}},\ \bibinfo {pages} {737} (\bibinfo {year} {1970})}\BibitemShut
  {NoStop}%
\bibitem [{\citenamefont {Moncrief}(1974)}]{Moncrief:1974am}%
  \BibitemOpen
  \bibfield  {author} {\bibinfo {author} {\bibfnamefont {V.}~\bibnamefont
  {Moncrief}},\ }\href {\doibase 10.1016/0003-4916(74)90173-0} {\bibfield
  {journal} {\bibinfo  {journal} {Annals Phys.}\ }\textbf {\bibinfo {volume}
  {88}},\ \bibinfo {pages} {323} (\bibinfo {year} {1974})}\BibitemShut
  {NoStop}%
\bibitem [{\citenamefont {Baker}\ \emph {et~al.}(2002)\citenamefont {Baker},
  \citenamefont {Campanelli},\ and\ \citenamefont {Lousto}}]{Baker:2001sf}%
  \BibitemOpen
  \bibfield  {author} {\bibinfo {author} {\bibfnamefont {J.~G.}\ \bibnamefont
  {Baker}}, \bibinfo {author} {\bibfnamefont {M.}~\bibnamefont {Campanelli}}, \
  and\ \bibinfo {author} {\bibfnamefont {C.~O.}\ \bibnamefont {Lousto}},\
  }\href {\doibase 10.1103/PhysRevD.65.044001} {\bibfield  {journal} {\bibinfo
  {journal} {Phys. Rev. D}\ }\textbf {\bibinfo {volume} {65}},\ \bibinfo
  {pages} {044001} (\bibinfo {year} {2002})},\ \Eprint
  {http://arxiv.org/abs/arXiv:gr-qc/0104063} {arXiv:gr-qc/0104063} \BibitemShut
  {NoStop}%
\bibitem [{\citenamefont {Damour}\ \emph {et~al.}(2008)\citenamefont {Damour},
  \citenamefont {Nagar}, \citenamefont {Hannam}, \citenamefont {Husa},\ and\
  \citenamefont {Br{\"u}gmann}}]{Damour2008}%
  \BibitemOpen
  \bibfield  {author} {\bibinfo {author} {\bibfnamefont {T.}~\bibnamefont
  {Damour}}, \bibinfo {author} {\bibfnamefont {A.}~\bibnamefont {Nagar}},
  \bibinfo {author} {\bibfnamefont {M.}~\bibnamefont {Hannam}}, \bibinfo
  {author} {\bibfnamefont {S.}~\bibnamefont {Husa}}, \ and\ \bibinfo {author}
  {\bibfnamefont {B.}~\bibnamefont {Br{\"u}gmann}},\ }\href@noop {} {\bibfield
  {journal} {\bibinfo  {journal} {Physical Review D}\ }\textbf {\bibinfo
  {volume} {78}},\ \bibinfo {pages} {044039} (\bibinfo {year}
  {2008})}\BibitemShut {NoStop}%
\bibitem [{\citenamefont {Reisswig}\ and\ \citenamefont
  {Pollney}(2011)}]{Reisswig:2011notes}%
  \BibitemOpen
  \bibfield  {author} {\bibinfo {author} {\bibfnamefont {C.}~\bibnamefont
  {Reisswig}}\ and\ \bibinfo {author} {\bibfnamefont {D.}~\bibnamefont
  {Pollney}},\ }\href@noop {} {\bibfield  {journal} {\bibinfo  {journal}
  {Classical and Quantum Gravity}\ }\textbf {\bibinfo {volume} {28}},\ \bibinfo
  {pages} {195015} (\bibinfo {year} {2011})}\BibitemShut {NoStop}%
\bibitem [{\citenamefont {Baiotti}\ \emph {et~al.}(2009)\citenamefont
  {Baiotti}, \citenamefont {Bernuzzi}, \citenamefont {Corvino}, \citenamefont
  {De~Pietri},\ and\ \citenamefont {Nagar}}]{Baiotti:2009gravitational}%
  \BibitemOpen
  \bibfield  {author} {\bibinfo {author} {\bibfnamefont {L.}~\bibnamefont
  {Baiotti}}, \bibinfo {author} {\bibfnamefont {S.}~\bibnamefont {Bernuzzi}},
  \bibinfo {author} {\bibfnamefont {G.}~\bibnamefont {Corvino}}, \bibinfo
  {author} {\bibfnamefont {R.}~\bibnamefont {De~Pietri}}, \ and\ \bibinfo
  {author} {\bibfnamefont {A.}~\bibnamefont {Nagar}},\ }\href@noop {}
  {\bibfield  {journal} {\bibinfo  {journal} {Physical Review D}\ }\textbf
  {\bibinfo {volume} {79}},\ \bibinfo {pages} {024002} (\bibinfo {year}
  {2009})}\BibitemShut {NoStop}%
\bibitem [{\citenamefont {Berti}\ \emph {et~al.}(2007)\citenamefont {Berti},
  \citenamefont {Cardoso}, \citenamefont {Gonzalez}, \citenamefont {Sperhake},
  \citenamefont {Hannam}, \citenamefont {Husa},\ and\ \citenamefont
  {Br{\"u}gmann}}]{Berti:2007inspiral}%
  \BibitemOpen
  \bibfield  {author} {\bibinfo {author} {\bibfnamefont {E.}~\bibnamefont
  {Berti}}, \bibinfo {author} {\bibfnamefont {V.}~\bibnamefont {Cardoso}},
  \bibinfo {author} {\bibfnamefont {J.~A.}\ \bibnamefont {Gonzalez}}, \bibinfo
  {author} {\bibfnamefont {U.}~\bibnamefont {Sperhake}}, \bibinfo {author}
  {\bibfnamefont {M.}~\bibnamefont {Hannam}}, \bibinfo {author} {\bibfnamefont
  {S.}~\bibnamefont {Husa}}, \ and\ \bibinfo {author} {\bibfnamefont
  {B.}~\bibnamefont {Br{\"u}gmann}},\ }\href@noop {} {\bibfield  {journal}
  {\bibinfo  {journal} {Physical Review D}\ }\textbf {\bibinfo {volume} {76}},\
  \bibinfo {pages} {064034} (\bibinfo {year} {2007})}\BibitemShut {NoStop}%
\bibitem [{\citenamefont {Nakano}\ \emph {et~al.}(2015)\citenamefont {Nakano},
  \citenamefont {Healy}, \citenamefont {Lousto},\ and\ \citenamefont
  {Zlochower}}]{Nakano:2015perturbative}%
  \BibitemOpen
  \bibfield  {author} {\bibinfo {author} {\bibfnamefont {H.}~\bibnamefont
  {Nakano}}, \bibinfo {author} {\bibfnamefont {J.}~\bibnamefont {Healy}},
  \bibinfo {author} {\bibfnamefont {C.~O.}\ \bibnamefont {Lousto}}, \ and\
  \bibinfo {author} {\bibfnamefont {Y.}~\bibnamefont {Zlochower}},\ }\href@noop
  {} {\bibfield  {journal} {\bibinfo  {journal} {Physical Review D}\ }\textbf
  {\bibinfo {volume} {91}},\ \bibinfo {pages} {104022} (\bibinfo {year}
  {2015})}\BibitemShut {NoStop}%
\bibitem [{\citenamefont {Br{\"u}gmann}\ \emph {et~al.}(2008)\citenamefont
  {Br{\"u}gmann}, \citenamefont {Gonz{\'a}lez}, \citenamefont {Hannam},
  \citenamefont {Husa}, \citenamefont {Sperhake},\ and\ \citenamefont
  {Tichy}}]{Brugmann:2008calibration}%
  \BibitemOpen
  \bibfield  {author} {\bibinfo {author} {\bibfnamefont {B.}~\bibnamefont
  {Br{\"u}gmann}}, \bibinfo {author} {\bibfnamefont {J.~A.}\ \bibnamefont
  {Gonz{\'a}lez}}, \bibinfo {author} {\bibfnamefont {M.}~\bibnamefont
  {Hannam}}, \bibinfo {author} {\bibfnamefont {S.}~\bibnamefont {Husa}},
  \bibinfo {author} {\bibfnamefont {U.}~\bibnamefont {Sperhake}}, \ and\
  \bibinfo {author} {\bibfnamefont {W.}~\bibnamefont {Tichy}},\ }\href@noop {}
  {\bibfield  {journal} {\bibinfo  {journal} {Physical Review D}\ }\textbf
  {\bibinfo {volume} {77}},\ \bibinfo {pages} {024027} (\bibinfo {year}
  {2008})}\BibitemShut {NoStop}%
\bibitem [{\citenamefont {Alic}\ \emph {et~al.}(2013)\citenamefont {Alic},
  \citenamefont {Kastaun},\ and\ \citenamefont {Rezzolla}}]{Alic:2013xsa}%
  \BibitemOpen
  \bibfield  {author} {\bibinfo {author} {\bibfnamefont {D.}~\bibnamefont
  {Alic}}, \bibinfo {author} {\bibfnamefont {W.}~\bibnamefont {Kastaun}}, \
  and\ \bibinfo {author} {\bibfnamefont {L.}~\bibnamefont {Rezzolla}},\ }\href
  {\doibase 10.1103/PhysRevD.88.064049} {\bibfield  {journal} {\bibinfo
  {journal} {Phys. Rev.}\ }\textbf {\bibinfo {volume} {D88}},\ \bibinfo {pages}
  {064049} (\bibinfo {year} {2013})},\ \Eprint {http://arxiv.org/abs/1307.7391}
  {arXiv:1307.7391 [gr-qc]} \BibitemShut {NoStop}%
\bibitem [{\citenamefont {Ruiz}\ \emph {et~al.}(2011)\citenamefont {Ruiz},
  \citenamefont {Hilditch},\ and\ \citenamefont {Bernuzzi}}]{Ruiz:2010qj}%
  \BibitemOpen
  \bibfield  {author} {\bibinfo {author} {\bibfnamefont {M.}~\bibnamefont
  {Ruiz}}, \bibinfo {author} {\bibfnamefont {D.}~\bibnamefont {Hilditch}}, \
  and\ \bibinfo {author} {\bibfnamefont {S.}~\bibnamefont {Bernuzzi}},\ }\href
  {\doibase 10.1103/PhysRevD.83.024025} {\bibfield  {journal} {\bibinfo
  {journal} {Phys. Rev.}\ }\textbf {\bibinfo {volume} {D83}},\ \bibinfo {pages}
  {024025} (\bibinfo {year} {2011})},\ \Eprint {http://arxiv.org/abs/1010.0523}
  {arXiv:1010.0523 [gr-qc]} \BibitemShut {NoStop}%
\bibitem [{\citenamefont {Hotokezaka}\ \emph
  {et~al.}(2015{\natexlab{a}})\citenamefont {Hotokezaka}, \citenamefont
  {Kyutoku}, \citenamefont {Okawa},\ and\ \citenamefont
  {Shibata}}]{hotokezaka:2015exploring}%
  \BibitemOpen
  \bibfield  {author} {\bibinfo {author} {\bibfnamefont {K.}~\bibnamefont
  {Hotokezaka}}, \bibinfo {author} {\bibfnamefont {K.}~\bibnamefont {Kyutoku}},
  \bibinfo {author} {\bibfnamefont {H.}~\bibnamefont {Okawa}}, \ and\ \bibinfo
  {author} {\bibfnamefont {M.}~\bibnamefont {Shibata}},\ }\href@noop {}
  {\bibfield  {journal} {\bibinfo  {journal} {Physical Review D}\ }\textbf
  {\bibinfo {volume} {91}},\ \bibinfo {pages} {064060} (\bibinfo {year}
  {2015}{\natexlab{a}})}\BibitemShut {NoStop}%
\bibitem [{\citenamefont {Kurganov}\ and\ \citenamefont
  {Tadmor}(2000)}]{kurganov:2000new}%
  \BibitemOpen
  \bibfield  {author} {\bibinfo {author} {\bibfnamefont {A.}~\bibnamefont
  {Kurganov}}\ and\ \bibinfo {author} {\bibfnamefont {E.}~\bibnamefont
  {Tadmor}},\ }\href@noop {} {\bibfield  {journal} {\bibinfo  {journal}
  {Journal of Computational Physics}\ }\textbf {\bibinfo {volume} {160}},\
  \bibinfo {pages} {241} (\bibinfo {year} {2000})}\BibitemShut {NoStop}%
\bibitem [{\citenamefont {Palenzuela}\ \emph {et~al.}(2015)\citenamefont
  {Palenzuela}, \citenamefont {Liebling}, \citenamefont {Neilsen},
  \citenamefont {Lehner}, \citenamefont {Caballero}, \citenamefont
  {O’Connor},\ and\ \citenamefont {Anderson}}]{palenzuela:2015effects}%
  \BibitemOpen
  \bibfield  {author} {\bibinfo {author} {\bibfnamefont {C.}~\bibnamefont
  {Palenzuela}}, \bibinfo {author} {\bibfnamefont {S.~L.}\ \bibnamefont
  {Liebling}}, \bibinfo {author} {\bibfnamefont {D.}~\bibnamefont {Neilsen}},
  \bibinfo {author} {\bibfnamefont {L.}~\bibnamefont {Lehner}}, \bibinfo
  {author} {\bibfnamefont {O.}~\bibnamefont {Caballero}}, \bibinfo {author}
  {\bibfnamefont {E.}~\bibnamefont {O’Connor}}, \ and\ \bibinfo {author}
  {\bibfnamefont {M.}~\bibnamefont {Anderson}},\ }\href@noop {} {\bibfield
  {journal} {\bibinfo  {journal} {Physical Review D}\ }\textbf {\bibinfo
  {volume} {92}},\ \bibinfo {pages} {044045} (\bibinfo {year}
  {2015})}\BibitemShut {NoStop}%
\bibitem [{\citenamefont {Radice}\ \emph {et~al.}(2015)\citenamefont {Radice},
  \citenamefont {Rezzolla},\ and\ \citenamefont {Galeazzi}}]{Radice:2015nva}%
  \BibitemOpen
  \bibfield  {author} {\bibinfo {author} {\bibfnamefont {D.}~\bibnamefont
  {Radice}}, \bibinfo {author} {\bibfnamefont {L.}~\bibnamefont {Rezzolla}}, \
  and\ \bibinfo {author} {\bibfnamefont {F.}~\bibnamefont {Galeazzi}},\
  }\href@noop {} {\  (\bibinfo {year} {2015})},\ \Eprint
  {http://arxiv.org/abs/1502.00551} {arXiv:1502.00551 [gr-qc]} \BibitemShut
  {NoStop}%
\bibitem [{Note1()}]{Note1}%
  \BibitemOpen
  \bibinfo {note} {The EOB waveform has been obtained through the publicly
  available LALSUIT LIGO/Virgo software
  (``git://versions.ligo.org/lalsuite.git'') using the utility:
  ``./lalsim-inspiral -a EOBNRv2 [options]''.}\BibitemShut {Stop}%
\bibitem [{\citenamefont {Stergioulas}\ \emph {et~al.}(2011)\citenamefont
  {Stergioulas}, \citenamefont {Bauswein}, \citenamefont {Zagkouris},\ and\
  \citenamefont {Janka}}]{stergioulas:2011gravitational}%
  \BibitemOpen
  \bibfield  {author} {\bibinfo {author} {\bibfnamefont {N.}~\bibnamefont
  {Stergioulas}}, \bibinfo {author} {\bibfnamefont {A.}~\bibnamefont
  {Bauswein}}, \bibinfo {author} {\bibfnamefont {K.}~\bibnamefont {Zagkouris}},
  \ and\ \bibinfo {author} {\bibfnamefont {H.-T.}\ \bibnamefont {Janka}},\
  }\href@noop {} {\bibfield  {journal} {\bibinfo  {journal} {Monthly Notices of
  the Royal Astronomical Society}\ }\textbf {\bibinfo {volume} {418}},\
  \bibinfo {pages} {427} (\bibinfo {year} {2011})}\BibitemShut {NoStop}%
\bibitem [{\citenamefont {Hotokezaka}\ \emph
  {et~al.}(2015{\natexlab{b}})\citenamefont {Hotokezaka}, \citenamefont
  {Kyutoku}, \citenamefont {Okawa},\ and\ \citenamefont
  {Shibata}}]{hotokezaka:2013mass}%
  \BibitemOpen
  \bibfield  {author} {\bibinfo {author} {\bibfnamefont {K.}~\bibnamefont
  {Hotokezaka}}, \bibinfo {author} {\bibfnamefont {K.}~\bibnamefont {Kyutoku}},
  \bibinfo {author} {\bibfnamefont {H.}~\bibnamefont {Okawa}}, \ and\ \bibinfo
  {author} {\bibfnamefont {M.}~\bibnamefont {Shibata}},\ }\href {\doibase
  10.1103/PhysRevD.91.064060} {\bibfield  {journal} {\bibinfo  {journal} {Phys.
  Rev.}\ }\textbf {\bibinfo {volume} {D91}},\ \bibinfo {pages} {064060}
  (\bibinfo {year} {2015}{\natexlab{b}})},\ \Eprint
  {http://arxiv.org/abs/1502.03457} {arXiv:1502.03457 [gr-qc]} \BibitemShut
  {NoStop}%
\bibitem [{\citenamefont {Bauswein}\ \emph {et~al.}(2014)\citenamefont
  {Bauswein}, \citenamefont {Stergioulas},\ and\ \citenamefont
  {Janka}}]{bauswein:2014revealing}%
  \BibitemOpen
  \bibfield  {author} {\bibinfo {author} {\bibfnamefont {A.}~\bibnamefont
  {Bauswein}}, \bibinfo {author} {\bibfnamefont {N.}~\bibnamefont
  {Stergioulas}}, \ and\ \bibinfo {author} {\bibfnamefont {H.-T.}\ \bibnamefont
  {Janka}},\ }\href@noop {} {\bibfield  {journal} {\bibinfo  {journal}
  {Physical Review D}\ }\textbf {\bibinfo {volume} {90}},\ \bibinfo {pages}
  {023002} (\bibinfo {year} {2014})}\BibitemShut {NoStop}%
\bibitem [{\citenamefont {Bauswein}\ and\ \citenamefont
  {Janka}(2012)}]{bauswein:2012measuring}%
  \BibitemOpen
  \bibfield  {author} {\bibinfo {author} {\bibfnamefont {A.}~\bibnamefont
  {Bauswein}}\ and\ \bibinfo {author} {\bibfnamefont {H.-T.}\ \bibnamefont
  {Janka}},\ }\href@noop {} {\bibfield  {journal} {\bibinfo  {journal}
  {Physical review letters}\ }\textbf {\bibinfo {volume} {108}},\ \bibinfo
  {pages} {011101} (\bibinfo {year} {2012})}\BibitemShut {NoStop}%
\bibitem [{\citenamefont {Bauswein}\ \emph {et~al.}(2015)\citenamefont
  {Bauswein}, \citenamefont {Stergioulas},\ and\ \citenamefont
  {Janka}}]{bauswein:2015exploring}%
  \BibitemOpen
  \bibfield  {author} {\bibinfo {author} {\bibfnamefont {A.}~\bibnamefont
  {Bauswein}}, \bibinfo {author} {\bibfnamefont {N.}~\bibnamefont
  {Stergioulas}}, \ and\ \bibinfo {author} {\bibfnamefont {H.-T.}\ \bibnamefont
  {Janka}},\ }\href@noop {} {\  (\bibinfo {year} {2015})},\ \Eprint
  {http://arxiv.org/abs/1508.05493} {arXiv:1508.05493} \BibitemShut {NoStop}%
\bibitem [{\citenamefont {Hotokezaka}\ \emph {et~al.}(2013)\citenamefont
  {Hotokezaka}, \citenamefont {Kiuchi}, \citenamefont {Kyutoku}, \citenamefont
  {Muranushi}, \citenamefont {Sekiguchi}, \citenamefont {Shibata},\ and\
  \citenamefont {Taniguchi}}]{hotokezaka:2013remnant}%
  \BibitemOpen
  \bibfield  {author} {\bibinfo {author} {\bibfnamefont {K.}~\bibnamefont
  {Hotokezaka}}, \bibinfo {author} {\bibfnamefont {K.}~\bibnamefont {Kiuchi}},
  \bibinfo {author} {\bibfnamefont {K.}~\bibnamefont {Kyutoku}}, \bibinfo
  {author} {\bibfnamefont {T.}~\bibnamefont {Muranushi}}, \bibinfo {author}
  {\bibfnamefont {Y.-i.}\ \bibnamefont {Sekiguchi}}, \bibinfo {author}
  {\bibfnamefont {M.}~\bibnamefont {Shibata}}, \ and\ \bibinfo {author}
  {\bibfnamefont {K.}~\bibnamefont {Taniguchi}},\ }\href@noop {} {\bibfield
  {journal} {\bibinfo  {journal} {Physical Review D}\ }\textbf {\bibinfo
  {volume} {88}},\ \bibinfo {pages} {044026} (\bibinfo {year}
  {2013})}\BibitemShut {NoStop}%
\bibitem [{\citenamefont {Read}\ \emph {et~al.}(2013)\citenamefont {Read},
  \citenamefont {Baiotti}, \citenamefont {Creighton}, \citenamefont {Friedman},
  \citenamefont {Giacomazzo}, \citenamefont {Kyutoku}, \citenamefont
  {Markakis}, \citenamefont {Rezzolla}, \citenamefont {Shibata},\ and\
  \citenamefont {Taniguchi}}]{read:2013matter}%
  \BibitemOpen
  \bibfield  {author} {\bibinfo {author} {\bibfnamefont {J.~S.}\ \bibnamefont
  {Read}}, \bibinfo {author} {\bibfnamefont {L.}~\bibnamefont {Baiotti}},
  \bibinfo {author} {\bibfnamefont {J.~D.}\ \bibnamefont {Creighton}}, \bibinfo
  {author} {\bibfnamefont {J.~L.}\ \bibnamefont {Friedman}}, \bibinfo {author}
  {\bibfnamefont {B.}~\bibnamefont {Giacomazzo}}, \bibinfo {author}
  {\bibfnamefont {K.}~\bibnamefont {Kyutoku}}, \bibinfo {author} {\bibfnamefont
  {C.}~\bibnamefont {Markakis}}, \bibinfo {author} {\bibfnamefont
  {L.}~\bibnamefont {Rezzolla}}, \bibinfo {author} {\bibfnamefont
  {M.}~\bibnamefont {Shibata}}, \ and\ \bibinfo {author} {\bibfnamefont
  {K.}~\bibnamefont {Taniguchi}},\ }\href@noop {} {\bibfield  {journal}
  {\bibinfo  {journal} {Physical Review D}\ }\textbf {\bibinfo {volume} {88}},\
  \bibinfo {pages} {044042} (\bibinfo {year} {2013})}\BibitemShut {NoStop}%
\bibitem [{\citenamefont {Bernuzzi}\ \emph {et~al.}(2015)\citenamefont
  {Bernuzzi}, \citenamefont {Dietrich},\ and\ \citenamefont
  {Nagar}}]{bernuzzi:2015modeling}%
  \BibitemOpen
  \bibfield  {author} {\bibinfo {author} {\bibfnamefont {S.}~\bibnamefont
  {Bernuzzi}}, \bibinfo {author} {\bibfnamefont {T.}~\bibnamefont {Dietrich}},
  \ and\ \bibinfo {author} {\bibfnamefont {A.}~\bibnamefont {Nagar}},\ }\href
  {\doibase 10.1103/PhysRevLett.115.091101} {\bibfield  {journal} {\bibinfo
  {journal} {Phys. Rev. Lett.}\ }\textbf {\bibinfo {volume} {115}},\ \bibinfo
  {pages} {091101} (\bibinfo {year} {2015})},\ \Eprint
  {http://arxiv.org/abs/1504.01764} {arXiv:1504.01764 [gr-qc]} \BibitemShut
  {NoStop}%
\bibitem [{\citenamefont {Kastaun}\ and\ \citenamefont
  {Galeazzi}(2015)}]{kastaun:2015properties}%
  \BibitemOpen
  \bibfield  {author} {\bibinfo {author} {\bibfnamefont {W.}~\bibnamefont
  {Kastaun}}\ and\ \bibinfo {author} {\bibfnamefont {F.}~\bibnamefont
  {Galeazzi}},\ }\href@noop {} {\bibfield  {journal} {\bibinfo  {journal}
  {Physical Review D}\ }\textbf {\bibinfo {volume} {91}},\ \bibinfo {pages}
  {064027} (\bibinfo {year} {2015})}\BibitemShut {NoStop}%
\bibitem [{\citenamefont {Bauswein}\ and\ \citenamefont
  {Stergioulas}(2015)}]{bauswein:2015unified}%
  \BibitemOpen
  \bibfield  {author} {\bibinfo {author} {\bibfnamefont {A.}~\bibnamefont
  {Bauswein}}\ and\ \bibinfo {author} {\bibfnamefont {N.}~\bibnamefont
  {Stergioulas}},\ }\href {\doibase 10.1103/PhysRevD.91.124056} {\bibfield
  {journal} {\bibinfo  {journal} {Phys. Rev.}\ }\textbf {\bibinfo {volume}
  {D91}},\ \bibinfo {pages} {124056} (\bibinfo {year} {2015})},\ \Eprint
  {http://arxiv.org/abs/1502.03176} {arXiv:1502.03176 [astro-ph.SR]}
  \BibitemShut {NoStop}%
\bibitem [{\citenamefont {Dietrich}\ \emph {et~al.}(2015)\citenamefont
  {Dietrich}, \citenamefont {Bernuzzi}, \citenamefont {Ujevic},\ and\
  \citenamefont {Brügmann}}]{dietrich:2015numerical}%
  \BibitemOpen
  \bibfield  {author} {\bibinfo {author} {\bibfnamefont {T.}~\bibnamefont
  {Dietrich}}, \bibinfo {author} {\bibfnamefont {S.}~\bibnamefont {Bernuzzi}},
  \bibinfo {author} {\bibfnamefont {M.}~\bibnamefont {Ujevic}}, \ and\ \bibinfo
  {author} {\bibfnamefont {B.}~\bibnamefont {Brügmann}},\ }\href {\doibase
  10.1103/PhysRevD.91.124041} {\bibfield  {journal} {\bibinfo  {journal} {Phys.
  Rev.}\ }\textbf {\bibinfo {volume} {D91}},\ \bibinfo {pages} {124041}
  (\bibinfo {year} {2015})},\ \Eprint {http://arxiv.org/abs/1504.01266}
  {arXiv:1504.01266 [gr-qc]} \BibitemShut {NoStop}%
\bibitem [{\citenamefont {Ashtekar}\ \emph {et~al.}(2000)\citenamefont
  {Ashtekar}, \citenamefont {Beetle}, \citenamefont {Dreyer}, \citenamefont
  {Fairhurst}, \citenamefont {Krishnan}, \citenamefont {Lewandowski},\ and\
  \citenamefont {Wisniewski}}]{Ashtekar:2000sz}%
  \BibitemOpen
  \bibfield  {author} {\bibinfo {author} {\bibfnamefont {A.}~\bibnamefont
  {Ashtekar}}, \bibinfo {author} {\bibfnamefont {C.}~\bibnamefont {Beetle}},
  \bibinfo {author} {\bibfnamefont {O.}~\bibnamefont {Dreyer}}, \bibinfo
  {author} {\bibfnamefont {S.}~\bibnamefont {Fairhurst}}, \bibinfo {author}
  {\bibfnamefont {B.}~\bibnamefont {Krishnan}}, \bibinfo {author}
  {\bibfnamefont {J.}~\bibnamefont {Lewandowski}}, \ and\ \bibinfo {author}
  {\bibfnamefont {J.}~\bibnamefont {Wisniewski}},\ }\href {\doibase
  10.1103/PhysRevLett.85.3564} {\bibfield  {journal} {\bibinfo  {journal}
  {Phys. Rev. Lett.}\ }\textbf {\bibinfo {volume} {85}},\ \bibinfo {pages}
  {3564} (\bibinfo {year} {2000})},\ \Eprint
  {http://arxiv.org/abs/gr-qc/0006006} {arXiv:gr-qc/0006006 [gr-qc]}
  \BibitemShut {NoStop}%
\bibitem [{\citenamefont {Ashtekar}\ \emph {et~al.}(2002)\citenamefont
  {Ashtekar}, \citenamefont {Beetle},\ and\ \citenamefont
  {Lewandowski}}]{Ashtekar:2001jb}%
  \BibitemOpen
  \bibfield  {author} {\bibinfo {author} {\bibfnamefont {A.}~\bibnamefont
  {Ashtekar}}, \bibinfo {author} {\bibfnamefont {C.}~\bibnamefont {Beetle}}, \
  and\ \bibinfo {author} {\bibfnamefont {J.}~\bibnamefont {Lewandowski}},\
  }\href {\doibase 10.1088/0264-9381/19/6/311} {\bibfield  {journal} {\bibinfo
  {journal} {Class. Quant. Grav.}\ }\textbf {\bibinfo {volume} {19}},\ \bibinfo
  {pages} {1195} (\bibinfo {year} {2002})},\ \Eprint
  {http://arxiv.org/abs/gr-qc/0111067} {arXiv:gr-qc/0111067 [gr-qc]}
  \BibitemShut {NoStop}%
\bibitem [{\citenamefont {Ashtekar}\ and\ \citenamefont
  {Krishnan}(2004)}]{Ashtekar:2004cn}%
  \BibitemOpen
  \bibfield  {author} {\bibinfo {author} {\bibfnamefont {A.}~\bibnamefont
  {Ashtekar}}\ and\ \bibinfo {author} {\bibfnamefont {B.}~\bibnamefont
  {Krishnan}},\ }\href {\doibase 10.12942/lrr-2004-10} {\bibfield  {journal}
  {\bibinfo  {journal} {Living Rev. Rel.}\ }\textbf {\bibinfo {volume} {7}},\
  \bibinfo {pages} {10} (\bibinfo {year} {2004})},\ \Eprint
  {http://arxiv.org/abs/gr-qc/0407042} {arXiv:gr-qc/0407042 [gr-qc]}
  \BibitemShut {NoStop}%
\bibitem [{\citenamefont {Dreyer}\ \emph {et~al.}(2003)\citenamefont {Dreyer},
  \citenamefont {Krishnan}, \citenamefont {Shoemaker},\ and\ \citenamefont
  {Schnetter}}]{Dreyer:2002mx}%
  \BibitemOpen
  \bibfield  {author} {\bibinfo {author} {\bibfnamefont {O.}~\bibnamefont
  {Dreyer}}, \bibinfo {author} {\bibfnamefont {B.}~\bibnamefont {Krishnan}},
  \bibinfo {author} {\bibfnamefont {D.}~\bibnamefont {Shoemaker}}, \ and\
  \bibinfo {author} {\bibfnamefont {E.}~\bibnamefont {Schnetter}},\ }\href
  {\doibase 10.1103/PhysRevD.67.024018} {\bibfield  {journal} {\bibinfo
  {journal} {Phys. Rev. D}\ }\textbf {\bibinfo {volume} {67}},\ \bibinfo
  {pages} {024018} (\bibinfo {year} {2003})},\ \Eprint
  {http://arxiv.org/abs/arXiv:gr-qc/0206008} {arXiv:gr-qc/0206008} \BibitemShut
  {NoStop}%
\bibitem [{\citenamefont {Thornburg}(2004)}]{Thornburg:2003sf}%
  \BibitemOpen
  \bibfield  {author} {\bibinfo {author} {\bibfnamefont {J.}~\bibnamefont
  {Thornburg}},\ }\href {\doibase 10.1088/0264-9381/21/2/026} {\bibfield
  {journal} {\bibinfo  {journal} {Class. Quantum Grav.}\ }\textbf {\bibinfo
  {volume} {21}},\ \bibinfo {pages} {743} (\bibinfo {year} {2004})},\ \Eprint
  {http://arxiv.org/abs/arXiv:gr-qc/0306056} {arXiv:gr-qc/0306056} \BibitemShut
  {NoStop}%
\bibitem [{Note2()}]{Note2}%
  \BibitemOpen
  \bibinfo {note} {See the University of Parma Gravity group WEB page: \protect
  \href
  {http://www.fis.unipr.it/gravity/Research/BNS2015.html}{www.fis.unipr.it/gravity/Research/BNS2015.html}.}\BibitemShut
  {Stop}%
\end{thebibliography}
\end{document}